\documentclass{siamltex1213}
\usepackage{hyperref}
\usepackage{graphicx}
\usepackage{amsfonts}
\usepackage{amsmath}
\usepackage[usenames,dvipsnames]{xcolor}
\usepackage{soul}

\DeclareGraphicsExtensions{.pdf,.png}

\newcommand{\eps}{\varepsilon}
\newcommand{\la}{\lambda}

\newcommand{\K}{\mathrm{K}}
\newcommand{\E}{\mathrm{E}}

\newcommand{\sgn}{\mathrm{sgn}}
\newcommand{\beq}{\begin{equation}}
\newcommand{\eeq}{\end{equation}}
\newcommand{\ud}{\,\mathrm{d}}
\newcommand{\R}{\mathbb{R}}
\def\barr{\begin{array}}
\def\earr{\end{array}}
\def\half{{\textstyle {1\over 2}}}

\newcommand{\Mnote}[1]{{#1}}
\newcommand{\MH}[1]{{#1}}

\title{DISPERSIVE AND DIFFUSIVE-DISPERSIVE SHOCK WAVES FOR NON-CONVEX
 CONSERVATION LAWS\thanks{This work was supported by a Royal Society
   International Exchanges Scheme  Grant IE131353.  The authors thank G. Esler, E. Ferapontov and A. Kamchatnov for valuable discussions.}}

\author{G.~A. El\footnotemark[2], M.~A.
  Hoefer\footnotemark[3],
  and M. Shearer\footnotemark[4]}

\begin{document}
\maketitle

\renewcommand{\thefootnote}{\fnsymbol{footnote}}

\footnotetext[2]{Department of Mathematical Sciences, Loughborough
  University, UK}
%\ead{g.el@lboro.ac.uk}
\footnotetext[3]{Department of Applied Mathematics, University of
  Colorado, Boulder, USA; supported by NSF CAREER award DMS-1255422}
%\ead{hoefer@colorado.edu}
\footnotetext[4]{Department of Mathematics, North Carolina State
  University, Raleigh, NC 27695, USA; supported by NSF grant }
%\ead{shearer@ncsu.edu}

\pagestyle{myheadings}
\thispagestyle{plain}
\markboth{G.~A. EL, M.~A. HOEFER, AND M. SHEARER}{DISPERSIVE AND
  DIFFUSIVE-DISPERSIVE SHOCK WAVES} 

\begin{abstract}
  We consider two physically and mathematically distinct
  regularization mechanisms of scalar hyperbolic conservation laws.
  When the flux is convex, the combination of diffusion and dispersion
  are known to give rise to monotonic and oscillatory traveling waves
  that approximate shock waves.  The zero-diffusion limits of these
  traveling waves are dynamically expanding dispersive shock waves
  (DSWs).  A richer set of wave solutions can be found when the flux
  is non-convex.  This review compares the structure of solutions of
  Riemann problems for a conservation law with non-convex, cubic flux
  regularized by two different mechanisms: 1) dispersion in the
  modified Korteweg--de Vries (mKdV) equation; and 2) a combination of
  diffusion and dispersion in the mKdV-Burgers equation. In the first
  case, the possible dynamics involve two qualitatively different
  types of DSWs, rarefaction waves (RWs) and kinks (monotonic
  fronts). In the second case, in addition to RWs, there are traveling
  wave solutions approximating both classical (Lax) and non-classical
  (undercompressive) shock waves. Despite the singular nature of the
  zero-diffusion limit and rather differing analytical approaches
  employed in the descriptions of dispersive and diffusive-dispersive
  regularization, the resulting comparison of the two cases reveals a
  number of striking parallels.  In contrast to the case of convex
  flux, the mKdVB to mKdV mapping is not one-to-one.  The mKdV kink
  solution is identified as an undercompressive DSW. Other prominent
  features, such as shock-rarefactions, also find their purely
  dispersive counterparts involving special contact DSWs, which
  exhibit features analogous to contact
  discontinuities.     This review describes an important link between two major areas of
  applied mathematics, hyperbolic conservation laws and nonlinear
  dispersive waves.
\end{abstract}

%\pacs{}
% \ams{MSC}

% \submitto{\NL}

\section{Introduction}

Shock waves represent one of the most recognizable features of
nonlinear wave systems.  In their simplest form, shock waves can be
modeled as discontinuous, weak solutions of scalar conservation laws
in one space dimension $x$ and time $t$ 
\begin{equation}
  \label{cons1}
  u_t+f(u)_x=0 .
\end{equation}
When the flux function $f$ is convex, shock admissibility rests upon
an entropy condition that uniquely selects the physically relevant
weak solution \cite{lax2}.  The existence of entropy presupposes some
irreversible dissipative mechanism, e.g., diffusion, in the underlying
physical medium.  A simple example of a diffusively regularized
conservation law is Burgers' equation 
\begin{equation}
  \label{eq:14}
  u_t + \frac{1}{2}(u^2)_x = \nu u_{xx},
\end{equation}
where $\nu > 0$ is a constant representing the diffusion strength.

However, in conservative physical systems whose dominant shock
regularizing mechanism is dispersion, there is no corresponding notion
of entropy.  Discontinuous, weak solutions of \eqref{cons1} no longer
accurately model wavebreaking dynamics.  An alternative theory of
dispersive shock waves (DSWs) \cite{scholarpedia} seeks to examine the
asymptotic structure of solutions of dispersive regularizations of
(\ref{cons1}), the simplest example being the Korteweg--de Vries (KdV)
equation
\begin{equation}
  \label{kdv1} 
  u_t+\frac{1}{2}(u^2)_x=\mu u_{xxx},
\end{equation}
in which $\mu\neq 0$ is a constant.

The Burgers' \eqref{eq:14} and KdV \eqref{kdv1} equations represent
two distinctly different ways to regularize the conservation law
\eqref{cons1} with the convex flux $f(u) = \frac{1}{2}u^2$.  When the
flux in \eqref{cons1} is non-convex, new wave features emerge
including composite wave solutions and undercompressive shocks that do
not satisfy the Lax entropy condition \cite{LeFloch}.  % While
% the theory of Lax and Oleinik addresses shock wave admissibility in
% terms of analysis and geometry,
Many authors have approached the issue of shock wave admissibility
when the flux is non-convex by introducing criteria based on mechanics
or physics, notably through kinetic relations and nucleation
conditions \cite{abeyaratne,shearer_lefloch}.  An additional approach
includes regularization of the equation by higher order terms that are
dissipative and dispersive, representing more of the physics than is
contained in the conservation law (\ref{cons1}) alone
\cite{bertozzi_thinfilms,LeFloch}.  The analysis of shock dynamics in
dispersive equations is in many respects quite different from the
theory of conservation laws, as exemplified by the famous book of
Whitham \cite{wh65}, the pioneering work of Gurevich and Pitaevskii
\cite{gp74} and the series of papers of Lax and Levermore
\cite{lax_levermore}.  Much less is known about dispersive shock waves
in equations with non-convex flux.

The aim of this paper is to review and compare results of these two
regularization approaches for shock wave solutions of
(\ref{cons1}). The comparison is well known for the case of convex
flux functions (see, e.g., \cite{ha2006}), but it reveals a number of
new nontrivial parallels and contrasts if $f$ is non-convex.  This
comparison serves two purposes: 1) it identifies the dispersive theory
as a singular limit of the diffusive-dispersive theory, 2) it serves
as a bridge between dispersive nonlinear waves and hyperbolic
conservation laws, two major areas of applied mathematics.  As far as
possible, dispersive shock waves are described in the language of
hyperbolic conservation laws (characteristics, admissibility criteria,
non-classical, undercompressive shocks) and, similarly,
diffusive-dispersive shocks are explored in the language of dispersive
shock theory (shock polarity and orientation).  Most of the results
presented here exist in the literature.  The purpose of this review is
to distill them into a cohesive narrative that contrasts and unifies
two areas of applied mathematics.

We focus on the modified KdV-Burgers (mKdVB) equation
\begin{equation}
  \label{mkdvb1} 
  u_t+ (u^3)_x=\nu u_{xx}+\mu u_{xxx},
\end{equation}
in which $\nu\geq 0$ and $\mu\neq 0$ are parameters.  When $\nu=0,$
the equation is purely dispersive, and we refer to the equation using
the abbreviation mKdV. For $\nu>0,$ the equation has both diffusion
and dispersion and we refer to it as the mKdVB equation.  The modified
equation is considered because the combination of non-convex flux with
dispersion yields a rich collection of wave solutions. Furthermore,
eq.~\eqref{mkdvb1} can be viewed as a universal model of
diffusive-dispersive nonlinear wave dynamics under a certain
criticality condition.  We demonstrate this by deriving \eqref{mkdvb1}
from a class of diffusive-dispersive Euler equations.

The applied significance of purely dispersive conservation law
regularization is twofold. Firstly, DSW dynamics describe the initial
stage of shock development in media with both regularization
mechanisms present (subject to an appropriate scale separation of
diffusive and dispersive effects). However, the most interesting and
rich applications of DSW theory arise in media whose regularization
dynamics are dominated by dispersion alone.  Prominent examples of
such media are Bose-Einstein condensates and nonlinear optical media
(see, e.g., \cite{ha2006}, \cite{rothenberg1989}, \cite{wan_etal_2007}
for relevant experimental and theoretical DSW studies), although
conservative nonlinear-dispersive dynamics are also common in water
waves and the atmosphere, where DSW-like structures, termed undular
bores, have been observed \cite{smyth_holloway}, \cite{christie89}.

\subsection{Scalar conservation laws}
\label{sec:scal-cons-laws}

The study of conservation laws goes back to Riemann's analysis of the
classic shock tube problem in which a membrane separating gas under
different pressures is ruptured, generating a shock wave and a
rarefaction wave propagating in opposite directions away from the
membrane. This problem is simply represented by the nonlinear system
(\ref{eulerd1}) of isentropic gas dynamics (see
\S\ref{sec:kdvb-mkdvb-equations}), in the diffusion and
dispersion-free limit ($\nu=0, D=0$).  The study of such hyperbolic
systems in one space dimension and time has progressed rapidly since
the 1950's, with early work of Lax, Oleinik and others establishing
existence of solutions of Riemann problems, shock formation and
long-time behavior of solutions, for initial data with small
variation. The seminal paper of Glimm \cite{glimm} injected profound
new ideas that were used to prove existence of solutions of initial
value problems.  As in Glimm's paper, the method now referred to as
wave front tracking \cite{bressan,dafermos} relies on solutions of
Riemann problems to approximate weak solutions of hyperbolic
systems. Development of the theory in more than one space dimension
has been more difficult, but papers of Majda \cite{majda1,majda2} have
led the way to further refinements \cite{zumbrun}.
    
 The theory of scalar conservation laws is much more complete than for systems.  For convex fluxes, the theory of initial value problems in one space dimension is classical  \cite{lax1}. The existence and uniqueness results of Kruzhkov \cite{kruzkov} in any number of dimensions rely on entropy conditions. 
 
A function $u=u(x,t)$ in $L^\infty(\R\times\R^+)$  is a weak solution 
of the initial value problem 
$$
u_t+f(u)_x=0, \quad t>0, \ x\in \R \qquad u(x,0)=g(x), \quad x\in \R,
$$ 
if for every test function $\phi\in C^1(\R\times\R^+)$ with compact support, we have
\beq\label{weak_soln}
\int_0^\infty\int_{-\infty}^\infty (u\phi_t+f(u)\phi_x) \, dx dt +\int_{-\infty}^\infty g(x)\phi(x,0)\,dx= 0.
\eeq
 Here, $\R^+=[0,\infty),$ so that the support of $\phi$ can include a
 portion of the initial line $\R\times\{t=0\},$ and $\phi$ can be
 non-zero there. When $f$ is smooth and uniformly convex (i.e., there
 is $\theta>0 $ such that $f''(u)\geq\theta$ for all $u\in \R$), the
 initial value problem has a unique weak solution whose shocks satisfy the
 Rankine-Hugoniot conditions (see eq.~\eqref{rh} below), but also an entropy condition of Oleinik \cite{Oleinik1, Oleinik2}: there is a constant $C>0$ with the property that 
$$%\label{entropy}
 u(x+z,t)-u(x,t)\leq C(1+1/t)z \quad \mbox{for almost all}\ \ x\in \R, \ z>0, \ t>0.
$$
A readable account of the proof of this result is given in the text of Evans \cite{evans}. DiPerna \cite{diperna} proved asymptotic results for large $t$ for entropy solutions of systems of equations, following earlier analysis of Glimm and Lax \cite{glimm2}. The DiPerna results for scalar equations are proved succinctly  in Evans' text.

Another   analysis approach is to modify the equations with terms that make initial value problems well posed in suitable function spaces.
%in spaces of smooth functions. 
Typically this regularization will be controlled by a small parameter, $\epsilon,$ say; when $\epsilon=0$, the equations revert to the original hyperbolic equations. This is generally a singular limit and the convergence is to a function that is less smooth than the approximating solutions. The analysis therefore proceeds in two steps, the first being to show convergence in a suitable topology, the second to show that the limiting function is indeed a weak solution. Additional steps may be needed to show uniqueness of the limit, and to show that the solution satisfies appropriate side conditions such as entropy conditions. Even for scalar equations, the nature of the regularization, and convexity or lack of convexity of the flux, 
play crucial  roles in this approach. 

In this paper, we focus on two kinds of regularization, involving either purely dispersive terms, or a combination of dispersive and diffusive terms. For scalar equations with convex flux, these two cases yield strikingly different approximations of shock waves; the differences become even more profound as the flux becomes non-convex. However, there is also considerable correspondence between the structure of regularized shocks in the purely dispersive case and the structure in the case of mixed dispersive-diffusive regularization.

\subsection{DSWs and modulation equations}
\label{sec:dsws-modul-equat}

\begin{figure}
  \centering
  \includegraphics{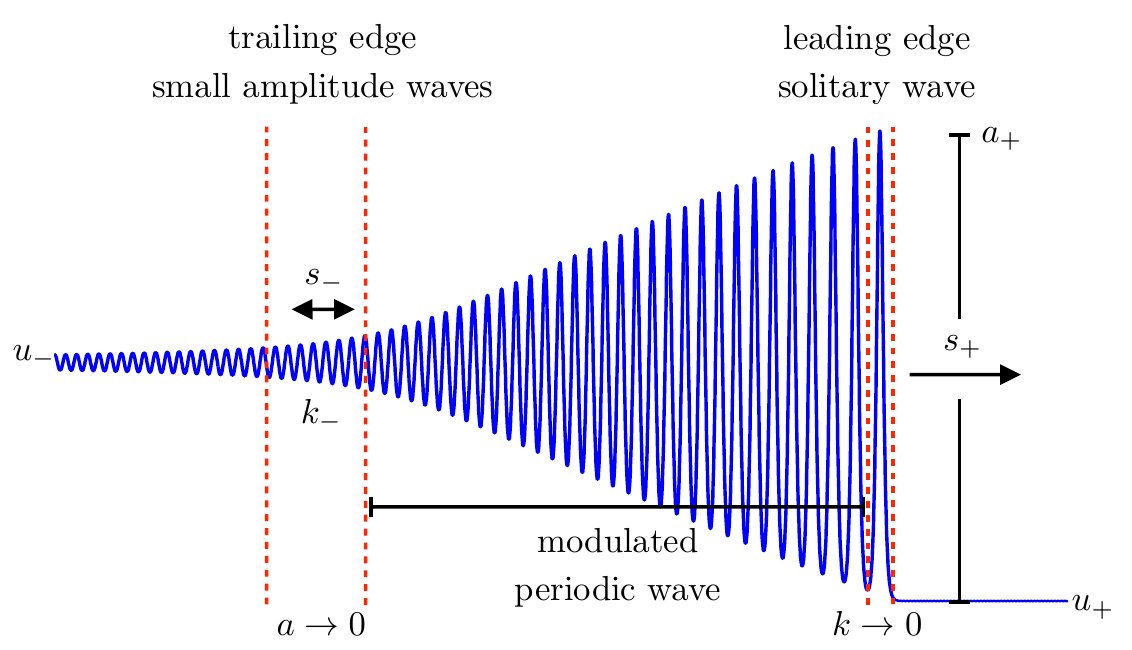}
  \caption{The structure of a DSW with negative dispersion. The amplitude and wave number within the modulated wave are denoted $a(x,t), k(x,t),$ respectively, with limits indicated at the leading and trailing parts of the DSW. The wave speeds $s_+, s_-$ refer respectively to the leading solitary wave, and to the trailing waves. The speed $s_-$ may be positive or negative.
  %Structure of DSW in a medium with negative dispersion
  }
  \label{fig:dsw_schematic}
\end{figure}

A DSW manifests as a nonlinear, expanding wavetrain connecting two
disparate states of slowly varying (or constant) flow.  When viewed
locally, i.e., over a small region of space and time, DSWs display
periodic, or quasiperiodic structure, forming due to the interplay
between nonlinear and dispersive effects. However, over a larger
region covering multiple wave oscillations, the DSW wavetrain reveals
slow modulation of the wave's parameters (amplitude, frequency, mean
etc.), and this modulation itself behaves as a nonlinear hyperbolic
wave.  This kind of ``dispersive-hyperbolic'' duality of modulated
waves is familiar from linear wave theory, but DSWs prominently
display it over a full range of nonlinearity, from a weakly nonlinear
regime to solitary waves realized as an integral part of the modulated
wave train (see figure \ref{fig:dsw_schematic}). The striking
manifestation of nonlinearity in a DSW is that it is characterized by
at least two distinct speeds of propagation, those of its leading and
trailing edges.  In contrast, energy transport by a linear wave packet
is described by a single group velocity.  The unsteady dynamics of
DSWs have far-reaching physical and mathematical implications, among
which are the principal inapplicability of the classical
Rankine-Hugoniot relations and inseparability of the macroscopic DSW
dynamics from the analysis of its nonlinear oscillatory structure.

The mathematical description of DSWs involves a synthesis of methods
from hyperbolic quasi-linear systems, asymptotics, and soliton theory.
One of the principal tools is nonlinear wave modulation theory, often
referred to as Whitham averaging (Whitham 1965).  Whitham theory was
first applied to DSWs in the framework of the KdV equation by Gurevich
and Pitaevskii \cite{gp74}. The Gurevich-Pitaevskii approach rests
upon the fundamental assumption that the DSW can be asymptotically
represented as a slowly modulated periodic traveling wave solution of
the original nonlinear dispersive equation, e.g., the KdV equation
\eqref{kdv1}, where the spatio-temporal modulation scale (variations
of the wave amplitude, wavelength, mean etc.) is assumed to be much
greater than the dispersion length, the wavelength scale of the
carrier, locally periodic traveling wave (see figure
\ref{fig:dsw_schematic}). This scale separation enables one to
effectively split the DSW description problem into two separate tasks
of different complexity: the relatively easy problem of the periodic
traveling wave solution description and the harder problem of finding
an appropriate modulation which provides a match between the modulated
wave train (the DSW) with the smooth, slowly varying external flow.

Following Whitham \cite{wh65}, the equations describing slow
modulations of periodic nonlinear waves can be obtained by averaging
conservation laws of the original dispersive equations over the family
of periodic traveling wave solutions.  Given an $N$-th order nonlinear
evolution equation, implementation of the Whitham method requires the
existence of a $N$-parameter family of periodic traveling wave
solutions $\phi(\theta; {\bf u})$ with phase $\theta=x-Ut$, phase
velocity $U({\bf u})$, and parameters ${\bf u} \in \mathbb{R}^N$.
Additionally, the evolution equation must admit at least $N-1$
conserved densities $\mathcal{P}_i[\phi]$ and fluxes
$\mathcal{Q}_i[\phi]$, $i=1, \dots N-1$ corresponding to the local
conservation laws
\begin{equation}
  \label{1cl}
  \frac{\partial}{\partial t}\mathcal{P}_i + \frac{\partial}{\partial
    x}\mathcal{Q}_i =0\, , \quad i=1, \dots, N-1 . 
\end{equation}
We note that in the majority of known, physically relevant
dispersive-hydrodynamic systems, the number of parameters
characterizing periodic waves is $N=3$ or $N=4$, although systems
yielding $N>4$ can arise in some applications (see
\cite{el_hoefer_2016}).  For the KdV and mKdV equations, $N=3$ while
for the more general case of Eulerian dispersive hydrodynamics
considered in Section \ref{sec:kdvb-mkdvb-equations}, $N=4$.

Assuming slow evolution of the wave's parameters ${\bf u} = {\bf
  u}(x,t)$ on spatio-temporal scales much larger than the wave's
wavelength $L({\bf u})$ and period $T(\mathbf{u}) =
L/U$, the conservation laws are then averaged
over a wavelength resulting in the modulation equations
\begin{equation}
  \label{1wh}
  \left(\frac{1}{L} \int_0^L \mathcal{P}_i[\phi(\theta;{\bf u})] d
    \theta\right)_t + \left(\frac{1}{L} \int_0^L
    \mathcal{Q}_i[\phi(\theta;{\bf u})] d \theta \right)_x = 0\, ,
  \quad i=1, \dots, N-1 \, . 
\end{equation}

The $N-1$ averaged local conservation equations (\ref{1wh}) are
completed by the addition of the conservation of waves
\begin{equation}\label{1cw}
  k_t+\omega_x=0\, ,
\end{equation}
where $k({\bf u})=2\pi/L $ and $\omega({\bf u})=kU = 2\pi/T$ are the
wavenumber and frequency of the nonlinear traveling wave,
respectively.  Equation (\ref{1cw}) represents a consistency condition
for the application of modulation theory.  Note that the system
(\ref{1wh}), (\ref{1cw}), unlike (\ref{1cl}), is a system of {\it
  non-dispersive} conservation laws, which, assuming non-vanishing of
the relevant Jacobians, can be represented in the standard form of a
system of first order quasilinear equations
\begin{equation}\label{1modeq}
 {\bf u}_t +  \mathrm{A}({\bf u})  {\bf u}_x  =0\, .
\end{equation}
The matrix $\mathrm{A}({\bf u})$ encodes information about both the
nonlinear and dispersion properties of the original evolution
equation.  An alternative version of the Whitham method, leading to the
same set of modulation equations (\ref{1modeq}), employs an
averaged variational principle \cite{wh74}.

In some special but important cases, the system (\ref{1modeq}) can be
reduced to diagonal form
\begin{equation}
  \label{Riemann}
  \frac{\partial  r_i}{\partial t} + V_i({\bf r}) \frac{\partial
    r_i}{\partial x}  =0 \, , \quad i=1, \dots, N \, , 
\end{equation}
where $r_i(u_1, \dots, u_N)$ are Riemann invariants, and the
characteristic speeds $V_i({\bf r})$ are eigenvalues of the Jacobian
matrix $\mathrm{A}({\bf u ({r})})$. The existence of Riemann
invariants dramatically simplifies the modulation equation analysis
but is only guaranteed if $N \le 2$.  Remarkably, Whitham found
Riemann invariants for the modulation system associated with periodic
solutions of the KdV equation \eqref{kdv1} where $N=3$ \cite{wh65}. A
similar set of Riemann invariants was found in \cite{dron76} for the
mKdV equation connected with the KdV equation by the Miura transform
\cite{miura}.  Later it was shown that the availability of Riemann
invariants for the KdV-Whitham system is intimately linked to the
integrable structure of the KdV equation via the inverse spectral
transform \cite{flaschka1980}, \cite{krichever1988}. Subsequently, Riemann invariants were
found for other modulation systems associated with integrable
equations such as the nonlinear Schr\"odinger equation
\cite{forest1986}, \cite{pavlov1987}, the Kaup-Boussinesq system
\cite{el2001} and others. We note that the notion of hyperbolicity
plays an important role in the analysis of modulation systems: for
strictly hyperbolic systems, the mapping ${\bf u} \mapsto {\bf r}$ is
invertible and the characteristic speeds are distinct $V_1 < V_2 <
\dots < V_N$. We also note a useful representation of the
characteristic speeds \cite{gurevich91}, \cite{kudashev1991}, \cite{kamch2000}
\begin{equation}
  \label{potentialV}
  V_i= \left[ 1-\frac{L}{\frac{\partial L}{\partial
      r_i}}\frac{\partial}{\partial r_i} \right ] U({\bf r}) \, ,  
\end{equation}
which follows from the consideration of the wave conservation equation
(\ref{1cw}) as a consequence of the Riemann system (\ref{Riemann}).

The Whitham method can be viewed as a field-theoretic analogue of the
Krylov-Bogoliubov averaging method known in the theory of ordinary
differential equations (ODEs). As a matter of fact, the modulation
equations (\ref{1modeq}) can also be derived via a rigorous
multiple-scale perturbation procedure, an appropriate nonlinear analog
of the classical WKB method involving an asymptotic expansion in a
formal small parameter characterizing the scale separation. The
modulation equations then naturally arise as the solvability
conditions, in the space of periodic functions, for the first-order
approximation.  The leading order approximation yields the periodic
solution itself.  This approach, however, lacks directness and
explicitness of the averaging method and can involve rather cumbersome
calculations, especially in non-integrable cases.

It should be stressed that the Whitham system (\ref{1modeq}) describes
modulations of {\it nonlinear} periodic waves. In the hyperbolic case,
the matrix $\mathrm{A}({\bf u})$ generically has $N$ distinct real
eigenvalues, which are the system's characteristic speeds,
sometimes referred to as {\it nonlinear group speeds}. In the
limit of vanishing amplitude, two of these characteristic speeds
merge together and degenerate into the {\it linear group velocity}
$\omega_0'(k)$, where $\omega=\omega_0(k)$ is the dispersion relation
of the linearized original evolution equation. The splitting of the
linear group velocity for modulated waves of finite amplitude is one of the most important results of the Whitham theory
(see \cite{wh74} Ch.~15.4).  This effect leads to the expanding
structure of DSWs.

In the majority of dispersive hydrodynamic problems involving DSW
formation, the modulation systems are hyperbolic, ensuring
modulational stability of the DSW wavetrain.  Hyperbolicity of the
modulation system also enables use of the well developed theory of
quasilinear systems, such as the method of characteristics, which is
not available for the original, dispersive evolution equation.

By applying the Whitham method, the asymptotic description of a DSW
effectively reduces to the integration of the quasilinear modulation
system (\ref{1modeq}) with certain initial/boundary conditions.  These
are formulated in terms of a continuous matching of the averaged, mean
flow in the DSW region with the smooth external flow at boundaries
that are determined along with the solution \cite{gp74,el05}.  This
nonlinear free boundary problem and its extensions are now commonly
known as the Gurevich-Pitaevskii (GP) matching regularization.
Importantly, the Whitham equations subject to the GP matching
conditions admit a global solution describing the modulations in an
expanding DSW. The simplest, yet very important and representative,
example of such a solution was obtained in the original paper
\cite{gp74} for a prototypical problem of the dispersive
regularization of an initial step, the Riemann problem, for the KdV
equation. The DSW modulation solution was found to be a rarefaction
wave solution of the Whitham equations.  For a more general case of
monotone decreasing initial data, the existence of a global modulation
solution describing the KdV DSW was proved in \cite{tian93}.

The GP asymptotic DSW construction procedure for the case of the
Riemann problem has been extended to non-integrable systems by a
priori assuming the existence of a rarefaction wave solution to the
Whitham modulation equations \cite{el05}.  This assumption and the GP
matching conditions yield the key, physical features of the DSW
including the leading and trailing edge speeds, solitary wave edge
amplitude, and small amplitude wave edge modulation wavenumber (see
figure \ref{fig:dsw_schematic}).

Thus, within the context of the Whitham modulation approach, the
dispersive regularization of wave breaking in a hydrodynamic system is
implemented by the introduction of additional hyperbolic conservation
laws (\ref{1wh}), (\ref{1cw}) augmented by certain free boundary
matching conditions.  This procedure is a dispersive counterpart to
the classical  regularization of viscous shocks, where the
inviscid conservation laws are augmented by the Rankine-Hugoniot
relations and entropy conditions.

The modulation description of a DSW is formal in the sense that it is
based on certain assumptions about the asymptotic structure of the DSW
solution and thus, does not resolve the initial DSW development. Its
rigorous validation requires the availability of the exact
solution, and thus, is only possible for completely integrable
systems.  The modulation regularization procedure has been rigorously
justified for fundamental integrable models such as the KdV and
defocusing Nonlinear Schr\"{o}dinger equations in the framework of the
zero dispersion limit of the Inverse Scattering Transform (IST)
developed by Lax, Levermore and Venakides
\cite{lax_levermore,venakides1985,jin1999}.  In non-integrable cases,
the validation of modulation theory is usually made via comparison
with numerical solutions of the original dispersive initial or
boundary value problem.  We note that, even for integrable equations,
the direct application of modulation theory to the description of a
DSW often has a number of advantages to the complex IST based
approaches.  Being quasilinear hyperbolic equations, the modulation
equations do not have any global restrictions imposed on the solution,
e.g., rapid decay as $|x| \to \infty$ or spatial periodicity.  The
application of rigorous methods based on the IST usually involve such
limitations.  Also, the applicability of the Whitham method to
non-integrable equations makes the modulation regularization
especially attractive for physical applications.

Yet, the modulation approach has its own limitations. In particular,
the results of modulation theory have an inherent long-time asymptotic
character and do not yield important information about the initial
stages of DSW development.  A short time asymptotic theory has
recently been developed by Dubrovin \cite{dubrovin2006}, which seeks
to identify the universal oscillatory structure of dispersive
hydrodynamic wave breaking.  Another limitation to modulation theory
is the fundamental assumption about the DSW structure as a locally
periodic wave, which can be violated in some regions of the
$(x,t)$-plane, in particular near the DSW edges \cite{grava_klein2007,
  grava_klein2008}.  Other approaches such as rigorous Riemann-Hilbert
steepest descent methods \cite{deift1997}, \cite{teschl_2013} as well
as the combination of IST with matched asymptotic expansions
\cite{ablowitz2013} can yield additional information about the
detailed structure of the oscillations.  However, for the majority of
applications, the results of modulation theory usually provide the
sought for qualitative and quantitative information of DSW behavior.
The principal feature of the modulation approach is that it provides a
unified framework for the analysis of dispersive regularization of
nonlinear wave breaking singularities within the well established
context of the theory of one-dimensional quasilinear systems.

\subsection{Overview of this work}
\label{sec:outline-this-work}

In this paper, we compare solutions of the purely dispersive mKdV
equation to solutions of the diffusive-dispersive mKdVB equation with
comparable initial data. We find quite remarkable similarities, but
also striking differences in the structure of solutions. It is also of
interest that the comparison depends on the sign of the dispersion
coefficient $\mu$; for both mKdV and mKdVB equations, the sign of
$\mu$ affects the structure of some solutions.  In each case ($\mu>0,
\ \mu<0$), we classify solutions of initial value problems with jump
initial data: \beq\label{RP} u(x,0)=\left\{\begin{array}{ll}
    u_-, \quad & x<0\\[6pt]
    u_+, &x>0
\end{array}
\right.
\eeq
according to the values of the constants $u_-$, $u_+.$ We refer to
these special initial value problems as {\em Riemann problems}. In our
comparison of solutions of Riemann problems, we consider times long
enough that sustained wave patterns have emerged. These are the
long-time asymptotics available explicitly due to the simplicity of
the mKdVB equation. 

In the purely dispersive case, solutions typically involve DSWs,
expanding, slowly modulated wavetrains that provide for a dynamic
transition between two different constant values of $u$.  As described
in the previous section, the asymptotic description of DSWs can be
achieved by an averaging procedure that results in a system of
quasi-linear first-order {\it partial} differential equations
supplemented with initial/boundary conditions. DSWs resulting from
Riemann problems are described by similarity solutions of the
modulation equations.  In contrast, solutions of the
diffusive-dispersive equation involve traveling waves (TWs), solutions
of an \emph{ordinary} differential equation. The TWs approximate shock
waves, which are discontinuous weak solutions of the conservation law
\beq\label{inviscid_pde} u_t+(u^3)_x=0.  \eeq
This equation has characteristic speed $3u^2,$ the speed of small disturbances, analogous 
to the sound speed in a  compressible gas.  Shock waves travel at the speed
$s=u_+^2+u_+u_-+u_-^2$ depending on the left and right limits, given by the Rankine-Hugoniot condition. Consequently, weak shocks (for which $|u_+-u_-| $ is small), travel  at roughly characteristic speed. 

One of the most striking comparisons between the two equations is that
for $\mu>0$, the mKdV and mKdVB equations exhibit special TWs termed
{\em undercompressive} (in the language of
\cite{jacobs-mckinney-shearer95}) because the corresponding shock
wave 
\beq\label{shock1}
u(x,t)=\left\{\begin{array}{ll}
u_-, \quad & x<st\\[6pt]
u_+, &x>st
\end{array}
\right.  \eeq is subsonic both ahead of and behind the wave:
\beq\label{ucw} s<3u^2_\pm .  \eeq    The undercompressive TWs for mKdV have been
known in the literature as \textit{kinks}.

Undercompressive shocks are non-classical in the sense that they do
not satisfy the Lax entropy condition. They are called
undercompressive by analogy with gas dynamics, in which classical or
Lax shocks involve a jump in pressure associated with compression of
the gas as a shock wave passes a fixed observer.  When a shock fails
to satisfy the Lax entropy condition, it is termed non-classical
provided it has some relevance for the equation or application under
consideration. For scalar equations, the only non-classical shocks are
undercompressive in the sense of being either supersonic (see
\cite{spayd}) or subsonic on both sides of the wave.  However, for
systems of conservation laws, such as a system of magnetohydrodynamics
\cite{wu}, there can be a rich collection of non-classical shocks that
have physical relevance, including both undercompressive and
overcompressive shocks.  See \cite{schaeffer1} for examples of
$2\times 2$ systems with both undercompressive and overcompressive
shock waves.

For $\mu<0$ there are no undercompressive shocks. In this case, the
effect of the non-convex flux on the diffusive-dispersive dynamics
(\ref{mkdvb1}) is the occurrence of a double-wave structure known as a
shock-rarefaction. In the limit $\nu \to 0$, the shock wave in the
shock-rarefaction represents a single-sided contact discontinuity
propagating with characteristic speed \cite{dafermos}. Contact
discontinuities find their purely dispersive ($\nu=0$) counterparts in
the form of non-standard, contact DSWs whose modulations are described
by special, double-characteristic expansion-fan solutions of the mKdV
modulation equations. Such solutions become possible owing to
nonstrict hyperbolicity of the mKdV-Whitham modulation system.

The shock wave theory for the mKdVB equation has been constructed in
\cite{jacobs-mckinney-shearer95} by analyzing traveling wave
solutions.  Generalizations to different equations and to the Cauchy
problem for equation (\ref{cons1}) using wave front tracking are
detailed in the book by LeFloch \cite{LeFloch}.  

There have been several recent developments in the modulation theory
for mKdV DSWs. The two types of DSW solutions to the mKdV equation
with $\mu<0$ were constructed in \cite{march2008} using the modulation
equations obtained in \cite{dron76}.  These included the usual,
KdV-type (convex flux) DSW solutions as well as non-standard
sinusoidal undular bores, which we identify here as contact DSWs.  The
Riemann problem for the closely related defocusing complex mKdV
equation was considered in \cite{kod2008}.  Elements of the DSW theory
for the mKdV equation with $\mu>0$ appear in \cite{chanteur1987} in
the context of collisionless plasma physics.  A complete
classification of the Riemann problem solutions for the Gardner
equation, a combined KdV-mKdV equation reducible to mKdV, was
constructed in the recent paper \cite{kamch2012} for both signs of
$\mu$.  Application of the results for the Gardner equation to the
resonant generation of internal undular bores in stratified fluid
flows over topography was considered in \cite{kamch2013}.

Throughout this review, we incorporate direct numerical solutions in
order to provide a visual rendering of the underlying mathematics.
Here we fully resolve the small scale oscillatory behavior by
utilizing a sufficiently small spatio-temporal discretization (see
Appendix \ref{sec:numerical-method}).  Note that numerical methods
producing large scale, weak solutions of diffusive-dispersively
regularized non-convex conservation laws that avoid the small scale
discretization are an active area of research \cite{lefloch_mishra}.

In \S 2, we execute a derivation of the KdVB and mKdVB equation from a
general system of Eulerian equations for isentropic gas dynamics with
non-convex pressure law regularized by a combination of viscous
dissipation and a generic, third-order dispersion, using nonlinear
multiple scale expansions. Similar derivations for particular cases
can be found in \cite{johnson_book,ablowitz, watanabe84, ruderman2008}.  The main purpose of our
derivation is to identify the generic conditions under which the mKdVB
model dynamics persist. The condition for the asymptotic mKdVB
dynamics to dominate is shown to coincide with the condition for the
loss of genuine nonlinearity of the Eulerian system in the
non-diffusive, dispersionless limit. The important result of this
section is the expression for the dispersion coefficient in the
resulting mKdVB equation in terms of the pressure law and linear
dispersion relation of the original non-convex Eulerian system. The
sign of this coefficient is given by (\ref{signmu}) and plays the
definitive role in determining the qualitative structure of
solutions of the Riemann problem for both mKdV and mKdVB equations.

Section 3 is devoted to the comparison of classical Lax shocks with their
purely dispersive counterparts by reviewing and comparing Riemann
problem solutions of the KdVB and KdV equations. The main feature of a
DSW, which is a key element of the KdV Riemann problem solution, is
its expanding nature characterized by two distinct speeds, in
sharp contrast with the steady structure of 
diffusive-dispersive shocks propagating with a single shock speed and
described by TW solutions of the KdVB equation. The notion of a
classical DSW is introduced via an analog of the Lax entropy condition
formulated in terms of the characteristic speeds of the Whitham
modulation system. It is shown that, for the KdV equation, the DSW
admissibility condition coincides with the classical shock entropy
condition, although the associated Rankine-Hugoniot condition is not
applicable to the purely dispersive case due to the unsteady,
expanding nature of the DSW.
  
In \S 4, a unified description of the Riemann problem solutions for the
mKdVB equation with positive and negative dispersion is presented
following the results of \cite{jacobs-mckinney-shearer95}.  The
detailed structure of TW solutions approximating classical and
non-classical (undercompressive) shocks is studied, distinguishing
between oscillatory and non-oscillatory regimes.

In \S 5, the full classification of Riemann problem solutions to the
mKdV equations with positive and negative dispersion terms is
constructed in the framework of Whitham modulation theory, following
the results of the previous study \cite{kamch2012} performed for the
Gardner equation. Families of classical and non-classical DSWs are
then constructed and the admissibility conditions (\ref{adm-}) for
classical DSWs are introduced. These conditions include the
entropy-analog causality condition and the convexity condition implying
a single-wave regularization. In contrast with the KdV case, the
classical DSW admissibility conditions for the mKdV equation differ from
the counterpart conditions obtained for mKdVB classical shocks in
the zero-diffusion limit.  The distinct cases, where one or both of
the classical DSW admissibility conditions fail, fully determine the
boundaries in the $u_-$-$u_+$ plane of initial data between regions of
solutions with qualitatively different behaviors.  Non-strict
hyperbolicity of the modulation equations is shown to play the key
role in the occurrence and structure of non-classical DSW solutions
such as kinks and contact DSWs.

Section 6 is central and provides a bridge between the two theories
reviewed in \S \ref{sec:mkdvb-non-classical} and \S
\ref{sec:riemann-problem-mkdv}.  It is devoted to a detailed
comparison of the Riemann problem solutions to the mKdVB equation with
the solutions to the mKdV equation with comparable initial data. This
comparison reveals a number of non-trivial parallels and contrasts
between the solutions of the two equations. We produce characteristic
diagrams for diffusive-dispersive and dispersive
regularizations. These diagrams, in particular, clarify the complex
structure of non-classical mKdV solutions and further elucidate the
singular nature of the zero-diffusion limit.  One of the
manifestations of this singular nature is that the mapping between
solutions of the mKdVB and mKdV Riemann problems is not one-to-one.
The ambiguities are resolved in the small-diffusion regime by
identifying relevant time scales, including a transient window during
which the dispersion-dominated dynamics evolve to accommodate the
diffusive-dispersive balance. Finally, we show how the two contrasting
analytical approaches to the description of diffusive-dispersive
shocks and DSWs can be reconciled for the regime of small diffusion in
the framework of a perturbed modulation theory.

In \S 7, we draw conclusions from our study and outline future
perspectives.  The appendix contains a description of the numerical
method used to compute solutions.

\section{KdVB and mKdVB equations as weakly nonlinear approximations of diffusive-dispersive Eulerian hydrodynamics}
\label{sec:kdvb-mkdvb-equations}

The KdVB and mKdVB equations are prototypical equations, which often arise as unidirectional, weakly nonlinear long-wave approximations of more general diffusive-dispersive equations.  For example, the Euler equations for isentropic gas dynamics in one space dimension, regularized by a combination of viscous dissipation  and a generic third-order dispersion take the form,
 \begin{eqnarray} 
  && \rho_t + (\rho v)_x = 0 \,  \label{eulerd1}, \\
 && v_t + v v_x + \rho^{-1}[P(\rho)]_x = \tilde \nu v_{xx}+ [D(\rho,v)]_x  \, . \nonumber
\end{eqnarray}
Here $\rho \ge 0$ is the density and $v$ the velocity of the fluid, although these labels do not necessarily reflect the actual meaning of the corresponding physical entities (e.g. in nonlinear optics). 
In Eulerian dispersive fluids, the pressure law $P(\rho)$  is  an increasing function, $P'(\rho)>0$. The parameter $\tilde \nu>0$ is the viscosity   and $D(\rho, v)$ is a second order differential operator. 
We mention that the inviscid version of system (\ref{eulerd1}) for certain classes of the  operators $[D(\rho,v)]_x$ is sometimes called the Euler-Korteweg system \cite{benzoni2013, benzoni2014}.

The linear dispersion relation for system (\ref{eulerd1}) has the form
\begin{equation}\label{ldr0}
\omega(k,\rho_0,v_0)=v_0 k - \frac{i\tilde \nu}{2} k^2 \pm \omega_0(k, \rho_0),
\end{equation}
obtained by linearizing (\ref{eulerd1}) about the uniform background
state $\rho=\rho_0$, $v=v_0$ and seeking the solution in the form of a
traveling wave proportional to $ \exp[i(kx - \omega t)]$.  { The $\pm$
  signs in (\ref{ldr0}) correspond to the right- and left-propagating
  wave modes respectively. }
%The appropriate branch of the dispersion relation is fixed by the $\pm$ sign in (\ref{ldr0}) with
%$\omega_0(k, \rho_0) \ge 0$ for $k \ge 0$. \\

%\FIX{This all seems a bit vague: for example, $\omega_0(k, \rho_0) \ge 0$ for $k \ge 0$ could be false if $\tilde \mu = \mathrm{sgn} \{\omega_0''(k, \rho_0)\}<0.$ If we are only considering long waves, then that has to be included in  the condition $\omega_0(k, \rho_0) \ge 0$ for $k \ge 0.$ Is that what we want here?} \FIX{  I suggest to remove this sentence and  mention instead that the $\pm$ signs in (\ref{ldr0}) correspond to the right- and left-propagating modes respectively (and this is actually used in the long-wave derivations below). The sign $\omega_0>0$ for $k>0$ in the long-wave regime is  implied by (\ref{ldr}) below so does not need  to be explained.}\\
%

The key property of the dispersion relation we need is determined by
the long-wave expansion of $\omega_0(k, \rho_0)$, which we assume
generically has the form
\begin{equation}\label{ldr}
  \omega_0(k, \rho_0) = c_0 k  + \tilde \mu k^3 + o(k^3), \quad \ 0 < k
  \ll 1,  \quad \tilde \mu \ne 0,
\end{equation}
where $c_0=c(\rho_0)$, $c(\rho)=\sqrt{P'(\rho)}$ being the long wave
speed of sound.  The sign $\mathrm{sgn}\, \tilde \mu = \mathrm{sgn} \{\omega_0''(k, \rho_0)\}$ is called the sign of dispersion and plays an important role in our consideration. The aim of this section is to outline the derivations of the KdVB and mKdVB equations from the generic system (\ref{eulerd1}) of isentropic Euler equations modified by a combination of viscous and dispersive corrections in order to identify the conditions under which the model dynamics, described in the subsequent sections, persist.

We consider uni-directional, weakly nonlinear long-wave approximations of the diffusive-dispersive Euler equations (\ref{eulerd1}) by introducing the following multiple scales expansions (see, e.g., \cite{johnson_book,ablowitz}). 
%\FIX{  The reference to the long-wave multiple scales procedure is added and also the connection with the sign in the dispersion relation is made explicit}
\begin{equation}
  \label{expand}
  \begin{split}
    \rho &= \rho_0 + \eps \rho_1 + \eps^2 \rho_2+ \dots , \\
    v &= v_0 + \eps v_1 + \eps^2 v_2+ \dots,
  \end{split}
\end{equation}
where $\rho_i=\rho_i(\chi, \tau)$, $v_i= v_i(\chi, \tau), i\geq 1$, and
\begin{equation}\label{stretch}
\chi = \eps^{p}[x-U t], \quad \tau =\eps^{q}t, \quad 0<\eps \ll 1\, , \quad p, q>0 .
\end{equation}
To keep nonlinearity, dispersion and dissipation in balance  we also need to require $\tilde \nu =O(\eps^p)$. To this end, along with (\ref{stretch}) we introduce
\begin{equation}
\nu = \frac{\tilde{\nu}}{2\eps^{p}}.\end{equation}
{ To be definite we consider rightward-propagating waves (the plus sign in (\ref{ldr0})) whose speed  in the long-wave limit  is $U=v_0+c_0$.}
We also assume, \Mnote{for $i\geq 1,$ that $\rho_i, v_i \to 0$ as $\chi \to +\infty$.}

\medskip
{\it 1. KdVB approximation}

\smallskip
The choice $p=1/2$, $q=3/2$ yields, assuming the long-wave dispersion behavior (\ref{ldr}), the KdVB equation for the first order corrections $v_1$, $\rho_1$,
\begin{equation}\label{kdvb0}
(v_1)_\tau + \beta v_1 (v_1)_\chi = \nu (v_1)_{\chi \chi} +  \tilde \mu (v_1)_{\chi \chi \chi}  \, ,
\end{equation}
\begin{equation}\label{beta}
\hbox{where} \qquad \beta= 1+\frac{\rho_0 P''(\rho_0)}{2P'(\rho_0)}=1+ \frac{ \rho_0c'(\rho_0)}{c_0}\, .
\end{equation}
Also, we get
\begin{equation*}
 \rho_1 = \frac{\rho_0}{c_0}v_1\, .
\end{equation*}

\medskip
{\it 2. mKdVB approximation}

\smallskip
The KdVB approximation (\ref{kdvb0}) is valid as long as
$|\beta|=O(1)$, $\eps \to 0$ in (\ref{beta}), which is always the case for systems with convex pressure law, $P''(\rho)>0$. However, if $P''(\rho)<0$  for some interval of $\rho$, then there is the possibility
that the nonlinear coefficient $\beta$ could be zero.  This occurs if the background density $\rho_0$ satisfies the criticality condition
\begin{equation}\label{critical}
\left . (\rho^2 P'(\rho))' \right |_{\rho = \rho_0}=0 \, , \quad \hbox{or, equivalently} \quad
 \left . (\rho c(\rho))' \right |_{\rho = \rho_0} =0\, \ .
 \end{equation}
%The converse of   (\ref{critical}) represents the known 
On the other hand, if $\beta$ remains non-zero, then the hyperbolic
limit of equations \eqref{eulerd1} are genuinely nonlinear in the
sense of Lax \cite{lax2} (see, e.g., \cite{dafermos}). Genuine nonlinearity is most succinctly formulated as convexity of the pressure  expressed as a function of the specific volume $w=\rho^{-1}$: $\mathcal{P}''(w) \ne 0$ where $\mathcal{P}(w) \equiv P(1/w)$.   Our main interest will be in {\it non-convex systems}, for which the genuine nonlinearity condition is violated  at {\it exactly} one point.
In that case, third-order terms  become necessary in the expansions (\ref{expand}) to account for the higher-order nonlinear term in the resulting equation for $v_1$.  Then balance is achieved with new scaling parameters  $p=1$ and  $q=3$ in the transformation (\ref{stretch}) (see, e.g., \cite{watanabe84}, \cite{ruderman2008}).  Substituting the new multiple scales expansions (\ref{expand}), (\ref{stretch}) into (\ref{eulerd1}) and equating like powers of $\eps$ up to $\eps^{4}$, we obtain the mKdVB equation for the first order corrections,
\begin{equation*}%\label{mkdv1}
(v_1)_\tau + \gamma (v_1)^2(v_1)_\chi  = \nu (v_1)_{\chi \chi}
+ \tilde \mu (v_1)_{\chi \chi \chi} \, , \quad \rho_1 = \frac{\rho_0}{c_0}v_1.
\end{equation*}
Here,
\begin{equation}\label{ms}
\chi = \eps[x-(v_0+c_0)t], \quad \tau =\eps^{3}t \, ,
\end{equation}
and
\begin{equation}\label{}
\gamma = \frac{\rho_0}{4 c_0^3} \kappa'''(\rho_0)\, , \quad \hbox{where} \quad \kappa(\rho)= \rho P(\rho) \, .
\end{equation}
Introducing
\begin{equation}\label{}
u=\sqrt{\frac{|\gamma|}{3}}v_1\, , \quad \mu = \tilde \mu\, \mathrm{sgn} \, \gamma ,
\end{equation}
and replacing $\tau$ by $t$, $\chi $ by $\mathrm{sgn}(\gamma) x$, we arrive at the standard form (\ref{mkdvb1}) of the mKdVB equation. 

%\FIX{  We could mention here that, if $\beta$ (\ref{beta}) is not strictly zero but $O(\eps)$ the long-wave multiple scales expansions (\ref{ms}) lead to the Gardner-Burgers equation (mixed KdVB-mKdVB, with both quadratic and cubic nonlinearities).}

If $\beta$ in (\ref{beta}) is not strictly zero but $O(\eps)$, the
slightly modified long-wave multiple scales expansions \eqref{stretch} with
$p=1$, $q=3$ lead to the mixed, KdV-mKdV-Burgers equation, with both
quadratic and cubic nonlinearities present. The zero-diffusion version
of this equation is often called the Gardner equation.  See \cite{ruderman2008} and \cite{kamch2014} for the multiple scales derivations of the Gardner equation for ion-acoustic waves in collisionless plasmas with negative ions and  nonlinear polarization waves in two-component Bose-Einstein condensates respectively.

As  mentioned above,  the structure of some solutions to the Riemann problem for the mKdV and mKdVB equations  depends crucially on the sign of the dispersion coefficient $\mu$, which is defined in terms of the key nonlinear and dispersive characteristics of the original Eulerian system (\ref{eulerd1}) as
\begin{equation}\label{signmu}
  \mathrm{sgn}\,\mu = \hbox{sgn} [\kappa'''(\rho_0) \omega_{kk}(k,
  \rho_0) ] \, .
\end{equation}
In systems with {\it non-convex dispersion law} \ $\omega_0(k,\rho_0)$, in
addition to non-convexity of the pressure law, there is the
possibility of a sign change in $\mu$ due to a range of values of $k$
involved in the Riemann problem solutions, which can lead to further
complexity in their structure.  See \cite{conduit} for the Riemann
problem analysis of a dispersive hydrodynamic equation with convex
hydrodynamic flux but non-convex dispersion law.  However, the
possible change of dispersion sign is likely to occur for sufficiently
short waves and thus, may not affect the long-wave approximation.

%due to a change of the medium's dispersion sign 
%Since the Riemann problem's solutions involve a range of values of $k$, the  it will be necessary to use a combination of solutions to the $\mu>0$ and $\mu<0$ 
%%``plus'' and the ``minus'' 
%mKdV equations to construct the full Riemann problem classification. \FIX{did I change the intended meaning here? - MS } \FIX{\color{red} Fine}
%

\section{Lax shocks and their dispersive counterparts}
\label{sec:lax-shocks-their}

The shock wave (\ref{shock1}) is a weak solution of the conservation law (\ref{cons1}) if it satisfies \eqref{weak_soln}, from which the Rankine-Hugoniot condition
\begin{equation}
  \label{rh}
-s(u_+-u_-)+f(u_+)-f(u_-)=0
\end{equation}
can be deduced. Such a solution is a {\em Lax shock} if it satisfies the Lax entropy condition relating the characteristic speeds $f'(u_\pm) $ to the shock speed $s=(f(u_+)-f(u_-))/(u_+-u_-):$
\beq\label{lax1}
f'(u_+)<s<f'(u_-).
\eeq
For the inviscid Burgers equation
\beq\label{iburgers}
u_t+\half(u^2)_x=0,
\eeq
Lax shocks are characterized by $s=\half(u_++u_-),$ with $u_+<u_-.$
In the next subsection, we show that shock wave solutions of (\ref{iburgers}) are Lax shocks if and only if there are traveling wave solutions of the KdVB equation
\beq\label{kdvb1}
u_t+\half(u^2)_x=\nu u_{xx}+\mu u_{xxx}
\eeq
with $\nu>0$.

In \S \ref{sec:kdv-equation-dsws}, we describe the purely dispersive counterpart of the Lax shock for $\nu = 0$ in (\ref{kdvb1}), the KdV equation.  The main point we wish to emphasize here is that solutions of the KdV equation with initial data a Lax shock are quite different in character from those of the KdVB equation, and possess a structure quite distinct from the Rankine-Hugoniot condition or the traveling wave solutions of the KdVB equation (\ref{kdvb1}) with $\nu>0$.

\subsection{Traveling wave solutions approximating shocks}\label{tw_shocks}
We wish to consider (\ref{kdvb1}) with $\mu\neq 0.$ Since the
change of variables $x\to -x, u\to -u$ leaves the equation unchanged
if $\mu$ is replaced by $-\mu,$ we can take $\mu<0$ without loss of
generality. Then if we seek TW solutions in the form $u(x,t)=
\tilde{u}((x-st)/\sqrt{|\mu|})$ with
$\tilde{u}(\pm\infty)=u_\pm,$ we see that such solutions approach the
shock wave (\ref{shock1}) as $|\mu|, \ \nu \to 0^+$ and it remains only to
demonstrate the existence of the TW solutions.

\begin{figure}
\centerline{\includegraphics{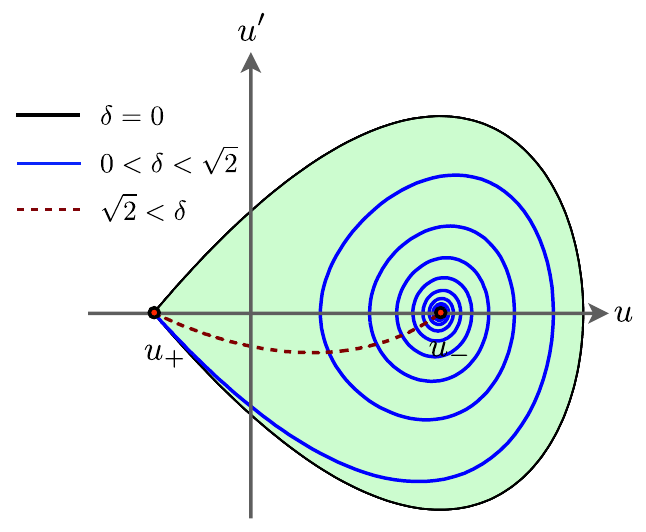} }
\caption{Phase plane depicting KdVB TW solutions connecting $u_+ <
  u_-$ when $\mu < 0$.  The dashed curve is a monotone TW with strong
  diffusion.  The
  solid, spiraling curve is an oscillatory TW with weak diffusion.
  Both are contained inside the homoclinic orbit of the zero diffusion
  KdV solitary wave solution.  The filled region is the union of all
  KdV periodic orbits spanned across a DSW.}
\label{fig:kdvbPhasePlane}
\end{figure}

We find after integrating once that TW solutions satisfy a second order equation
\beq\label{tw2}
-s(u-u_-) +\half(u^2-u_-^2)=\delta u'- u'', \quad \delta=\nu/\sqrt{|\mu|},
\eeq
where $\xi=(x-st)/\sqrt{|\mu|}, \ '=d/d\xi,$ and we have dropped the tilde from $\tilde{u}.$
If $u(\xi)$ is a solution for a fixed $\delta,$ then as $\mu\to 0,$ we have $\nu\to 0$ also. Hence, the traveling wave approaches a shock wave, which is in fact a weak solution of the conservation law (\ref{iburgers}), i.e.,  (\ref{kdvb1}) with $\mu=\nu=0.$
Specifically, assuming that $u'$ and $u''$ approach zero at $\pm\infty,$ the Rankine Hugoniot condition is satisfied: $s=\half(u_++u_-).$
The second order equation (\ref{tw2}) can be written as a first order system
$$ %\label{system1a}
u'=v, \quad v'=\delta v -\half(u^2-u_-^2)+s(u-u_-).
$$
For $\delta>0,$ the phase portrait of this system has two equilibria
$(u,v)=(u_\pm, 0),$ one of which is a stable saddle point and the other is 
an unstable node or spiral, depending on whether the eigenvalues of
the linearized system
$$
\lambda_\pm(u)=\half\left(\delta\pm\sqrt{\delta^2+4(u-s)}\right), \ s=\half(u_++u_-)
$$
are real ($\delta \ge \sqrt{2\Delta}$) or complex ($\delta <
\sqrt{2\Delta}$), respectively, where $\Delta = |u_- - u_+|$ is the
jump across the TW (see figure \ref{fig:kdvbPhasePlane}).  An analysis
of the phase portrait reveals that the saddle point is connected to
the second unstable equilibrium by a heteroclinic orbit, and this
corresponds to the TW we seek if and only if $u_+<u_-,$ namely the Lax
entropy condition.  (Note that for $\mu > 0$, the previously unstable
node or spiral is stable). The structure of the TW changes from a monotonic
to oscillatory profile as $\delta$ is decreased across the critical
value $\sqrt{2\Delta}$.  Further analysis shows that for $0 < \delta
\ll 1$, the leading amplitude of the oscillatory profile is
approximately $a = \frac{3}{2}\Delta$ \ \cite{johnson70}. For
$\delta=0$,  however, the unstable equilibrium becomes a center, i.e.,
with imaginary eigenvalues, and a TW connecting $u_+$ and $u_-$ no
longer exists. Instead, there is a homoclinic orbit connecting the
saddle point to itself, and the corresponding TW is a KdV solitary
wave.  Traveling wave solutions to the KdVB equation were analyzed in
\cite{grad_hu,johnson70,bona}.

When $u_+>u_-,$ the inviscid Burgers equation (\ref{iburgers}) has a rarefaction wave solution
\beq\label{rare1}
u(x,t)=\left\{\barr{ll}
u_-, \quad & x<u_-t\\[6pt]
{\displaystyle\frac{x}{t}},&u_-t<x<u_+t \\[6pt]
u_+, & x>u_+t.
\earr\right.
\eeq
This solution requires that $u_+>u_-.$ Although rarefaction waves have
corners at the edges $x=u_\pm t,$ these irregularities are smoothed by
the dissipative and dispersive terms, so that   apart from this small
correction, the rarefaction waves appear for the full KdVB equation. 

\begin{figure}
\centerline{\includegraphics{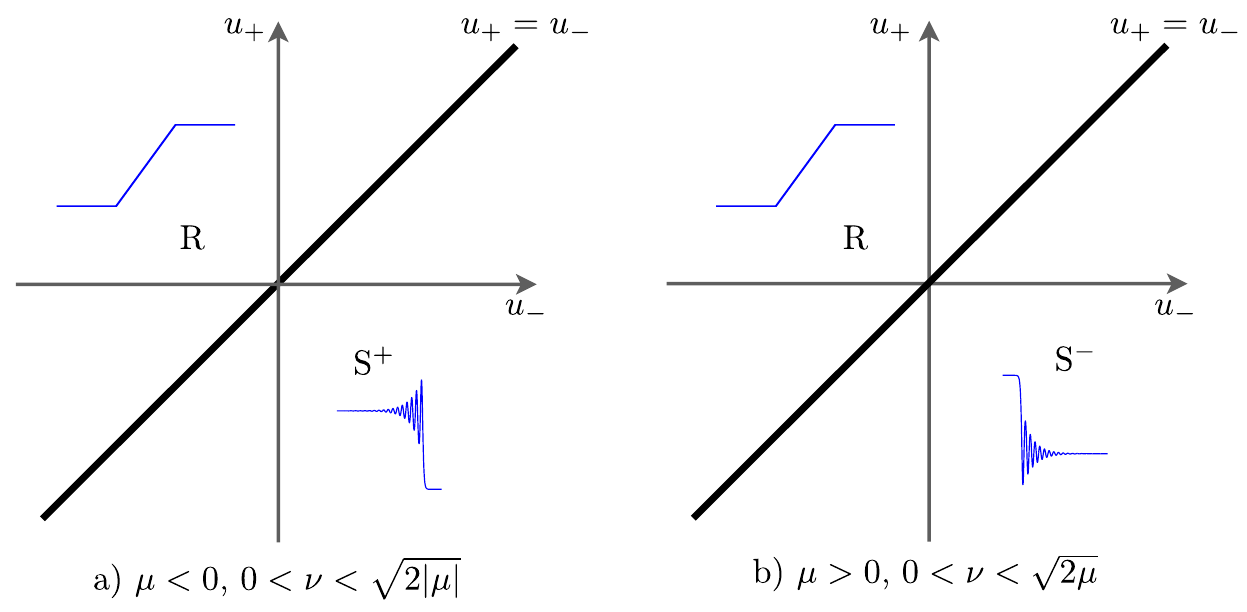}}
\caption{Classification of KdVB Riemann problem solutions.  (a)
  Negative dispersion; (b) Positive dispersion.}
\label{fig:kdvbclass}
\end{figure}

We can now classify the long-time behavior of solutions of the Riemann
problem for (\ref{kdvb1}).  For $u_->u_+,$ the solution approaches (as
$t\to \infty$) the traveling wave connecting $u_-$ to $u_+,$ whereas
for $u_-<u_+,$ the solution is approximately a rarefaction wave.  This
classification is depicted in figure \ref{fig:kdvbclass}.

\subsection{KdV equation and DSWs}
\label{sec:kdv-equation-dsws}
As remarked at the end of \S\ref{tw_shocks}, when $\nu=0$ in (\ref{kdvb1}), there is no traveling wave corresponding to a heteroclinic orbit in the phase portrait, hence the Rankine-Hugoniot condition for a shock wave is not satisfied. Instead, there is a homoclinic orbit connecting the saddle point to itself (a soliton), and there are periodic orbits representing periodic traveling wave solutions of (\ref{kdv1}).
 In particular, the  approximation  of a shock by traveling waves is invalid when $\nu=0.$  This could lead one to the conclusion that a certain amount of dissipation is necessary to sustain a shock wave transition. Indeed,
 the classical theory of undular bores by Benjamin and Lighthill
 \cite{benjamin_lighthill} and the theory of collisionless shocks in
 plasma due to Sagdeev \cite{sagdeev62, sagdeev66} and others (see,
 e.g., \cite{grad_hu}) are based on dissipative-dispersive models, essentially reducible to the KdVB equation (\ref{kdvb1}). It was not until Gurevich and Pitaevskii's pioneering work \cite{gp74} that 
it was realized that there is a qualitatively different type of shock
wave that develops in the total absence of dissipation. These
conservative shocks are called dispersive shock waves (DSWs) and
represent expanding, modulated nonlinear wavetrains connecting the
constant states $u_-$ and $u_+$. 

The dispersion relation for the KdV equation (\ref{kdv1}) on the
background $u_0$ is $\omega(k,u_0) = u_0k + \mu k^3$.  It will be
convenient to first consider the case of negative dispersion, $\mu <
0$.   DSW
solutions of the KdV equation are asymptotically described by periodic
solutions (cnoidal waves)
\begin{equation}\label{cnoidal}
  u(x,t)= \tilde u (\xi)=r_1+r_2-r_3+2(r_3-r_1)\hbox{dn}^2\left(\sqrt{\frac{r_3-r_1}{6}} \xi,m \right),
\end{equation}
where $r_1 \le r_2 \le r_3$ are parameters, which remain approximately constant on the wavelength/period scale $\Delta x, \Delta t  \sim \sqrt{|\mu|}$; $\hbox{dn}(\xi, m)$ is a Jacobi elliptic function,
\begin{equation}\label{vm}
  \xi=\frac{x-Ut+x_0}{\sqrt{|\mu|}}\, , \quad U=\frac{1}{3}(r_1+r_2+r_3)\, , \quad m=\frac{r_2-r_1}{r_3-r_1}.
\end{equation}
Here $x_0/\sqrt{|\mu|}$ is the initial phase and the modulus $m$ satisfies $0 \le m \le 1$.  As $m$ is varied from 0 to
1, the corresponding periodic orbits sweep out a phase plane region
enclosed by the solitary wave homoclinic orbit when $m \to 1^-$ (see
figure \ref{fig:kdvbPhasePlane}). The limit $m \to 0^+$ corresponds to
the linear, harmonic wave limit.   Slow, spatiotemporal modulations
$r_i(x,t)$ of the function (\ref{cnoidal}) occur on a scale
substantially larger than $\sqrt{|\mu|}$ and provide the required
transition between the two distinct constants $u_-$ and $u_+$ so that
$m=0$ at one edge of the DSW and $m=1$ at the other edge. Thus the DSW
completely fills the phase plane region enclosed by the solitary wave
orbit.  In figure \ref{fig:kdvbPhasePlane}, we see that the behavior
of the KdVB TW for weak diffusion $0 < \delta < \sqrt{2}$ spirals
through, but does not fill, this same region.

The modulations are described by a system of three quasilinear, first
order (hydrodynamic type) equations derived originally by
Whitham \cite{wh65} by averaging three of the conservation laws of
(\ref{kdv1}) over the periodic family (\ref{cnoidal}) resulting in
  \begin{equation}\label{kdvW}
  \frac{\partial  r_i}{\partial t} + V_i(r_1, r_2, r_3) \frac{\partial r_i}{\partial x}  =0 \, , \quad i=1,2,3 \, ,
  \end{equation}
i.e., $r_1, r_2, r_3$ are Riemann invariants for the modulation equations. The characteristic speeds $V_1 \le V_2 \le V_3$ are found as (see (\ref{potentialV}), (\ref{vm}))
  \begin{equation}\label{vi}
  V_i= U-\frac{L}{3\frac{\partial L}{\partial r_i}} \, , 
  \end{equation}
  where
  \begin{equation}\label{Lkdv}
  L=2 \K(m)\sqrt{\frac{6}{r_3-r_1}},
  \end{equation}
  $\K(m)$ being the complete elliptic integral of the first kind. The
  quantity $L\sqrt{|\mu|}$ is the wavelength of the periodic wave
  (\ref{cnoidal}).  The solution ${\bf r}(x,t)$ of the Whitham system
  (\ref{kdvW}) with appropriate initial or boundary conditions is
  inserted into the cnoidal wave (\ref{cnoidal}) to obtain the
  asymptotic solution for a DSW.  We note that the initial phase $x_0$
  in (\ref{vm}) remains undetermined within Whitham theory so the
  outlined solution describes the local structure of a modulated
  wavetrain up to a spatial shift $\sim \sqrt{|\mu|}$ within the
  traveling wave period.

We now present the explicit expression for $V_2(r_1, r_2, r_3)$, which plays a particularly important role in the DSW theory.
\begin{equation*}
%\label{V2}
 V_2(r_1, r_2, r_3)=  \frac{1}{3}(r_1+r_2+r_3)-\frac{2}{3}(r_2-r_1)\frac{(1-m)\K(m)}{\E(m)-(1-m)\K(m)} \, ,
 \end{equation*}
where $\E(m)$ is the complete elliptic integral of the second kind. Of particular relevance are two limits of $V_2$ corresponding to the harmonic  and soliton  limits of the traveling wave (\ref{cnoidal}). In the harmonic limit, the modulus $m=0$ (i.e. $r_2=r_1$, see (\ref{vm})) and the characteristic speeds $V_2$ and $V_1$ merge together
\begin{equation}
\label{v2=v1}
m=0: \quad V_2(r_1, r_1, r_3)=V_1(r_1, r_1, r_3)= - r_3 + 2r_1 \equiv V_- \, .
\end{equation}
One can also show that in this limit the velocity $V_3(r_1, r_1, r_3)=r_3$. Thus, in the harmonic limit the Whitham system (\ref{kdvW}) reduces to a system of two equations, one of which is the inviscid Burgers (dispersionless KdV) equation (\ref{iburgers}) for $r_3$.

In the opposite, soliton limit $m=1$ (i.e., $r_2=r_3$), we have a similar degeneracy but now the merged characteristic speeds are $V_2$ and $V_3$
\begin{equation}
\label{v2=v3}
m=1: \quad V_2(r_1, r_3, r_3)  = V_3(r_1, r_3, r_3)=  \frac{1}{3} (2r_3 + r_1) \equiv V_+ \, ,
\end{equation}
while the remaining velocity $V_1(r_1, r_3, r_3)=r_1$ yields the inviscid Burgers equation for $r_1$.

The described reductions of the KdV-Whitham system (\ref{kdvW}) in the
harmonic and soliton limits enable one to construct the solution ${\bf
  r} (x,t)=(r_1(x,t), r_2(x,t), r_3(x,t))$ that provides a continuous
matching of the period-mean field,
 \begin{equation}
\label{meanu}
\bar u (r_1,r_2, r_3)= \frac{1}{L}\int_0^L \tilde u(\xi) d\xi = r_1+r_2-r_3 + 2(r_3-r_1) \frac{\E(m)}{\K(m)},
\end{equation}
with the distinct constant states $u_-, u_+.$ This results in $\bar u
(r_1, r_1, r_3) = r_3=u_-$ at some $x=x_-(t)$ (the DSW's trailing,
leftmost edge) and $\bar u (r_1, r_3, r_3)=r_1=u_+$ at some $x=x_+(t)$
(the leading, rightmost edge).  The positions of the DSW edges are
free boundaries defined by the merged characteristics
$dx_{\pm}/dt=V_\pm$ evaluated for the relevant modulation
solution. The outlined {\it matching regularization problem} is due to
Gurevich and Pitaevskii \cite{gp74}, and its solution ${\bf r} (x,t)$
describes slow modulations of the periodic solution (\ref{cnoidal}) in
a DSW. We note that, while the modulation theory defines the
``carrier'' cnoidal wave solution (\ref{cnoidal}) only up to an
arbitrary phase shift $\Delta x \sim \sqrt{|\mu|}$, macroscopic DSW
parameters (the amplitude, the wavenumber, the edge speeds, etc.) are
fully determined by the solution of the Gurevich-Pitaevskii problem.
The computation of the exact phase $x_0/\sqrt{|\mu|}$ in the DSW
solution would generally require invoking a higher-order
approximation, beyond the original leading-order Whitham
theory. Despite this, for some classes of initial data the phase can
be explicitly expressed in terms of the Whitham modulation parameters
$r_j$ \cite{deift1997}, \cite{grava_klein2007}, \cite{el_hoefer_2016}.
For the case of the Riemann problem considered in this paper, the DSW
phase was derived in \cite{ablowitz2013}.

\begin{figure}[h]
\centerline{\includegraphics{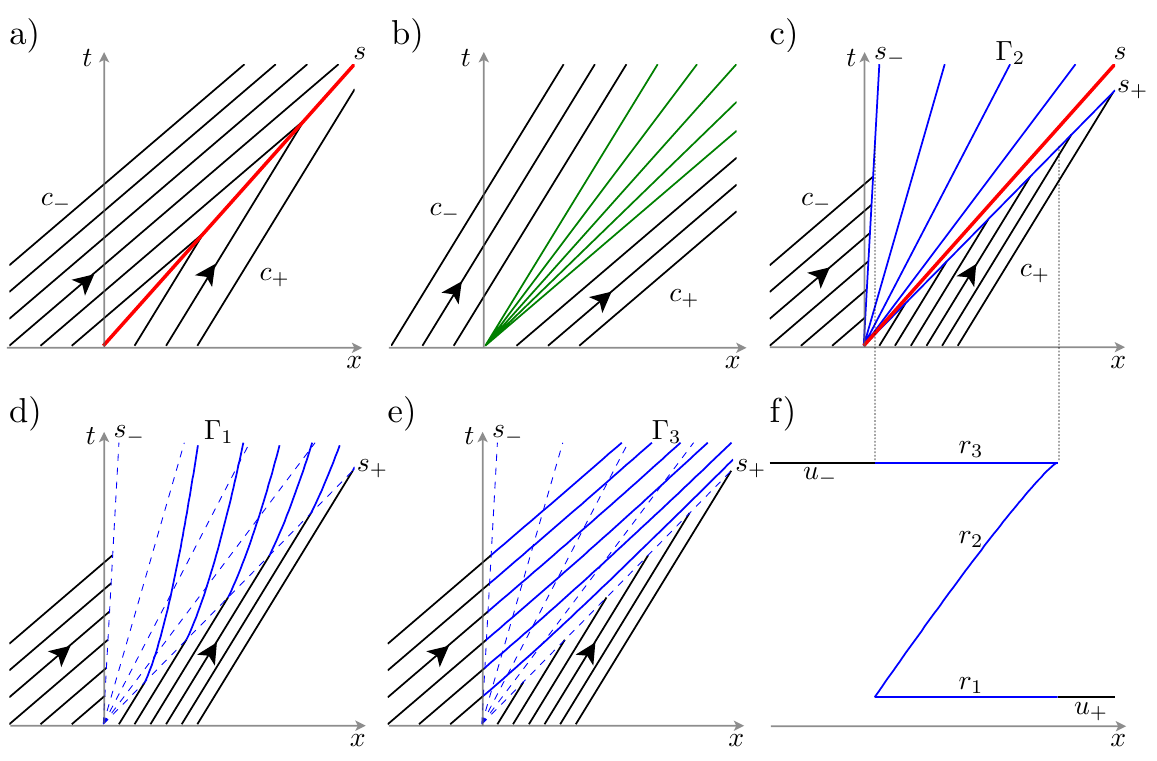} }
\caption{ Characteristics for the KdV Riemann problem solutions. (a,b) \ The inviscid Burgers equation $\mu=0.$ (a)
  $u_->u_+$, classical shock; 
  % (\ref{iburgers}); 
  (b) $u_-<u_+$,  rarefaction wave; 
  %of the inviscid Burgers equation (\ref{iburgers}); 
  \ (c-f) $u_->u_+$, $\mu
  \to 0$, rarefaction wave of the Whitham equations describing the DSW
  modulation. (c) The 2-characteristic family $\Gamma_2$ of the Whitham equations
   exhibiting the rarefaction fan. (The characteristic
  corresponding to the classical shock is shown in red.) (d) The   1-characteristics, labeled $\Gamma_1,$   approach $x=s_{_-}t$ as $t\to \infty$ (2-characteristics are shown as dashed lines for comparison).
  (e) The 3-characteristics, labeled   $\Gamma_3,$ approach
  $x=s_+t$ as $t \to \infty.$ 
  %the speeds $c_-$ and  $c_+$ are the characteristic speeds of the KdV dispersionless limit equation $u_t+uu_x=0$ corresponding to the left and right states $u_\pm.$
  (f) The  Whitham Riemann invariants $r_j, j=1,2,3$ at a specific time.}
 \label{KdVxt}
 \end{figure}

 The KdV-Whitham system (\ref{kdvW}) is strictly hyperbolic and genuinely nonlinear \cite{levermore88}, i.e. for all ${\bf r} \in \mathbb{R}^3$ one has: (i) $r_j \ne r_k \ \Longleftrightarrow \ V_j \ne V_k$  (strict hyperbolicity), and  (ii) $\partial V_j/\partial r_j \ne 0$, $j=1,2,3$ (genuine nonlinearity). Indeed,  the merged characteristic speeds (\ref{v2=v1}) and (\ref{v2=v3}) in the harmonic and soliton limits respectively define a regular characteristic since the order of the modulation system reduces to two in both limits, and thus strict hyperbolicity and genuine nonlinearity of the modulation system are preserved in these limits. 
 As a result, the dispersive regularization of the initial step (\ref{RP}) is described by a self-similar  simple-wave modulation of (\ref{cnoidal}) which is the {\it rarefaction wave solution of the Whitham equations},
 \begin{equation}\label{GPKdV}
\mu<0: \qquad r_1=u_+, \quad r_3=u_-, \quad V_2(u_+, r_2, u_-)=\frac{x}{t} .
 \end{equation}
Being a rarefaction fan (see figure \ref{KdVxt}c) solution (\ref{GPKdV}) exists for all $t>0$. It is defined in $s_-t \le x \le s_+t$, where $s_{-}$ and
 $s_+ $ are the speeds of the trailing and the leading DSW edges
 respectively (see figure \ref{KdVxt}). They are found from
 (\ref{GPKdV}) by setting $r_2=r_1$ ($m=0$ -- trailing, harmonic wave
 edge) and $r_2=r_3$ ($m=1$ -- leading, soliton edge) to obtain (see
 (\ref{v2=v1}), (\ref{v2=v3})) $s_-=V_2(u_+, u_+, u_-)=u_+ - \Delta$
 and $s_+= V_2(u_+, u_-, u_-)=u_++\frac{2}{3}\Delta$, where
 $\Delta=u_--u_+$ is the jump across the DSW \cite{gp74}.  It is
 instructive to note that the speed of the harmonic wave edge
 coincides with the linear group velocity $\omega_k$ evaluated at some
 $k>0$ while the soliton edge speed is determined by the phase
 velocity in the limit $k \to 0$ \cite{el05}. The wave number
 $k=2\pi/L$ (see (\ref{Lkdv})) in the DSW is a monotone function of
 $x$, which \textit{decreases} as the DSW is traversed from the
 harmonic to solitary wave edge. For $\mu<0$ one has $\partial k
 / \partial x <0$.

 Lax and Levermore \cite{lax_levermore} showed that this solution of
 the Whitham equations describes the weak limit solution $u(x,t,\mu)$
 of the KdV equation (\ref{kdv1}) when $\mu \to 0$, the weak limit
 being the period-mean (\ref{meanu}) with $r_1, r_2, r_3$ defined by
 (\ref{GPKdV}).  We stress that this weak limit {\it does not
   coincide} with the weak solution of the Riemann problem for the
 conservation law (\ref{iburgers}), which is the Lax shock
 (\ref{shock1}) with $s=(u_- + u_+)/2$. This is due to the
 oscillations in the DSW having wavelength with scale $\sim
 \sqrt{|\mu|}$ so the limit as $\mu \to 0$ is a singular
 one. Representations of the Riemann problem solutions in the $x$-$t$
 plane for the KdV equation with $\mu=0$ (the inviscid Burgers
 equation) and for $\mu \to 0$ (the Whitham equations) are shown in
 figure~\ref{KdVxt}.

 The general DSW admissibility conditions for a scalar dispersive equation with convex hyperbolic flux are \cite{el05}, \cite{el_hoefer_2016}:
  %(an analogue of Lax entropy conditions)
 \beq\label{laxdsw}
   s_-<f'(u_-), \quad s_+>f'(u_+), \quad s_- <s_+ \, .
 \eeq
  These conditions are also known as the {\it DSW  causality} conditions. Similar to the classical Lax entropy conditions (\ref{lax1}), inequalities (\ref{laxdsw}) ensure that the characteristics transfer initial data {\it into} the DSW region.  The generalization of the DSW admissibility conditions (\ref{laxdsw}) to bidirectional  Eulerian dispersive hydrodynamics can be found in \cite{el05}, \cite{hoefer14}. We shall refer to DSWs satisfying the Lax-type conditions (\ref{laxdsw}) as {\it classical DSWs}.
  % and  to the conditions (\ref{laxdsw}) as the {\it DSW  causality} conditions. (included immediately after (\ref{laxdsw}) ) 
  
For the KdV equation (\ref{kdv1}), the dispersionless ($\mu = 0$) characteristic speeds are $f'(u_-)=u_-$, $f'(u_+)=u_+$, while speeds of the leading and trailing DSW edges (for $\mu < 0$) are $s_+ = u_+ +\frac{2}{3} \Delta$ and $s_- = u_+ - \Delta$.   Thus, in the KdV case, the DSW admissibility conditions (\ref{laxdsw}) are equivalent to the entropy condition $u_+<u_-$ for Lax shocks. However, we stress again that the Rankine-Hugoniot condition $s=(u_-+u_+)/2=u_++\frac{1}{2}\Delta$ is not applicable to a DSW. Moreover, the Lax shock speed satisfies $s_-<s<s_+$ (see the red line in figure~\ref{KdVxt}(c)).

As noted for the KdVB equation earlier, the transformation $x \to -x$, $u \to -u$, $\mu \to -\mu$ leaves the KdV equation unchanged.  Therefore, a reflection of the spatial and amplitude axes of the KdV DSW with negative dispersion, $\mu < 0$, is sufficient to describe the DSW with positive dispersion.  Explicitly, the traveling wave solution and the modulation equations for the case of positive dispersion are obtained from (\ref{cnoidal}) -- (\ref{Lkdv}) by applying the transformations
\begin{equation}
\label{plusminus}
\mu< 0 \to \mu >0: \qquad u \to -u, \quad r_i \to -r_{4-i}, \quad V_i \to - V_{4-i}, \  i = 1, 2, 3.
\end{equation}
We note that under the transformations (\ref{plusminus}), the modulus of the elliptic function transforms  as $m \to 1-m$.

\begin{figure}[h]
\centerline{\includegraphics{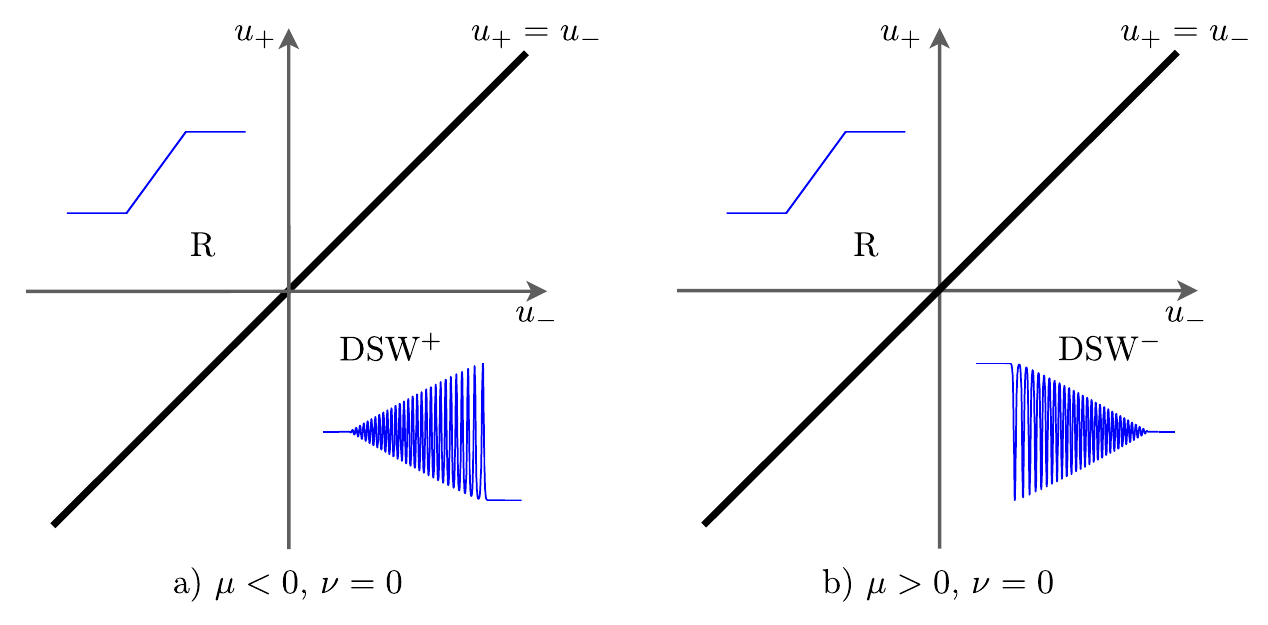}}
 \caption{ Classification of the KdV Riemann problem solutions. \
    (a) Negative dispersion; \ (b) Positive dispersion.}
 \label{KdVclass1}
 \end{figure}

The change $\mu <0 \to \mu>0$  alters the leading and trailing edge DSW speeds ($s_+ = V_2(u_+, u_-, u_-)=u_+ + 2\Delta$,
$s_- = V_2 (u_+, u_+, u_-) =  u_+ + \frac{1}{3} \Delta$ when $\mu > 0$) but, perhaps more strikingly, the DSW structure is changed.
The local description of the DSW according to (\ref{cnoidal}) for $\mu < 0$ takes the form of the KdV solitary wave
\begin{equation*}
  u = u_+ + 2 \Delta \textrm{sech}^2 \left ( \sqrt{\frac{\Delta}{6}}\xi \right ),
\end{equation*}
as the parameter $r_2 \to r_3$ ($m \to 1$).  Therefore, the leading edge of the DSW exhibits a positive, elevation wave with amplitude $a = 2 \Delta$.  Whereas for positive dispersion $\mu > 0$, the trailing edge of the DSW exhibits a negative, depression solitary wave.  Figure \ref{KdVclass1} displays the two cases.    The sign of the solitary wave, elevation  or depression, defines the \emph{polarity} $p$ of the DSW.  We denote a DSW that contains an elevation ($p=1$) or depression  ($p=-1$) solitary wave as DSW$^+$  or DSW$^-$, respectively. 

The position of the solitary wave, at either the leading edge or the trailing edge, defines the {\it orientation} $d$ of the DSW.  We say that  $d=1$ if the solitary wave is at the leading edge, otherwise $d=-1$. The orientation can be found by evaluating $d=-\mathrm{sgn}[{\partial k}/{\partial x}]$ at the harmonic edge.  This can be understood in very general terms as
follows  \cite{gurevich90}.  As already mentioned, 
 the modulation wavenumber $k$ decreases as the DSW
is traversed from the harmonic to solitary wave edge.  The orientation
of the DSW is determined by a well-ordering of the DSW in the vicinity
of the harmonic edge.  The harmonic edge is at the left (trailing
edge) if the group velocity increases with decreasing wavenumber,
i.e., $\omega_{kk} < 0$ ($\mu < 0$).  Conversely,
the harmonic edge is at the right (leading edge) if 
$\omega_{kk} > 0$ ($\mu > 0$).  In both cases $d=-\sgn\mathrm{\mu}$ i.e. the DSW orientation is uniquely defined by the sign of the dispersion. 

Thus for the KdV equation, both the DSW polarity and its orientation
are  determined by the sign of dispersion alone. As we shall see,
for the mKdV equation, a change in the DSW polarity (but not the
orientation) becomes possible without the change of sign of the
dispersion.  These properties are inherited from the corresponding
KdVB traveling wave solutions as indicated by a comparison between
figures \ref{fig:kdvbclass} and \ref{KdVclass1}.

 If $u_-<u_+$, the leading order solution of KdV with $|\mu| \ll 1$ represents a classical rarefaction wave (\ref{rare1}). In contrast to the Riemann problem solution of the KdVB equation (\ref{kdvb1}) with $u_-<u_+$, one corner (left if $\mu < 0$, right if $\mu > 0$) of the rarefaction wave is smoothed by linear dispersive oscillations whose wavelength $\sim \sqrt{|\mu|}$ and amplitude decays like $t^{-1/2}$, while the other corner is exponentially resolved to a constant \cite{leach-needham2008}.  The Riemann problem classification for the KdV equation is shown schematically in figure \ref{KdVclass1}.

% FIGURE

 % FIGURE

\begin{figure}
\centerline{\includegraphics{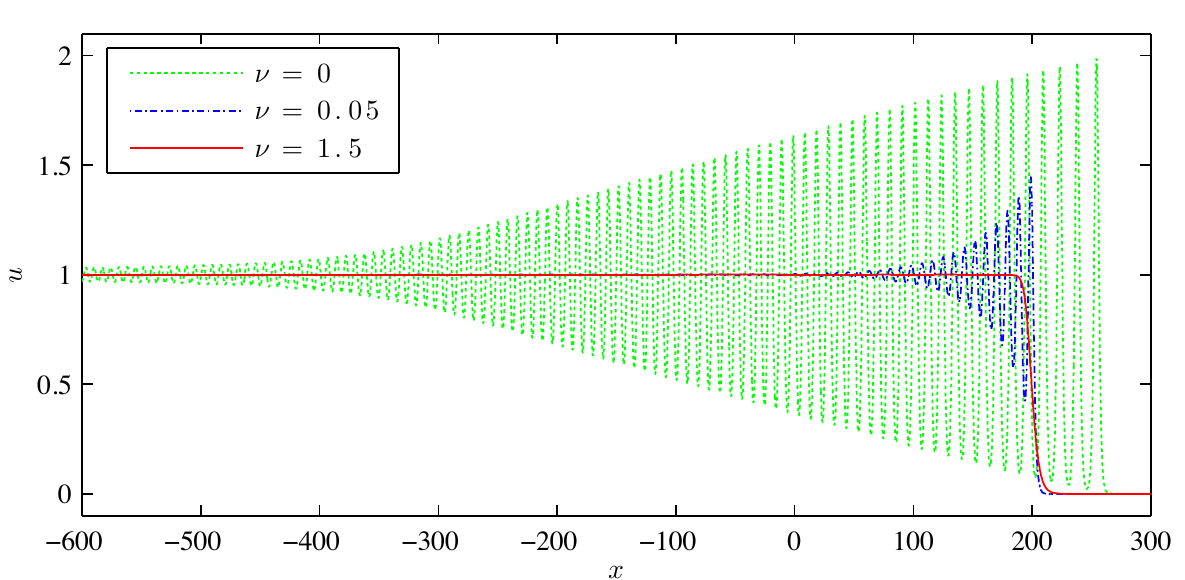}}%pdf}}
\caption{ Numerical solutions of the Riemann problem for KdV (dashed,
  green) and KdVB in the oscillatory (dash-dotted, blue) and monotonic
  (solid, red) TW regimes with the same step-like initial conditions
  and $\mu = -1$.}
 \label{comp_riemann}
 \end{figure}

Concluding this section, we summarize the contrasts between KdV DSWs and KdVB
traveling shock wave solutions (see figure~\ref{comp_riemann}) with
$0<\nu\ll \sqrt{2\Delta \mu}.$

% FIGURE

 \begin{enumerate}
 \item The most striking contrast between the two structures is that
   the KdV DSW is an unsteady, expanding structure, and is
   characterized by two speeds which are defined as the speeds of
   the leading and trailing edge characteristics  of the Whitham
   modulation equations. The KdVB TW describing the diffusively
   modified shock is characterized by a single classical shock speed
   determined by the Rankine-Hugoniot condition. 
 \item The amplitude of the leading soliton in the KdV DSW is $2\Delta$ while its counterpart in the shock solution of the KdVB equation with small $\nu>0$ has amplitude $3 \Delta/2$.
 \end{enumerate}

\section{mKdVB and non-classical shocks}
\label{sec:mkdvb-non-classical}
In this section, we restrict attention to the diffusive-dispersive mKdVB equation (\ref{mkdvb1}) with $\nu>0.$ Our primary interest is to relate traveling waves to shock waves (\ref{shock1}) satisfying the conservation law
\beq\label{cube1}
u_t+(u^3)_x=0.
\eeq
From (\ref{lax1}) we have that a shock wave solution (\ref{shock1})  of (\ref{cube1})  satisfies the Lax entropy condition if
\beq\label{laxm}
3u_+^2<s<3u_-^2.
\eeq
Here,
 \beq\label{rh2}
s=u_+^2+u_+u_-+u_-^2
\eeq
is the wave speed.
Thus, (\ref{shock1}) is a Lax shock \Mnote{(sometimes termed a {\em classical or compressive shock})} if and only if
\begin{equation}
\label{laxm1}
  -\mathrm{sgn} (u_-) \frac{1}{2} u_- <  \mathrm{sgn} (u_-) u_+  <  \mathrm{sgn} (u_-) u_- \, .
\end{equation}
Note the asymmetry in the inequalities resulting from the combination
of the Lax entropy condition and the Rankine-Hugoniot relation.

When $u_+=-u_-/2,$ the shock wave (\ref{shock1}) satisfies the Lax entropy condition in a limiting sense, since $s=3u_+^2.$ That is, the shock speed is characteristic on the right. Such a shock is sometimes called a one-sided contact discontinuity. 

\subsection{Traveling wave solutions}
To characterize traveling wave solutions $u=\tilde{u}(x-st)$ of (\ref{mkdvb1}) with speed $s,$ we impose the far field boundary conditions
\beq\label{bcs1}
\tilde{u}(\pm\infty)=u_\pm,\quad  \tilde{u}'(\pm\infty)=0, \quad \tilde{u}''(\pm\infty)=0.
\eeq
Then $\tilde{u}$  satisfies the ODE (dropping the tildes):
\beq\label{tw1}
-s(u-u_-) +u^3-u_-^3=\nu u'+\mu u''.
\eeq
In particular, the boundary condition (\ref{bcs1}) at $+\infty$ yields the Rankine-Hugoniot condition
\beq\label{rh1}
-s(u_+-u_-) +u_+^3-u_-^3=0,
\eeq
  giving the wave speed $s$  in (\ref{rh2}).

The following definition depends on fixing the parameter $\delta= \nu/\sqrt{|\mu|}$ and the sign of $\mu.$  A shock wave (\ref{shock1}) is termed {\em admissible} if there is a traveling wave solution of (\ref{mkdvb1}) satisfying the far-field conditions (\ref{bcs1}). Consequently, the shock wave is admissible if and only if the parameters $s,u_+,u_-$ are related by (\ref{rh2}) and there is a solution of the ODE (\ref{tw1}) satisfying $u(\pm\infty)=u_\pm.$ We note that the ODE is unchanged under the transformation $u\to -u, u_-\to -u_-.$ Consequently, if (\ref{shock1}) is an admissible shock with speed $s,$ then so is the shock with $u_\pm$ replaced by $-u_\pm,$ with the same speed $s.$

Just as for the KdVB equation, we have a second order ODE that describes traveling waves $u(\xi), \xi=(x-st)/\sqrt{|\mu|},$ except that for the MKdVB equation,    the sign of $\mu$ is significant:  
$$ %\label{mkdvbtw}
\mathrm{sgn}(\mu) u''+\delta u' =u^3-u_-^3-s(u-u_-), \ \delta= \nu/\sqrt{|\mu|}.
$$
 % for $\mu>0,\ \mu<0, $ respectively.
  %
The  corresponding  first order system  is
$$ %\label{system2}
u'=v, \quad \mathrm{sgn}(\mu) v'=-\delta v +(u^3-u_-^3)-s(u-u_-).
$$
For $\delta>0,$ the phase portrait of this system has between one and three equilibria including $(u,v)=(u_\pm, 0),$ where $u_+$ satisfies (\ref{rh1}). We are interested only in cases for which there are three equilibria, two of which may be a double root of (\ref{rh1}).   The nature of the equilibria  depends on   the eigenvalues 
\begin{equation}
  \label{eigenm}
  \lambda_\pm(u)=\half\left(-(\sgn{\mu})\delta\pm\sqrt{\delta^2+(\sgn{\mu})4(3u^2-s)}\right), 
  \ s=u^2+uu_-+u^2_-. 
\end{equation}
Consider the case of three equilibria with  $u=u_\pm, u_0$ satisfying $u_+<u_0<u_-.$ Then shocks from $u_-$ or $u_+$ (with $x<st$)  to $u_0$ (with $x>st$) satisfy the Lax entropy condition, since $3u_0^2 <s<3u_\pm^2.$ 

\begin{figure}
  \centering
  \includegraphics{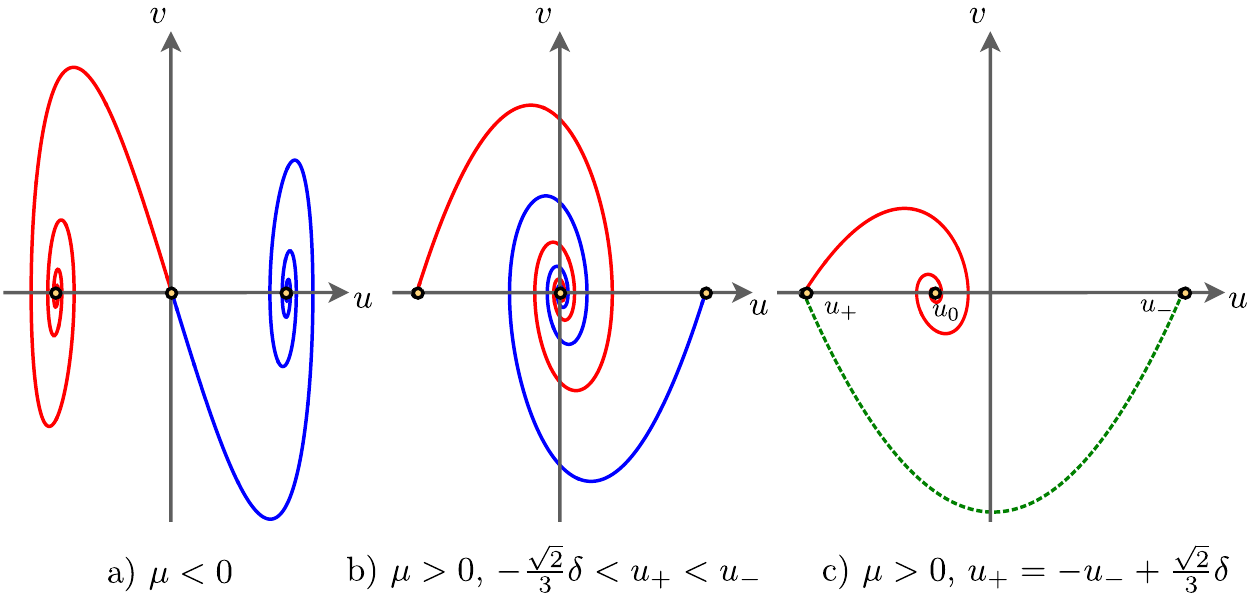}
  \caption{Representative phase portraits for traveling waves of the
    mKdVB equation.  The solid curves correspond to Lax shocks
    connecting $u_\pm$ to $u_0$.  The dashed curve in (c) represents
    an undercompressive shock connecting $u_-$ to $u_+$.}
  \label{fig:mkdvbPhasePlane}
\end{figure}

In terms of the equilibria, the eigenvalues (\ref{eigenm}) imply that
if $\mu<0, $ then $(u,v)=(u_\pm,0)$ are unstable nodes or spiral nodes
(if $\delta^2+4(3u_0^2-s)\geq 0$ or $\delta^2+4(3u_0^2-s) < 0,$
respectively), and $(u_0,0)$ is a saddle point. By tracing the stable
manifold of $(u_0,0)$ backwards in time, we can establish that there
is a trajectory from each of $(u_\pm,0)$ to $(u_0,0).$ Consequently,
for $\mu<0,$ all Lax shocks are admissible and there are no other
admissible shocks.  An example phase portrait is shown in figure
\ref{fig:mkdvbPhasePlane}(a).
 
On the other hand, if $\mu>0, $ then $(u,v)=(u_\pm,0)$ are saddle
points, and $(u_0,0)$ is a stable equilibrium, either a node or a
spiral. The saddle points at $(u_\pm,0)$ may or may not be connected
to $(u_0,0),$ and can be connected to each other.  See figure
\ref{fig:mkdvbPhasePlane}(b) for a phase portrait when both $u_\pm$
are connected to $u_0,$ corresponding to admissible Lax shocks.  Figure
\ref{fig:mkdvbPhasePlane}(c) is a phase portrait showing a connection for a Lax shock but also an orbit connecting $u_-$ to $u_+.$ In this case, there is no connection from $u_-$ to $u_0,$ even though  these constants define a Lax shock. 
It is proved in \cite{jacobs-mckinney-shearer95} that   an orbit joining saddle points $u_\pm$ exists if and only if it lies on an invariant parabola $v=\pm\frac{1}{\sqrt{2}}(u-u_-)(u-u_+);$ an explicit calculation shows that this occurs precisely for 
\beq\label{ucshocksa}
|u_-|> \frac{2\sqrt{2}}{3}\delta, \quad u_+= -u_-+\mathrm{sgn}(u_-)\frac{\sqrt{2}}{3}\delta,
\eeq
for which  $ s=u_-^2-\frac{\sqrt{2}}{3}\delta |u_-|+\frac{2}{9}\delta^2,$ and $u_0=-\mathrm{sgn}(u_-)\frac{\sqrt{2}}{3}\delta.$ % is surprisingly independent of $u_-.$
 These values  help determine  which Lax shocks are admissible for $\mu>0,$ and which are
not. The dynamics on the invariant parabola correspond to the traveling wave
  \beq\label{ucshocks2}
u(\xi) = \half\{u_-+u_+-|u_--u_+|\tanh(A\xi)\}, \ A=\frac{1}{2\sqrt{2}}(u_--u_+), \ \xi=(x-st)/\sqrt{\mu}.
\eeq

To summarize, for $|u_-|> \frac{2\sqrt{2}}{3}\delta,$ a shock wave
(\ref{shock1}) from $u_-$ to $u_+$ is admissible if either (i) \ $u_+$ is between
$-\mathrm{sgn}(u_-)\frac{\sqrt{2}}{3}\delta$ and $u_-$ or (ii) \ $u_+=
-u_-+\mathrm{sgn}(u_-)\frac{\sqrt{2}}{3}\delta.$ In case (i), the
shock satisfies the Lax entropy condition $3u_+^2<s<3u_-^2,$ and in
case (ii) the shock is undercompressive, in the sense that it has the property (\ref{ucw}) discussed in the introduction. 
In case (ii), other Lax shocks (for which $u_+$ lies between $-\half
u_-$ and $-\mathrm{sgn}(u_-)\frac{\sqrt{2}}{3}\delta$) are not
admissible.
  In the case $|u_-|\leq \frac{2\sqrt{2}}{3}\delta,$ there
is no undercompressive shock (for any $u_+$) and all Lax shocks (for
which $u_+$ lies between $u_-$ and $-\half u_-$) are admissible. 

When $u_+=-u_-/2,$ the shock wave (\ref{shock1}) is a one-sided contact discontinuity. It is  admissible if either $\mu<0,$ or $\mu>0$ and   $|u_-|\leq \frac{2\sqrt{2}}{3}\delta,$ the latter inequality arising from the existence of the saddle-saddle connection of Figure
\ref{fig:mkdvbPhasePlane}(c).
\begin{figure}[ht]
\centerline{\includegraphics{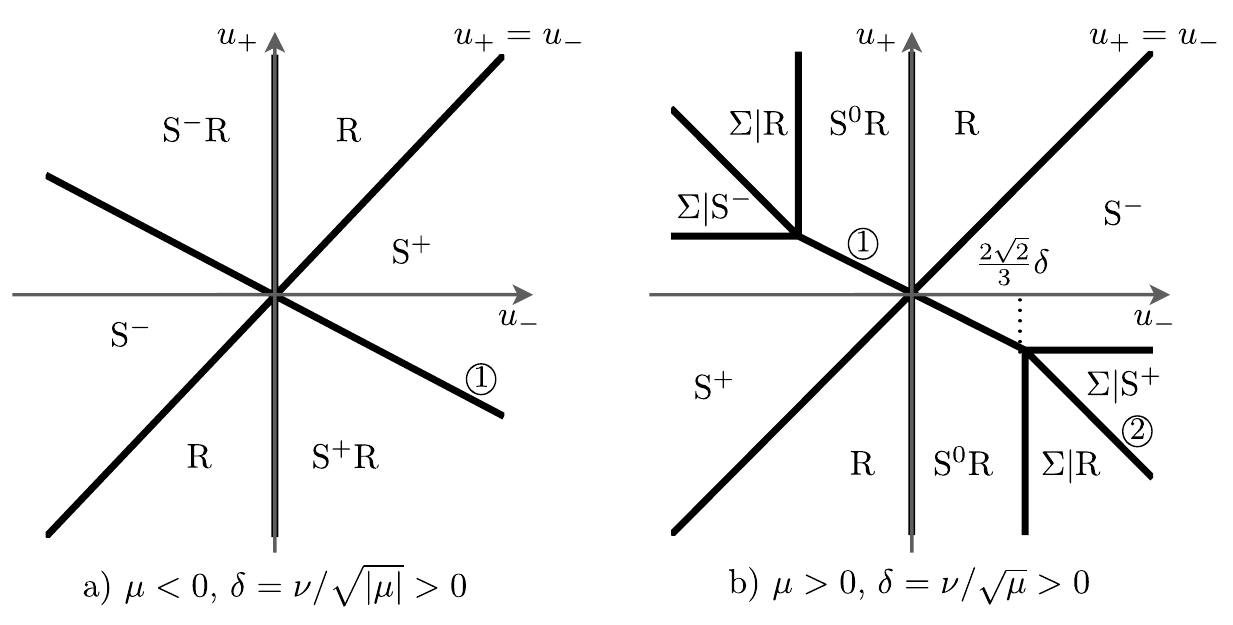}}%pdf}}
 \caption{ Solutions of the Riemann problem for the mKdVB equation with    $\delta=\nu/\sqrt{|\mu|}>0.$ Line \textcircled{\tiny 1}: $u_+=-u_-/2.$ Line \textcircled{\tiny 2}: $u_+=-u_-+\sqrt{2}\delta/3.$ }
 \label{riemann_fig1}
 \end{figure}

 \subsection{The Riemann problem}
\label{sec:riemann-problem}
To complete the description of solutions of the Riemann problem, we need to utilize not only constant solutions and admissible shocks, but {\em rarefaction waves.}  These are continuous self-similar solutions of (\ref{cube1}) that join constants $u_\pm,$ and are given by the explicit formula
\beq\label{rare2}
u(x,t)=\left\{\barr{ll}
u_-, \quad & x<3u_-^2t\\[6pt]
\pm \sqrt{\displaystyle\frac{x}{3t}},&3u_-^2t<x<3u_+^2t \\[6pt]
u_+, & x>3u_+^2t.
\earr\right.
\eeq
Rarefaction waves therefore propagate to the right, and $u(x,t)$ has a single sign throughout the wave. Consequently,   $|u_+|>|u_-|\geq 0.$ Note that the graph of $u=u(x,t)$ as a function of $x$ for fixed $t>0$ is continuous, and smooth apart from corners where the parabola meets the constants $u=u_\pm.$
Using this information about shock waves and rarefaction waves, we can construct solutions of the Riemann problem for any $u_+, u_-$ using only constant solutions, admissible shock waves (both classical and undercompressive), and rarefaction waves. The classification diagram for these solutions is presented in figure~\ref{riemann_fig1}. The construction labeled $S^jR,$ with $j=\pm, 0,$ indicates a {\em shock-rarefaction.} This is a composite wave in which a shock from $u_-$ to $-u_-/2$ has speed $s=3u_-^2/2$ that is characteristic on the right, at $u=-u_-/2.$ Consequently, to reach $u_+,$ beyond $-u_-/2$ (meaning $u_+<-u_-/2$ when $u_->0$ for example), a faster rarefaction wave can be attached to the shock wave, yielding the single  wave structure in which the rarefaction wave has a discontinuity at it's trailing edge. The discontinuity is sometimes called a {\em one-sided contact.}

In figure~\ref{riemann_fig1}(a), in which $\mu<0,$ we use the fact
that there are no undercompressive shocks and all Lax shocks are
admissible.  In figure~\ref{riemann_fig1}(b), we show the structure of
solutions when $\mu>0. $ The classification is more complicated, in
that there are new regions in the diagram, in which the Riemann
problem solution has two waves traveling with different speeds,
separated by a widening interval in which $u$ is constant. This
structure is indicated in the figure with a vertical line between the
wave letters.  One wave is an undercompressive wave, indicated with
the letter $\Sigma,$ the other being a shock S or rarefaction R.  Note
that figures \ref{riemann_fig1}(a,b) inherit the symmetry $u_+ \to
-u_+$, $u_- \to -u_-$ from the governing mKdVB equation.

Strictly speaking, figure~\ref{riemann_fig1} shows solutions of the Riemann problem for the conservation law  (\ref{inviscid_pde}), using information about traveling waves with $0<\delta<\infty$ to distinguish those shocks that are admissible. But the conservation law can be regarded as the expressing the leading order long-time behavior of solutions of the Riemann problem for the MKdVB equation as $\mu$ and $\nu$ approach zero with $\delta>0$ held constant. In figure~\ref{fig:mapping}, we show numerical solutions of the Riemann problem in representative cases of each of the regions shown in figure~\ref{riemann_fig1}. In that figure, shocks are replaced by smooth traveling waves, and rarefaction waves, though not precisely self-similar, are clearly visible.

It remains to explain the superscripts $\pm, 0$ associated with shock
waves. These indicate whether there may be oscillations ($\pm$) or not
($0$). As explained below in the context of purely dispersive shocks,
$S^+$ indicates that the traveling wave terminates exponentially at an
equilibrium, with an adjacent maximum. Similarly, $S^-$ indicates that
the largest oscillation involves a minimum before the traveling wave
terminates exponentially at the equilibrium.

To explain these structures, we refer to the ODE (\ref{tw1}). We can
write the equation as a first order autonomous system
\beq\label{system1} u'=v, \quad \mu v'=-\nu v+f(u;u_-,s), \quad
f(u;u_-,s)=u^3-u^3_--s(u-u_-).  \eeq See figure
\ref{fig:mkdvbPhasePlane} for representative phase portraits.
Traveling waves corresponding to Lax shocks are given by trajectories
for this system between a saddle point equilibrium and a second
equilibrium, either a node or spiral. Oscillations are associated with
a spiral equilibrium, at which the two eigenvalues are complex
conjugates. Since either $u_+$ or $u_-$ can be the saddle point, we
examine eigenvalues for both, indicating whether there are
oscillations behind or ahead of the wave. We find that the sign of
$\mu$ is also significant.

Intuitively, we expect that for small $\nu>0,$ there will be
oscillations, since this is similar to underdamping of
oscillations. For large $\nu$ however, there should be no
oscillations, corresponding to overdamping. We verify this distinction
with a quantitative prediction of the boundary between the presence of
oscillations in the traveling wave, and the absence of
oscillations. The prediction is then observed in numerical simulations
(see \S\ref{comparison}).

\begin{figure}[ht]
  \centering
  \includegraphics{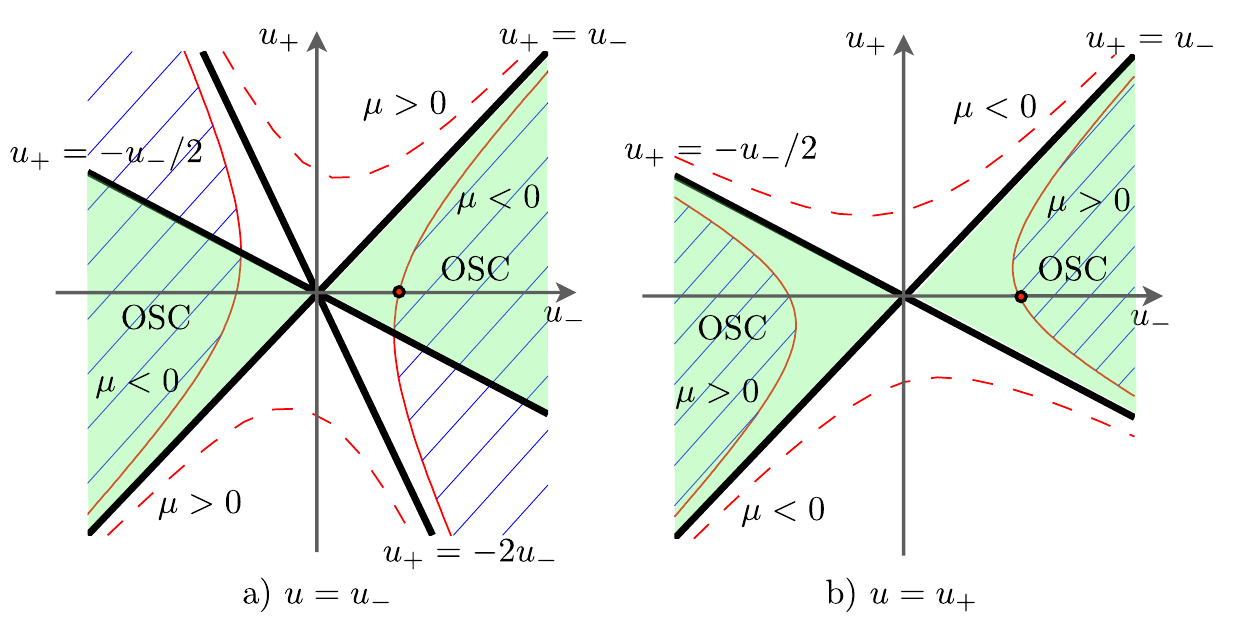}
  \caption{The hyperbolas delineating the boundary between oscillatory and non-oscillatory traveling waves. Lax shocks correspond to the shaded areas. (a) Region OSC where oscillations occur at $u_-,$ when $\mu<0.$ \ (b) Region OSC where oscillations occur at $u_+$ when $\mu>0.$ The dot on the $u_-$-axis is at $u_-=\nu/2\sqrt{2|\mu|}$ in (a), and at $u_-=\nu/2\sqrt{\mu}$ in (b). }
  \label{osc1}
\end{figure}

If we linearize system (\ref{system1}) about an equilibrium $u, $ we
find eigenvalues $\lambda$ given by the formula 
$$ %\label{eigen1}
\lambda=\half\left\{-\frac
  \nu\mu\pm\sqrt{\left(\frac{\nu}{\mu}\right)^2+4\frac{f_u}{\mu}}\right\}.
$$
Consequently, the boundary between real and complex eigenvalues
is $\frac{\nu^2}{\mu}+4f_u=0.$ Using the definition (\ref{system1}) of
$f(u,u_-,s)$ and the formula (\ref{rh2}) for $s$, we obtain hyperbolas
in the $u_-, u_+$ plane for each of $u=u_-, u=u_+:$ \beq\barr{lrcl}
u=u_-: & (2u_-+u_+)(u_--u_+)&=&-\frac{\nu^2}{4\mu}\\[6pt]
u=u_+: \quad &(2u_++u_-)(u_+-u_-)&=&-\frac{\nu^2}{4\mu}.  \earr \eeq
These are represented in figure~\ref{osc1}. The hyperbolas are
oriented differently depending on the sign of $\mu.$ Also in the
figure, we indicate the sectors corresponding to Lax shocks.

There are several conclusions to be drawn from these figures.

\begin{enumerate}
\item The dashed curves lie outside the sectors representing Lax
  shocks. Therefore, when $\mu>0,$ Lax shocks are non-oscillatory at
  $u_-$ and for $\mu<0,$ they are non-oscillatory at $u_+.$
\item For small enough $\nu,$ Lax shocks are oscillatory at $u_-$ if
  $\mu>0,$ and they are oscillatory at $u_+$ if $\mu<0.$
\end{enumerate}

Summarizing, we conclude that for $\nu\geq 0$ small enough, Lax shocks
are oscillatory on the left (at $u_-$) and non-oscillatory on the
right (at $u_+$) if $\mu>0,$ and the reverse is true if $\mu<0.$
Therefore we can identify the shock orientation $d =
-\mathrm{sgn}(\mu)$, analogous to the orientation of a DSW (recall
\S~\ref{sec:kdv-equation-dsws}).

However, there is a significant difference between the location of the solid line hyperbolas in the two figures. In figure~\ref{osc1}(a), the hyperbola crosses the line $u_+=-u_-/2,$ representing the shock in a shock-rarefaction wave, whereas in figure~\ref{osc1}(b), the hyperbola is contained within the sector representing Lax shocks. Correspondingly, for fixed $\mu>0,$ and $\nu>0,$ rarefaction-shocks have an oscillation on the left, whereas for $\mu<0,$ rarefaction-shocks are monotonic. In both cases, weak shocks (those for which $|u_+-u_-|$ is small) are monotonic.

The superscripts in figure~\ref{riemann_fig1} are related to the appearance of oscillations for small $\delta>0.$  For $\mu<0,$ there are oscillations on the left of the wave, so that for $u_+<u_-,$ the  final oscillation has a maximum, indicating an $S^+$ wave, whereas in an $S^-$ shock, we have $u+>u_-,$ for which the final oscillation necessarily has a minimum.  Note that in a solution designated $\Sigma | S^+,$ the shock is actually properly associated with the $S^+$ region with $u_+>u_-.$ A similar remark applies to the $\Sigma | S^-$ solutions. The $S^0$ shocks have monotonic traveling waves, in agreement with the remarks above concerning figure~\ref{osc1}(b).
%Figure

We have shown that there is an important distinction between Riemann
problem solutions to the mKdVB equation depending on dispersion.  When
$\mu < 0$, the effect of the dispersive term is the introduction of
oscillations when $\delta = \nu/\sqrt{|\mu|}$ is sufficiently small.
Otherwise, the solutions are classical in the sense that shocks
satisfy the Lax entropy condition.  The classification diagram in
figure~\ref{riemann_fig1}(a) is independent of $\delta > 0$.  While the
$\mu > 0$ case also yields oscillatory Lax shocks, this regime
supports non-classical, undercompressive shocks that do not satisfy the
Lax entropy condition.  Here, dispersion plays a more significant
role, which manifests as a $\delta$-dependent classification diagram
(figure~\ref{riemann_fig1}(b)).  Note that the limit $\delta \to \infty$
(i.e., $\mu\to 0,$ the purely diffusive case) of figure \ref{riemann_fig1}(b) leads to the
same Riemann problem classification as in figure
\ref{riemann_fig1}(a), i.e., the Riemann problem is independent of $\nu$ when 
$\mu\to 0.$

\section{Riemann problem for the mKdV equation}
\label{sec:riemann-problem-mkdv}
In this section, we present the classification of   Riemann problem solutions for the mKdV equation, i.e., (\ref{mkdvb1}) with $\nu=0$. Similar to the  KdVB equation discussed in \S \ref{sec:lax-shocks-their}, the transition to the inviscid case $\nu=0$ in the mKdVB equation is singular \Mnote{in the sense that although Lax shocks may have corresponding traveling waves for $\nu>0,$ they are not present for $\nu=0.$ Instead, the Lax shocks are replaced by DSW solutions.}
%one with respect to shock-like solutions, namely,  there are  no traveling waves corresponding to ``classical'' heteroclinic orbits in the phase portrait; so, one needs to use appropriate DSW solutions to replace the (Lax) shocks. 
In contrast to the KdVB/KdV case, non-convexity of the hyperbolic flux in the case of the mKdVB/mKdV equations results in a much richer variety of solutions, which makes establishing the parallels between the diffusive-dispersive and purely dispersive cases rather non-trivial.  One of the  pertinent questions arising in this connection is: what is a conservative, purely dispersive counterpart of an undercompressive shock  in the $\mu>0$ case? \Mnote{A second} question concerns the identification of dispersive counterparts of composite wave structures \Mnote{such as} shock-rarefactions, occurring \Mnote{as solutions of the mKdVB equation for both} $\mu>0$ and $\mu<0.$ 
%mKdVB Riemann problem classifications.

The full classification of   Riemann problem solutions for the \Mnote{Gardner equation, the} extended version of the mKdV equation containing both quadratic and cubic nonlinearities, 
%the  Gardner equation, 
was constructed in \cite{kamch2012}. In \Mnote{that paper,}
%\cite{kamch2012} 
the Gardner equation was taken in the standard form $w_{t'}+6 w w_{x'}-6\alpha w^2  w_{x'}+  w_{x' x' x'}=0$, which can be reduced to the mKdV equation in the form (\ref{mkdvb1}) with $\nu=0$, $\mu= \hbox{sgn}\, \alpha$ by the change of variables $u=\sqrt{2|\alpha|}[w - 1/(2\alpha)]$, $t=t'$, $x=-[x' - 3 t'/(2 \alpha)] \hbox{sgn}\, \alpha$.
 Here we adapt the classifications of \cite{kamch2012} for both signs of $\alpha$ to the pure mKdV case (with the \Mnote{corresponding} 
 %respective 
 signs of $\mu$) to enable detailed comparison with the counterpart  mKdVB classifications  outlined in \S 4.

\subsection{mKdV equation: DSW theory}
\label{sec:dsw_theory}

To produce the classification of asymptotic solutions to the Riemann problem for the mKdV equation, one needs first to construct elementary building blocks such as DSWs and rarefaction waves, and identify their admissibility conditions.    For this, we first describe mKdV periodic solutions and their limiting configurations, and then develop modulation theory for mKdV DSWs. Unlike the well-established theory of diffusive-dispersive shocks for the mKdVB equation, the counterpart mKdV theory is not directly available from a single source so we present here  a unified account   of previous results along with further development in the context of the theory of hyperbolic conservation laws.  
% In particular, we prove that the mKdV modulation equations are nonstrictly hyperbolic and not genuinely nonlinear.

\subsubsection{Traveling waves}
\label{sec:mkdv_trav}
Traveling wave solutions of the mKdV equation are available from a number of papers  (see, e.g.,
\cite{ksk-04,march2008}) and can also be obtained as reductions of
corresponding solutions to the Gardner equation \cite{kamch2012}.

Traveling wave solutions to the mKdV  equation are sought in the form $u=\tilde u(x-Ut)$, with speed $U$.  Note that we have omitted the initial phase 
in the TW ansatz here as this does not contribute to the modulation solution of the Riemann problem (see Section \ref{sec:kdv-equation-dsws} for the discussion of the DSW phase for the KdV case).

After integrating twice, one arrives at the ODE (dropping the tildes)
%\begin{equation}\label{ODE1}
%2\mu (u')^2 = u^4 - 2U u^2 +Au +B,
%\end{equation}
%where $A$ and $B$ are arbitrary constants. We introduce a new traveling variable $\eta=\frac{x-Ut}{\sqrt{2|\mu|}}$ and represent (\ref{ODE1}) in the form
\begin{equation}\label{eq3}
    u_{\eta}^2= \hbox{sgn}\, \mu (u-u_1)(u-u_2)(u-u_3)(u-u_4) \equiv Q(u),
\end{equation}
where  $\eta=\frac{x-Ut}{\sqrt{2|\mu|}}$ and   the roots of the polynomial $Q(u)$  satisfy the constraint
\begin{equation}\label{constr}
\sum_{i=1}^4u_i=0 \, .
\end{equation}
While only three roots $u_j$ are independent, it is still convenient to keep all four $u_j$'s in the
subsequent formulae to preserve  symmetry of the expressions. In the modulationally stable case of our interest, all $u_j$'s are real. We assume the ordering
\begin{equation*}%\label{eq4}
    u_1\leq u_2\leq u_3\leq u_4 \, .
\end{equation*}

\begin{figure}[h]
  \centering
  \includegraphics{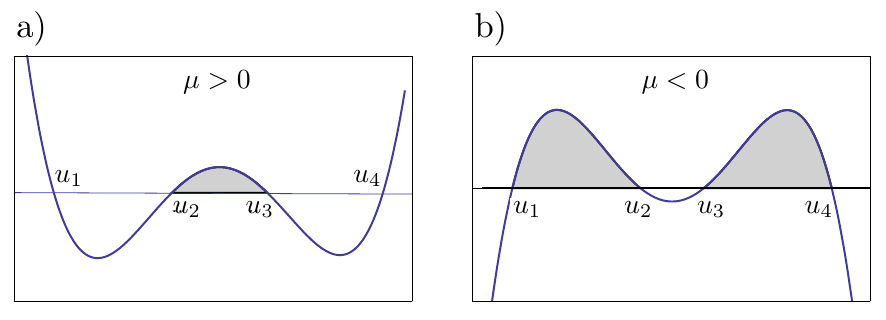}
  \caption{Potential curve $Q(u)$ for traveling wave solutions of the mKdV equation: (a) $\mu>0$; (b) $\mu <0$}
  \label{potential}
\end{figure}

The phase velocity $U$ is expressed in terms of the $u_j$'s as
\begin{equation}\label{el3}
  U=-\frac{1}{2} (u_1u_2+u_1u_3+u_1u_4+u_2u_3+u_2u_4+u_3u_4).
\end{equation}
The ODE (\ref{eq3}) can be viewed as a nonlinear undamped oscillator equation with the ``potential'' $-Q(u)$.
Since finite real-valued motion occurs only in the intervals of $u$, where $Q \ge 0 $ (see figure~\ref{potential}), the structure of the solutions crucially depends on $\hbox{sgn}\,\mu$.

Integration of the ODE (\ref{eq3}) for both orientations of the curve
$Q(u)$ shown in figure \ref{potential} leads generically to solutions
expressed in terms of elliptic functions.  The full
classification of solutions arising when all four roots $u_j$ are real
is presented in Appendix A.
 Some of
these solutions are qualitatively analogous to the cnoidal wave
solution (\ref{cnoidal}) of the KdV equation exhibiting two distinct
limits: a zero-amplitude harmonic wave and a soliton.  At the same
time, the fact that the potential $Q(u)$ is a quartic polynomial gives
rise to a number of qualitatively new features. For example, unlike the
KdV traveling waves, periodic solutions of the mKdV equation exhibit
solitons of both polarities for each sign of $\mu$. Moreover, there
are entirely new families of solutions not encountered in the KdV case. For
$\mu>0$ these are {\it kinks} -- smooth heteroclinic transitions
connecting two  equilibria $u_1=u_2$ and $u_3=u_4$ (note that $u_1+u_4=0$ due to (\ref{constr})),
\begin{equation}\label{kink0}
u= \pm \frac{1}{2} (u_4-u_1) \tanh[ \frac12 (u_4-u_1) \eta] \, .
\end{equation}
The traveling speed for both types of kinks is $U=u_1^2$, which agrees
with the classical shock speed (\ref{rh2}).

For $\mu<0$ there is a ``non-KdV'' family of {\it nonlinear
  trigonometric solutions} corresponding to the merger of two of the
roots, say $u_1$ and $u_2$, while the solution oscillates between the two
remaining troots $u_3 \ne u_4$, see (\ref{eq13c}).

\subsubsection{Modulation equations}
The purpose of this section is to explore the structure of the mKdV
modulation equations as a system of hyperbolic conservation laws.
Modulation equations for the mKdV equation (both signs of $\mu$) were
first derived by Driscoll and O'Neil \cite{dron76} using the original
Whitham averaging procedure applied to the first three mKdV
conservation laws. The resulting system describes the slow evolution
of the roots $u_1(x,t), u_2(x,t), u_3(x,t)$, uniquely defining the
(locally) periodic solution ((\ref{el1}) for $\mu>0$ or (\ref{eq11})
for $\mu<0$) and its physical properties (period-mean, wavelength,
amplitude etc.).  By employing a number of non-trivial relationships
between elliptic integrals it was shown in \cite{dron76} that the
linear combinations

\begin{equation}
\label{lu}
\lambda_1=\frac{1}{2} (u_1+u_2), \quad \lambda_2=\frac{1}{2}(u_1+u_3), \quad \lambda_3=\frac{1}{2}(u_2+u_3)
\end{equation}
are Riemann invariants of the mKdV modulation system. This result was inspired by the original finding of Whitham \cite{wh65} for the KdV equation.  It transpired later \cite{pav95}, \cite{ksk-04}, and more generally in \cite{kamch2012} using the methods of finite-gap spectral  theory \cite{kamch2000}, that the modulation system for the mKdV equation can be mapped {\it onto} the KdV modulation system (\ref{kdvW}).
% (note the different expressions for the characteristic velocities $V_j(r_1, r_2, r_3)$ for the   negative and positive dispersion cases, see (\ref{plusminus})). 
Specifically,
 the mKdV modulation system has the form
\begin{equation} \label{mkdvW}
  \frac{\partial  \la_i}{\partial t} + W_i(\boldsymbol {\la}) \frac{\partial \la_i}{\partial x}  =0 \, , \quad i=1,2,3 \, ,  \end{equation}
where  the characteristic speeds $W_3(\boldsymbol {\la})\le W_2(\boldsymbol {\la})\le W_1(\boldsymbol {\la})$  are related to the KdV-Whitham characteristic speeds (\ref{vi}), (\ref{plusminus}) by
\begin{equation}
\label{V-}
\mu< 0: \qquad  W_i(\boldsymbol{\la})  = V_{4-i}(\mathbf {r}) ,
\end{equation}
\begin{equation}
\label{V+}
\mu>0:  \qquad W_i(\boldsymbol{\la})  = V_{4-i} (\mathbf {r^*}), \quad \mathbf {r^*} = (r_3, r_2, r_1)\, .
\end{equation}
Here, the KdV-Whitham Riemann invariants $r_j$ are expressed in terms of the mKdV-Whitham Riemann invariants $\la_j$ as:
\begin{equation}\label{rla-}
\mu<0: \qquad r_1= 3\lambda_3^2, \quad r_2=3\lambda_2^2,  \quad r_3= 3\la_1^2,
\end{equation}
\begin{equation}\label{rla+}
\mu>0: \qquad r_1= -3\lambda_1^2, \quad r_2=-3\lambda_2^2,  \quad r_3= -3\la_3^2.
\end{equation}
 In each case $\mu>0, \mu<0,$ we have for the modulus $m_1, m_2$ respectively (see (\ref{el2}), (\ref{eq12})), the same formula
{
\beq\label{m12}
m_j=\frac{r_2-r_1}{r_3 - r_1 }, 
 \eeq
\Mnote{which is identical to} the KdV expression (\ref{vm}).}  The admissible sets of inverse formulae for $\lambda_j({\bf r})$  (and hence for $u_j({\bf r})$)  are selected by  the inequalities
\begin{equation*}
%\label{ ineq12}
\la_1 \le \la_2 \le \la_3 , \qquad r_1 \le r_2 \le r_3 \, .
\end{equation*}
Thus,  for $\mu<0$ one has either
\begin{equation}
\label{lar1-}
\la_1 = -\sqrt{r_3/3}, \  \la_2 = -\sqrt{r_2/3}, \  \la_3 = -\sqrt{r_1/3}
\end{equation}
or
\begin{equation}
\label{lar2-}
\la_1 = -\sqrt{r_3/3}, \  \la_2 = -\sqrt{r_2/3}, \  \la_3 = \sqrt{r_1/3} \, .
\end{equation}
We note that relations (\ref{lar1-}) correspond to  modulations of the periodic solution  (\ref{eq11}) occurring in the interval $u_3 \le u \le u_4$ while relations (\ref{lar2-}) correspond to modulations of the counterpart solution  occurring  in $u_1 \le u \le u_2$.  

Similarly, for $\mu>0$ one has either
\begin{equation}
\label{lar1+}
\la_1 = -\sqrt{-r_1/3}, \  \la_2 = -\sqrt{-r_2/3}, \  \la_3 = -\sqrt{-r_3/3}
\end{equation}
or
\begin{equation}
\label{lar2+}
\la_1 = -\sqrt{-r_1/3}, \  \la_2 = -\sqrt{-r_2/3}, \  \la_3 = \sqrt{-r_3/3}.
\end{equation}
The asymmetry in the expressions (\ref{lar1-}) and  (\ref{lar2-})  ((\ref{lar1+}) and (\ref{lar2+})) is due to the use of three variables $\la_j$ (\ref{lu}) replacing the four original quantities $u_j$ (see (\ref{lu})), related by the condition (\ref{constr}). 

Non-uniqueness in the determination of the vector $\boldsymbol {\la}=(\la_1, \la_2, \la_3)$ for a given  vector $\mathbf{r}=(r_1,r_2,r_3)$ is due to the invariant transformation 
\begin{equation}\label{invariance0}
u \to -u,  \quad u_i \to - u_{5-i}\, , \quad i=1,2,3,4
\end{equation}
 of the mKdV traveling wave solutions (see Appendix A) enabling two different sets of $\{ u_j\}$ to be mapped onto the same set $\{ r_j\}$.
Indeed, one can readily see that  the sets (\ref{lar1-}) and (\ref{lar2-}) (as well as  (\ref{lar1+}) and (\ref{lar2+})) are related via the transformation (\ref{invariance0}). The choice of the actual set $\{ \la_j\}$ is determined by the initial or boundary conditions for the Whitham equations.  It is also worth mentioning that the  mapping
$\{ \la_1, \la_2, \la_3 \} \mapsto \{ r_1,r_2,r_3\}$ can be viewed as the modulation theory counterpart of the Miura transformation connecting the KdV and mKdV equations \cite{miura}. 

The quadratic nature of the transformations (\ref{rla-}), (\ref{rla+})  implies that the mKdV-Whitham system (\ref{mkdvW}) is neither genuinely nonlinear nor strictly hyperbolic. Indeed one can see from 
(\ref{V-}), (\ref{lar2-}) for $\mu <0$ and  (\ref{V+}), (\ref{lar2+}) for $\mu>0$  that coinciding characteristic velocities   are admissible:  $W_2(\la_1, -\la_3, \la_3) = W_3(\la_1, -\la_3, \la_3)$, implying that system (\ref{mkdvW}) is non-strictly hyperbolic. Non-genuine nonlinearity then follows \cite{dafermos}.

It is instructive to derive the non-genuine nonlinearity property of the mKdV-Whitham system (\ref{mkdvW}) directly, without invoking non-strict hyperbolicity.

Let $\mu<0$. Consider the derivative
\begin{equation}
\label{der}
\frac{\partial W_j}{\partial \la_j} = 6\la_j \frac{\partial V_{4-j}}{\partial r_{4-j}} \, .
\end{equation}
Since $\frac{\partial V_j}{\partial r_j} \ne 0,  \ j=1,2,3$ due to genuine nonlinearity of the KdV-Whitham system,
the derivative (\ref{der}) can vanish only if  the value $\lambda_j=0$ is admissible for some $j \in \{ 1,2,3\}$. It follows from (\ref{lar1-}), (\ref{lar2-}) that one has $\la_j=0$ only when $r_j=0$.
From the inequality $r_1 \le r_2 \le r_3$  and  relations (\ref{lar1-}), (\ref{lar2-}) one can see that $r_1=0$  (i.e. $\la_3=0$) is admissible. Therefore the mKdV-Whitham system is not genuinely nonlinear for $\mu<0$. The proof for $\mu>0$ is analogous and involves admissibility of $r_3=0$ in (\ref{lar1+}), (\ref{lar2+}).

\medskip

{\bf Corollary}. The mKdV-Whitham system is genuinely nonlinear in any restricted $\boldsymbol{\la}$-domain not containing an open neighbourhood of  $\la_3=0$.

\medskip
One can see that non-convexity of the hyperbolic flux $f(u)=u^3$ in the mKdV equation gives rise to complex structure of the associated modulation system, a feature not present in the classical, simple-wave DSW theory typified by the KdV DSWs described in \S \ref{sec:kdv-equation-dsws}.
Indeed, the immediate consequence of non-convexity of $f(u)$ is non-genuine  nonlinearity  of the dispersionless limit equation (\ref{inviscid_pde}), for which the characteristic velocity  is $V(u)=3u^2$  and so the derivative $V'(u) = 6u$ vanishes at $u=0$.  This property is naturally inherited by the mKdV modulation system, simply because the dispersionless equation is an exact reduction of the modulation equations obtained both in the harmonic ($m=0$) and soliton ($m=1$) limits (see \cite{el05}). However, the mKdV modulation system (\ref{mkdvW}) exhibits further complexity by possessing  non-strict hyperbolicity. Non-strict hyperbolicity was also observed for the modulation systems associated  with the fifth order KdV equation (the second member of the integrable KdV hierarchy) \cite{pierce_tian2006}, the complex modified mKdV equation \cite{kod2008} and the Camassa-Holm equation \cite{grava_pierce_tian_CH}.  For all these equations, as well as for the mKdV equation (\ref{mkdvb1}) with  $\nu=0$, $\mu<0$, non-strict hyperbolicity is responsible for the occurrence of a new fundamental structure, a {\it contact DSW}, which will be described below.

To conclude this section we note that, in spite of the same non-genuinely nonlinear inviscid/dispersionless limit shared by the mKdVB and mKdV equations, non-strict hyperbolicity does not enter the  mKdVB TW shock theory due to the scalar nature of the problem.

\subsubsection{Classical mKdV DSW solutions and admissibility conditions}
\label{sec:DSW_admiss}
The description of classical DSW solutions occurring in the Riemann
problem for the mKdV equation requires obtaining self-similar
simple-wave solutions of the modulation system, in which all but one
of the Riemann invariants are constant. Such solutions exist as long
as the modulation system is genuinely nonlinear (see, e.g.,
\cite{lax2}). The mKdV DSW admissibility conditions will then include,
along with Lax-type causality conditions (\ref{laxdsw}), an extra
(convexity) condition ensuring that the modulation equations are
genuinely nonlinear in the relevant restricted domain of dependent
variables.  Due to the hyperbolic, albeit not necessarily strictly
hyperbolic, nature of the problem, the necessary and sufficient
condition for genuine nonlinearity is that the interval defined by the
initial discontinuity (\ref{RP}) must not contain an open neighborhood
of the point $u=0$.  In that case, the mapping (\ref{rla-}) between
the KdV and mKdV modulations is one-to-one and the mKdV DSW modulation
solution has the form of a self-similar expansion fan analogous to
(\ref{GPKdV}).  To avoid ambiguity, we assume below that $u_+ \ne 0$
and consider the special case $u_+=0$ separately.  Also, as we shall
see, the case $u_-=0$ does not involve DSW formation. Thus we shall be
assuming that $u_-u_+ \ne 0$ in our formulation of the DSW
admisibility conditions.

\begin{figure}
  \centering
  \includegraphics{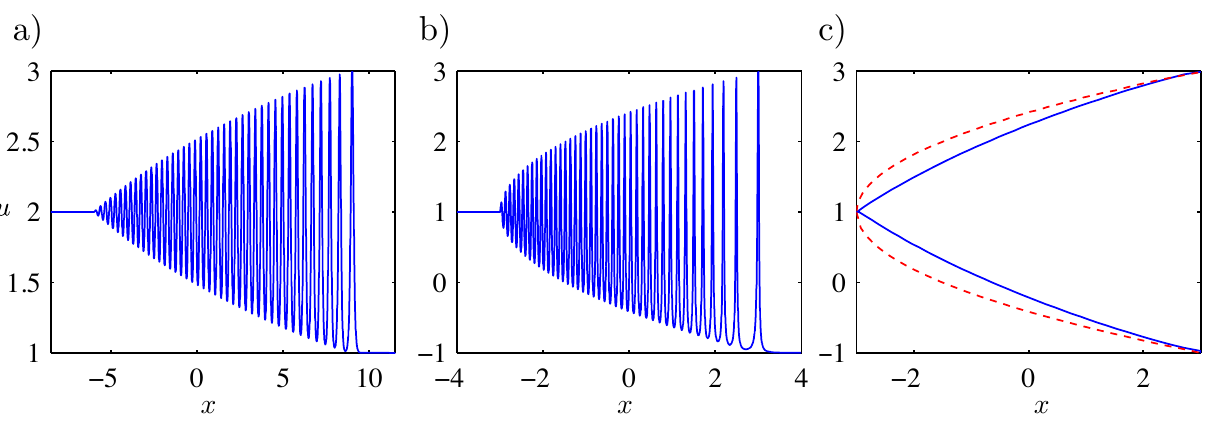}
  \caption{Dispersive shock waves for the mKdV equation,  constructed at $t = 1$ from mKdV modulation theory. (a) Classical DSW, $\mu =
    -2 \! \cdot \! 10^{-2}.$ (b) Non-Classical contact DSW, $\mu = -5 \cdot\! 10^{-4}$.       (c) Envelopes of
    the DSW (solid, martini glass shape, scaled for comparison) and
    the CDSW (dashed, bordeaux wine glass shape).   }
  \label{fig:bordeaux_martini}
\end{figure}

For $\mu <0$ we have
\begin{equation}\label{GPmKdV-}
  r_1 = 3u_+^2, \quad r_3=3u_-^2, \quad V_2(3u_+^2, r_2,
  3u_-^2)=\frac{x}{t} \, , 
\end{equation}
where the connection between $r_j$'s and the mKdV modulation parameters
$\la_j$ is given by (\ref{lar1-}) or (\ref{lar2-}) (see the criteria
below). The DSW edge speeds obtained from (\ref{GPmKdV-}) are:
\begin{equation} \label{mkdv-dsw_speed1}
s_-=3u_+^2 - 3\tilde \Delta,  \quad s_+ = 3u_+^2 +  2 \tilde \Delta,  \quad  \tilde \Delta = u_-^2 - u_+^2.
\end{equation}
This modulation solution, inserted into the periodic TW (\ref{eq11}) gives and approximate description of the mKdV DSW and
is shown in figure \ref{fig:bordeaux_martini}(a) for $(u_-,u_+) =
(2,1)$.  Similar to the KdV DSW case, the weak limit $u(x,t, \mu)$
as $\mu \to 0$  does not coincide with the Lax shock (\ref{shock1}) but is given by the period-mean value $\bar u(x,t)$.  Also, we re-iterate that the exact phase of the TW solution (\ref{eq11}) in the mKdV DSW
is undetermined within the leading-order modulation theory.  This does not affect the macroscopic characterization of the DSW, which is fully determined via the parameters $r_1, r_2, r_3$ by the modulation solution (\ref{GPmKdV-}).

The solution (\ref{GPmKdV-}) is subject to the admissibility conditions
\begin{equation}
\label{adm-}
|u_-|>|u_+| \, , \qquad  u_-u_+> 0 \, ,
\end{equation}
{The meaning of inequalities (\ref{adm-})  is elucidated when they are written in terms of the flux function $f(u)=u^3$}, 
\begin{equation}\label{adm1}
f'(u_-) > f'(u_+), \quad  f''(u_-) f''(u_+ )>0\, .
\end{equation} 
The first of  the conditions (\ref{adm-}) is a mKdV equivalent of the  general DSW causality conditions (\ref{laxdsw}), which guarantee that the dispersionless limit characteristics $x=c_{\pm}t+x_0$, where $c_{\pm}=f'(u_{\pm})=3u_{\pm}^2$, transfer initial data into the DSW region.  Note that this condition {\it does not} coincide with the counterpart Lax entropy condition (\ref{laxm1}) for the non-convex conservation law (\ref{cube1}),
which says that the wave is a Lax shock if and only if $\mathrm{sgn}(u_-) u_+$ lies between $-\mathrm{sgn}(u_-)u_-/2$ and $\mathrm{sgn}(u_-)u_-$.
This contrasts with the KdV case, where the DSW  causality condition does coincide with the Lax entropy condition (see \S \ref{sec:kdv-equation-dsws}).

The second condition in (\ref{adm-}) and more generally in
\eqref{adm1} is the {\it convexity condition}, which guarantees that
the initial step range does not include the inflection point $u=0$ of
the hyperbolic flux, thus ensuring genuine nonlinearity of the
modulation system for the solution involved.  Therefore, the convexity
condition guarantees a {\it single-wave} regularization of the
initial step. One can see that for the KdV equation, the second
condition (\ref{adm1}) is always satisfied so that any initial jump
can be regularized by a classical DSW, provided the DSW causality
condition holds.

When $\hbox{sgn} (u_-) = \hbox{sgn}(u_+)$, this sign indicates which
of the inverse mappings (\ref{lar1-}) or (\ref{lar2-}) should be used
to obtain the modulation $(\la_1, \la_2, \la_3)$ for the relevant
periodic wave (solution (\ref{eq11}) for DSW$^-$ or its reflection via
(\ref{invariance}) for DSW$^+$).  If $u_-> 0$ then one needs to use
relation (\ref{lar2-}) to describe modulations in the DSW$^+$, while
for $u_- < 0$ one has to use (\ref{lar1-}) to describe modulations in
the DSW$^-$.  The polarity $p$ of the DSW ( see \S
\ref{sec:kdv-equation-dsws}) satisfies
%\FIX{Derivation? Define polarity}  \FIX{\color{red}  The easiest way to `explain' the formula for $p$ is to refer to the classification in 5.2. There are just 4 relevant cases, so these can be readily summarised into (\ref{eq:9}). Otherwise we would need to explain part of the classification here.}
\begin{equation}
  \label{eq:9}
  p = - \mathrm{sgn}(\mu u_-),
\end{equation}
where $p = 1$ corresponds to an elevation wave and $p = -1$ corresponds to a depression wave at the DSW solitary wave edge (see \S \ref{sec:kdv-equation-dsws}). {Formula (\ref{eq:9})  follows from the analysis of the four admissible DSW configurations (two for each sign of $\mu$)  summarized in \S \ref{sec:riem-probl-class-plus}, \ref{sec:class_mu_neg} below.}

In more general terms, 
\begin{equation}
  \label{eq:12}
  p = - \mathrm{sgn}(\mu f''(u_-)) .
\end{equation}
The behaviors of characteristics for the modulation system in the DSW
region are analogous to those for the KdV case as in figure
\ref{KdVxt}.  However, KdV exhibits only the DSW$^+$ for $\mu<0$ (see
figure \ref{comp_riemann}).  In contrast, (\ref{eq:12}) demonstrates
that mKdV's non-convex flux allows for both DSW$^+$ and DSW$^-$
regularizations without changing the dispersion sign.  It is the
product of the dispersion convexity \textit{and} the hyperbolic flux
convexity in (\ref{eq:12}) that determines a DSW's polarity.  On the
other hand, the DSW orientation (see \S \ref{sec:kdv-equation-dsws})
is determined by the dispersion sign alone.  For the mKdV equation
with a fixed dispersion sign, DSW$^+$ and DSW$^-$ have the same
orientation, i.e. the same relative positions (trailing, leading) of
the solitary wave and harmonic edges.

The mKdV DSW modulation solution for the case $\mu>0$ is obtained from
(\ref{GPmKdV-}) by applying the transformation (\ref{plusminus}) and
then using one of the inverse mappings (\ref{lar1+}), (\ref{lar2+}) to
get the modulation $(\la_1, \la_2, \la_3)$ of the periodic solution
(\ref{el1}). One uses the mapping (\ref{lar1+}) for a DSW$^+$ and
(\ref{lar2+}) for a DSW$^-$. The DSW admissibility conditions have the
same form (\ref{adm-}) but the DSW polarity is opposite to that
occurring in the case $\mu<0$, see (\ref{eq:9}).  The speeds of the
trailing and leading edges for the DSWs of both polarities are (cf.\
(\ref{mkdv-dsw_speed1})): $s_- = 3 u_+^2 + \tilde \Delta$, $s_+= 3
u_+^2 + 6 \tilde \Delta$.

In all cases, the leading ($\mu<0$) or trailing ($\mu>0$) soliton
amplitude in the classical DSW is $a=2(|u_-| - |u_+|)$.
 
If one or both of the admissibility conditions (\ref{adm-}) fail for a
given pair $(u_-, u_+)$, then the regularization of a step (\ref{RP})
via a single classical DSW is not possible. If only the DSW causality
condition is violated but the convexity condition holds true, then the
dispersive resolution occurs via a rarefaction wave.  The mKdV
rarefaction wave solutions for $|u_+|>|u_-|$ for $t \gg 1$ have the
leading order asymptotic form (\ref{rare2}), same as for the mKdVB
equation (\ref{mkdvb1}).  In contrast to the counterpart rarefaction
wave solution of the mKdVB equation, the corners of the mKdV
rarefaction wave are smoothed by linear dispersive oscillations whose
wavelength $\sim \sqrt{|\mu|}$ and amplitude decays like $t^{-1/2}$
\cite{leach2012}, similar to the asymptotic behavior in the KdV
rarefaction wave described in Sec.~\ref{sec:kdv-equation-dsws}. Note
that only the $\mu>0$ case is discussed in \cite{leach2012}.

In the remaining two ($\mu < 0$ and $\mu > 0$) classical
DSW-inadmissible cases involving the breaking of the convexity
condition, ``non-classical'' dispersive regularizations occur; these
will be described in the following subsections.  The borderline case
when $u_+=0$ can be formally viewed as being of either classical or
non-classical regularization type, and will be discussed briefly.
 
\subsubsection{Non-classical DSWs}
\label{sec:nonclass_DSW}

We first consider the case $\mu>0$. Here, we already have an exact
heteroclinic solution to the mKdV equation -- the kink solution
(\ref{kink0}) -- that violates the convexity condition. Indeed, the
kink with the minus sign in (\ref{kink0}) connects two states $u_->0$
at $x \to - \infty$ and $u_+=-u_-$ at $x \to \infty$ so that
$u_-u_+<0$. The antikink with the plus sign in (\ref{kink0}) is the
reflection of the kink under the transformation (\ref{invariance}) and
has the same non-convexity property. One can see that kinks violate
both the DSW causality (\ref{adm-}) and Lax entropy (\ref{laxm})
conditions as well. Unlike the classical, expanding, DSW, the kink is
a moving front approximating a shock wave, so must be evaluated
against the Lax entropy condition.  The Lax entropy condition agrees
with the counterpart DSW causality condition (\ref{laxdsw}) if one
takes $s_- = s_+ =s_{\rm K}$.  However, the kink speed is $s_{\rm
  K}=u_-^2<f'(u_{\pm})=3u_-^2 = 3u_+^2$ so that kinks are
undercompressive as described in \S(\ref{ucw}). From the discussion of traveling waves corresponding to undercompressive shocks in \S~\ref{sec:mkdv_trav},
  we observe that the kink solution (\ref{kink}) is the
strong $\delta \to 0$ limit of the diffusive-dispersive
undercompressive shock solution (\ref{ucshocksa}), (\ref{ucshocks2}).  Thus we identify
kinks as {\it undercompressive DSWs}. Note that the undercompressive
nature of mKdV kinks was discussed in \cite{klu2007}.  In
Sec.~\ref{sec:riem-probl-class}, we will incorporate them into the
general context of the mKdV Riemann problem classification.

An exploration of solutions to the Riemann problem (\ref{RP}) for the mKdV equation with $\mu > 0$ was undertaken in \cite{chanteur1987}.  { The amplitude of the DSW's lead soliton was determined}  using an inverse scattering approach and the necessary and sufficient condition $u_- u_+ < 0$--breaking of the convexity condition (\ref{adm-})--for the formation of a kink (termed a double-layer solution) was identified.  Numerical simulations were used to identify kinks, classical DSWs, rarefaction waves, and double wave combination solutions.  In the next section, we will provide the full classification of the mKdV Riemann problem for both $\mu >0$ and $\mu < 0$.

The Riemann problem  (\ref{RP}) for the mKdV equation with $\mu>0$ and the particular initial data $u_+=-u_-$ was studied in \cite{leach2012}, where a detailed solution was obtained using matched asymptotic expansions. The long-time asymptotics of this solution is dominated by a kink, which agrees with the modulation theory approach where  kinks are used as one of the `building blocks'  in the construction of long-time asymptotic solutions in more general Riemann problems, see \cite{esler2011}, \cite{kamch2012}.
 
\medskip Now we turn to the case $\mu<0$. Instead of kinks, a peculiar
type of DSW occurs, which we term a {\it contact} DSW (CDSW). A CDSW,
similar to a kink, connects two conjugate states $u_->0$ at $x \to -
\infty$ and $u_+=-u_-$ at $x \to \infty$ (this is CDSW$^+$) but,
unlike the kink, has an oscillatory structure resembling that of a
classical DSW.  Contact DSWs represent the dispersive counterparts of
single-sided contact discontinuities (or sound shocks) known in the
theory of hyperbolic conservation laws (see, e.g., \cite{dafermos}).

The CDSW$^+$ solution for the mKdV equation is described by the
special modulation $\la_1, \la_2, \la_3$ in which $\la_2=\la_3$ so
that the elliptic parameter $m=0$ (see (\ref{m12})) throughout the
wave train, but the wave amplitude $a=u_4-u_3=-4\la_3 \ge 0$. This
sharply contrasts with KdV modulation theory where $m=0$ implies zero
wave amplitude.  The CDSW$^+$ modulation solution has the form
\begin{equation}
\label{cdsw_mod}
  \begin{split}
    -\lambda_1&=u_-, \quad \la_2=\la_3, \\  
    W_2(-u_-, \la_3, \la_3)&=W_3(-u_-, \la_3, \la_3)=\frac{x}{t} \, ,
  \end{split}
\end{equation}
or explicitly, using (\ref{V-}), (\ref{v2=v1}) and (\ref{rla-}),
\begin{equation*}
%\label{ }
\la_3=-\frac{1}{\sqrt{6}} \sqrt{\frac{x}{t} +3 u_-^2}  \, .
\end{equation*}
Since $\la_2=\la_3$ implies $u_1=u_2$ and the oscillations in the
CDSW$^+$ are confined to the interval $u_3\le u \le u_4$, we conclude
that the `carrier' wave in a CDSW$^+$ is the nonlinear trigonometric
solution (\ref{eq13c}), and that is why this type of DSW has been
termed a sinusoidal or trigonometric DSW elsewhere \cite{march2008},
\cite{kamch2012}.  At the leading edge of a CDSW$^+$,
$\la_1=\la_2=\la_3$ (i.e. $u_1=u_2=u_3$), which manifests a `bright'
algebraic soliton (\ref{eq13e}) with amplitude $a=-4\la_1=4u_-$ and
speed $U=3\la_1^2$. The algebraic soliton propagates on the background
$\bar u= \la_1$ which, according to the matching regularization
conditions, must be matched to the constant state $u=u_+$
upstream. Then (\ref{cdsw_mod}) yields that a CDSW$^+$ can only
connect the conjugate states $u=u_->0$ and $u=u_+=-u_-<0$.  At the
CDSW$^+$ trailing edge, the wave amplitude $a=0$, i.e.,
$\la_2=\la_3=0$. Thus, the speeds of the CDSW$^+$ trailing and leading
edges are: $s_-=W_2(-u_-, 0, 0)=-3u_-^2$, $s_+=W_2(-u_-, -u_-,
-u_-)=3u_-^2=3u_+^2$.  This modulation solution, inserted into the
periodic TW (\ref{eq13c}), is shown in figure
\ref{fig:bordeaux_martini}(b). Similar to the classical KdV and mKdV DSWs, the exact phase of the CDSW is not determined by the modulation solution (\ref{cdsw_mod}).
 
The fundamental difference between the modulation solution
(\ref{GPmKdV-}) for a classical DSW and the solution (\ref{cdsw_mod})
for a CDSW is that the solution of (\ref{cdsw_mod}) represents a {\it
  double characteristic} fan, the corresponding distinct Riemann
invariants being $\la_3$ and $-\la_2$. This solution is possible owing
to the nonstrict hyperbolicity of the mKdV modulation system.  Note
that the trigonometric structure of the underlying mKdV traveling wave
solution (\ref{eq13c}) is not the defining feature of CDSWs.  See
\cite{kod2008} where a qualitatively similar type of modulation occurs
for a wave with $m \ne 0$. The more fundamental contact property is
that the CDSW leading edge is a {\it triple characteristic},
tangential to the external dispersionless characteristic with
characteristic velocity $3u^2$. As a result, a CDSW$^+$ could be
matched with a rarefaction wave at the leading edge.  The composite
CDSW-RW solutions will be considered in \S \ref{sec:class_mu_neg}.  We
also mention the qualitative difference between the appearances of the
envelopes for classical and contact DSWs: the classical DSW typically
has a ``martini-glass'' shape due to the asymptotic behavior of the
amplitude $a \sim (x - x_-)$ near the harmonic edge at $x=x_-=s_-t,$
whereas the CDSW envelope defined by $a = -4\la_3 \sim \sqrt{x-x_-}$
has a ``Bordeaux-glass'' shape as shown in figure
\ref{fig:bordeaux_martini}(c).  We note that these envelope
distinctions may not be obvious in direct numerical simulations due to
contributions not included in the Whitham multiple-scale solution,
which could become important in the vicinity of the harmonic edge
defining the appearance of the envelope. See \cite{grava_klein2007} for
the corresponding KdV analysis of higher order corrections to the
harmonic edge.

The description of a CDSW$^-$, where the oscillations occur between
the other pair of roots, $u_1 \le u \le u_2$, is analogous and
involves the reflection of the traveling wave formulae (\ref{eq11}) --
(\ref{eq13e}) via the transformation (\ref{invariance}) and the
relationships (\ref{lar2-}) for the modulation variables.

 % Also, the characteristics behavior in another type of composite solutions representing a combination of a CDSW and a classical DSW reveals the properties of CDSWs, which are parallel to the %properties of diffusive-dispersive undercompressive shocks.  

\subsection{Riemann problem classification}
\label{sec:riem-probl-class}
\begin{figure}[ht]
\centerline{\includegraphics{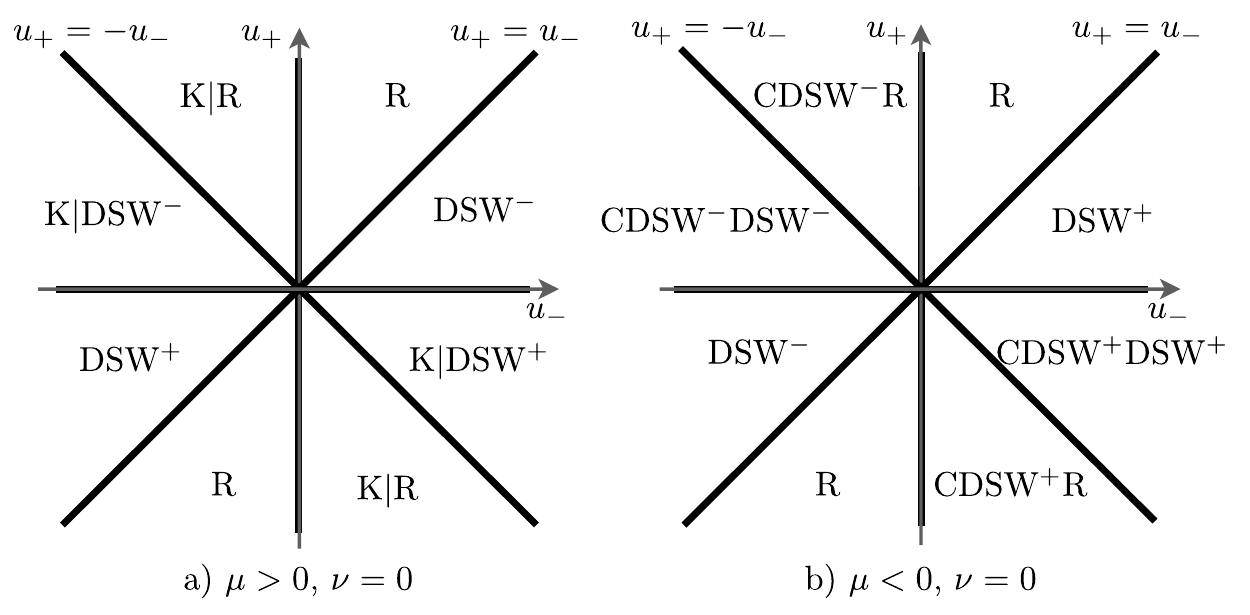}}
\caption{Solutions of the Riemann problem for the mKdV
  equation. Legend: (R) -- rarefaction wave; (DSW$^+$) -- `bright'
  DSW; (DSW$^-$) -- `dark' DSW; K -- kink; (CDSW$^\pm$) -- `contact'
  DSW$^\pm$; ($|$) -- intermediate constant state.}
 \label{riemann_mkdv}
 \end{figure}

The Riemann problem classification for the mKdV equation on the
$u_-$-$u_+$ plane of initial data is constructed by considering the
admissibility conditions (\ref{adm-}) and all the cases when one or
both of them fail. As a result, the $u_-$-$u_+$ plane is divided into
eight equi-spaced sectors, each corresponding to a distinct
fundamental wave regularization pattern (see figure
\ref{riemann_mkdv}). The separation lines are: $u_-=0$, $u_+=0$,
$u_+=u_-$ and $u_+ = -u_-$.  Due to the symmetry inherent in the
causality and convexity conditions \eqref{adm-}, the boundaries
between different solution types are symmetric under a $\pi/4$
rotation.  

% FIGURE

\subsubsection{Classification for $\mu >0$}
\label{sec:riem-probl-class-plus}

The classification is shown in figure \ref{riemann_mkdv}(a). We have
already considered solutions corresponding to the classical DSW
admissible regions defined by the conditions (\ref{adm-}).  The
adjacent regions with $|u_-| <|u_+|$, where the DSW causality condition
fails but the convexity condition holds, correspond to regularization
via rarefaction waves described by (\ref{rare2}). The line $u_+=-u_-$
corresponds to a step regularization via a single kink ($u_->0$) or
antikink ($u_-<0$). These have also been described in \S
\ref{sec:nonclass_DSW}. We now need to describe the remaining two
cases corresponding to the sectors $-u_-<u_+<0$ and
$u_+<-u_-<0$. Conjugate solutions corresponding to the opposite
sectors are obtained via the reflection transformation
(\ref{invariance}).

\medskip Let $-u_-<u_+<0$. The Riemann problem solution is a
double-wave K$|$DSW$^+$ consisting of a kink and a DSW$^+$ connected
by the constant state $u=-u_-$.  The kink speed $s_{\rm K}=u_-^2$, and
the speeds of the DSW$^+$ trailing and leading edges are $s_-= u_-^2 +
2u_+^2 $ and $s_+= 6u_-^2 - 3u_+^2 $ respectively. One can see that
$s_->s_{\rm K}$ so the interval between the kink and DSW$^+$ increases
with time.

If $u_+<-u_-<0$, then the initial step regularization occurs via a
kink and a rarefaction wave (formula (\ref{rare2}) with the minus
sign, and $u_- $ replaced with $ -u_-$) connected by a widening
interval in which $u=-u_-$.

It remains to explain the singular transition between the patterns
DSW$^-$ for $0<u_+<u_- $ and K$|$DSW$^+$ for $-u_-<u_+<0$, which
occurs when crossing the horizontal axis $u_+=0$. The trailing
oscillation in the DSW$^-$ represents a depression (dark) soliton
(\ref{eq7a}) having the amplitude $2(|u_-| - |u_+|)$ and propagating
with the speed $s_-=2u_+^2 + u_-^2$. When $u_+=0$ the soliton speed
becomes equal to the kink speed $s_{\rm K}=u_-^2$, and the left slope
of the dark soliton becomes equivalent to a kink. The further decrease
of $u_+$ results in a `peeling' off of the trailing kink from the
remaining wave train, which becomes a DSW$^+$.  See \cite{kamch2012}
for a series of numerical simulations illustrating the analogous
transition for the Gardner equation.

\subsubsection{Classification for $\mu<0$}
\label{sec:class_mu_neg}

The classification of the Riemann problem solution to the mKdV
equation with $\mu<0$ is shown in figure \ref{riemann_mkdv}(b).  The
classical DSW regularizations corresponding to two opposite sectors of
the $u_-$-$u_+$ plane where the admissibility conditions (\ref{adm-})
hold were described in \S \ref{sec:DSW_admiss}.  Similar to the
$\mu>0$ case, the adjacent regions with $|u_-| <|u_+|$, where the DSW
causality condition fails but the convexity condition holds,
correspond to regularization via rarefaction waves described by
(\ref{rare2}). The line $u_+=-u_-$ corresponds to a step
regularization via a single CDSW$^+$ ($u_->0$) or
CDSW$^-$($u_-<0$). These have also been described in \S
\ref{sec:nonclass_DSW}. We now need to describe the remaining two
cases corresponding to the sectors $-u_-<u_+<0$ and $u_+<-u_-<0$. The
conjugate solutions corresponding to the opposite sectors are obtained
via the reflection transformation (\ref{invariance0}).

\medskip Let $-u_-<u_+<0$. The Riemann problem solution is a compound
wave CDSW$^+$DSW$^+$ consisting of a partial CDSW$^+$ and DSW$^+$
attached to each other. The line separating the CDSW$^+$ and the
DSW$^+$ in the $x$-$t$ plane is the characteristic $s_*=6u_+^2 - 3
u_-^2$. The trailing edge of the compound wave CDSW$^+$DSW$^+$ is
$s_-= -3u_-^2$.  The leading edge is $s_+ = 2u_-^2 + u_+^2$.

If $u_+<-u_-<0$, then initial step regularization occurs via a
combination of a fully developed CDSW$^+$ connected at the leading
edge to a rarefaction wave.

\bigskip The classification of the regularization patterns in terms of
the admissibility/non-admissibility conditions for both signs of $\mu$
is presented in Table \ref{tab:cond_posneg}.
\begin{table}[h]
\centering
\begin{tabular}{|c|c|c|}
\hline
{Causality ($|u_-|>|u_+|$)} & Convexity ($u_-u_+>0$) & Regularization  $\mu>0$ ($\mu<0$)\\
\hline
Yes &  Yes &  DSW$^-$ (DSW$^+$) \\
\hline
No &  Yes & R  (R) \\
\hline
Yes & No & K$|$DSW$^+$  (CDSW$^+$$|$DSW$^+$)\\
\hline
No & No & K$|$R (CDSW$^+$$|$R) \\
\hline
\end{tabular}
\caption{Classification of regularization patterns in terms of satisfaction/violation of the admissibility conditions (\ref{adm-}) for $u_-u_+ \ne 0$. The DSW polarity is shown for $u_->0$.}
\label{tab:cond_posneg}
\end{table}%

\section{Comparison}\label{comparison}

As can be seen by the very different analytical approaches outlined in
\S\ref{sec:mkdvb-non-classical} and
\S\ref{sec:riemann-problem-mkdv}, the Riemann problem solutions in the
presence of dispersion with or without diffusion evolve in
fundamentally different ways.  However, the results do exhibit direct
parallels that provide a useful bridge connecting diffusive-dispersive
mKdVB dynamics to purely dispersive mKdV dynamics.  It is our aim in
this section to identify the key differences and commonalities between
these two models.

\subsection{Shock structure}

\begin{figure}
  \centering
  \includegraphics{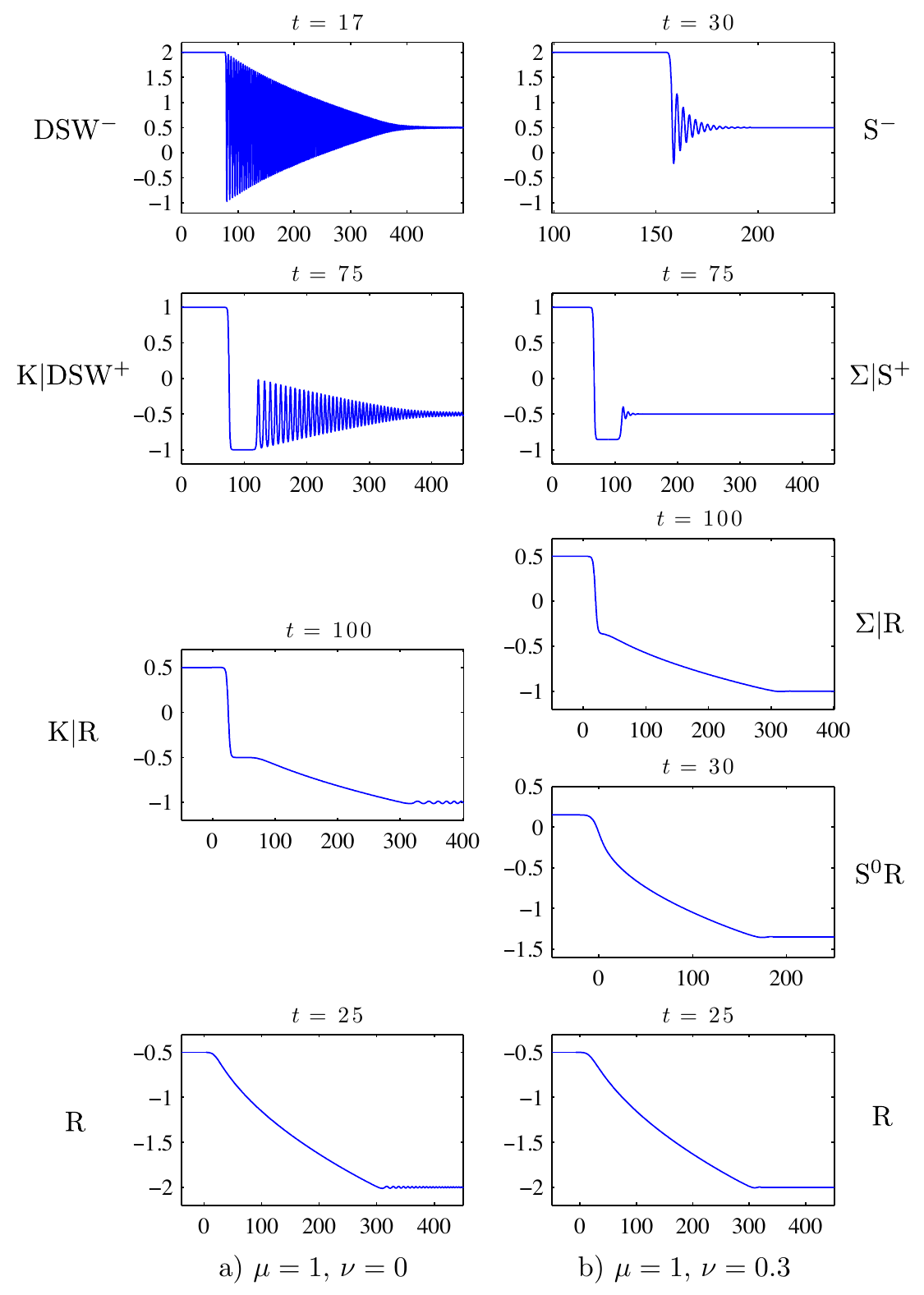}
  \caption{Numerical simulations of the   Riemann problem for
  (a)  mKdV  and (b) mKdVB, with positive dispersion.  The 
  %naming  conventions are determined by the theoretical classifications in
 panel labels identify the types of waves, as in figures \ref{riemann_fig1} and \ref{riemann_mkdv}.}
  \label{fig:numericsPos}
\end{figure}

\begin{figure}
  \centering
  \includegraphics{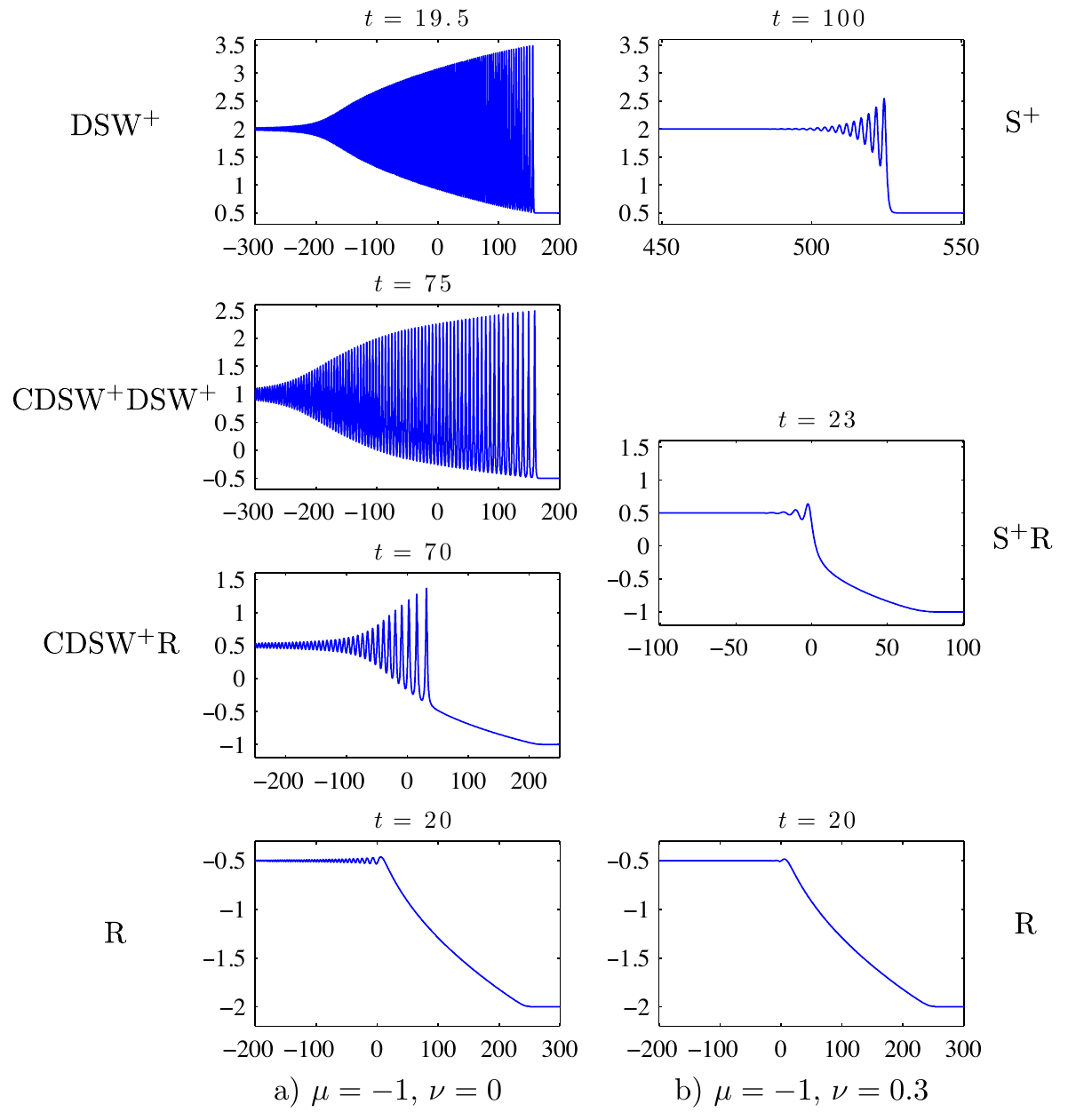}
  \caption{Numerical simulations of the smoothed Riemann problem for
   (a) mKdV  and (b) mKdVB, with negative dispersion.  The 
  % naming conventions are determined by the theoretical classifications in
   panel labels identify the types of waves, as in figures \ref{riemann_fig1} and \ref{riemann_mkdv}.}
  \label{fig:numericsNeg}
\end{figure}

The first and perhaps most obvious distinction between the two model
types can be seen in the spatial structure of the solutions
themselves.  Figures \ref{fig:numericsPos} and \ref{fig:numericsNeg}
display numerical solutions to smoothed versions of the Riemann
problem when $\mu = 1$ and $\mu = -1$, respectively. See the Appendix
for a description of the numerical method used.  The first column of
panels in each figure corresponds to purely dispersive dynamics with
$\nu=0$, whereas the second column corresponds to diffusive-dispersive
dynamics in which $\nu>0$.  The text labels adjacent to each panel
correspond to the classification in the $(u_-,u_+)$ regions of figures
\ref{riemann_fig1}, \ref{riemann_mkdv}.  Due to the invariance of
(\ref{mkdvb1}) under the transformation $u \to -u$, we restrict
attention to the choice $u_- < u_+$, understanding that the analogous
regimes when $u_- > u_+$ exhibit an amplitude reflection so $^+$ waves
become $^-$ waves and vice-versa.

%Generally, m
Many of the solutions exhibit oscillations due to the presence of
dispersion.   The presence of any diffusion damps the oscillation
amplitudes.  In contrast, the purely dispersive case leads to the
continual generation of large amplitude oscillations showcasing the
distinction between diffusive traveling waves approximating Lax shocks
and expanding, dynamic DSWs.  Even the rarefactions are subject to
larger oscillation in the absence of diffusion. Notice, however, that
both undercompressive shocks and kinks are monotone, exhibiting no
oscillation.

Due to the existence of two speeds, DSWs exhibit an orientation
distinguishing the large amplitude, soliton edge from the small
amplitude, harmonic wave edge.  Another DSW characteristic is
polarity, determined by the type of disturbance (elevation or
depression) occurring at the soliton edge.  Both orientation and
polarity are inherited from the corresponding diffusive-dispersive
wave when $\delta = \nu/\sqrt{|\mu|}$ is sufficiently small \MH{(recall
\S \ref{sec:riemann-problem})}, as shown
in figures \ref{fig:numericsPos} and \ref{fig:numericsNeg}. The DSW
orientation is determined by the sign of dispersion.  As in the KdV
equation, the only way to change the polarity of a DSW for a {\em convex}
conservation law is to change the sign of dispersion, recall
(\ref{eq:12}).  A {\em non-convex} flux introduces the possibility of
polarity change without changing the dispersion sign, as in figure
\ref{fig:numericsPos}(a).  This can have physically important
implications e.g. in internal wave dynamics ({the distinction} between
dam-break and lock-exchange undular bores \cite{esler2011}) or in
nonlinear optics/superfluids (\Mnote{the} occurrence of bright solitons in a
defocusing medium \cite{trillo}).

\subsection{Characteristic diagrams}
\label{sec:char-diagr}

A useful means for understanding different wave solutions is through
their characteristic diagrams.  For scalar conservation laws, there is
only one characteristic family.  For the diffusive-dispersive case,
this single characteristic family is sufficient to illustrate the TW
solutions.  However, in the purely dispersive case, shock solutions
are described by a modulated, periodic TW via the Whitham equations
with three distinct characteristic families.  The characteristic
description of single rarefactions, Lax shocks, and classical DSWs for
the mKdV(B) equation is similar to that for the KdV(B) equation shown
in figure \ref{KdVxt}.  

\begin{figure}
  \centering
  \includegraphics{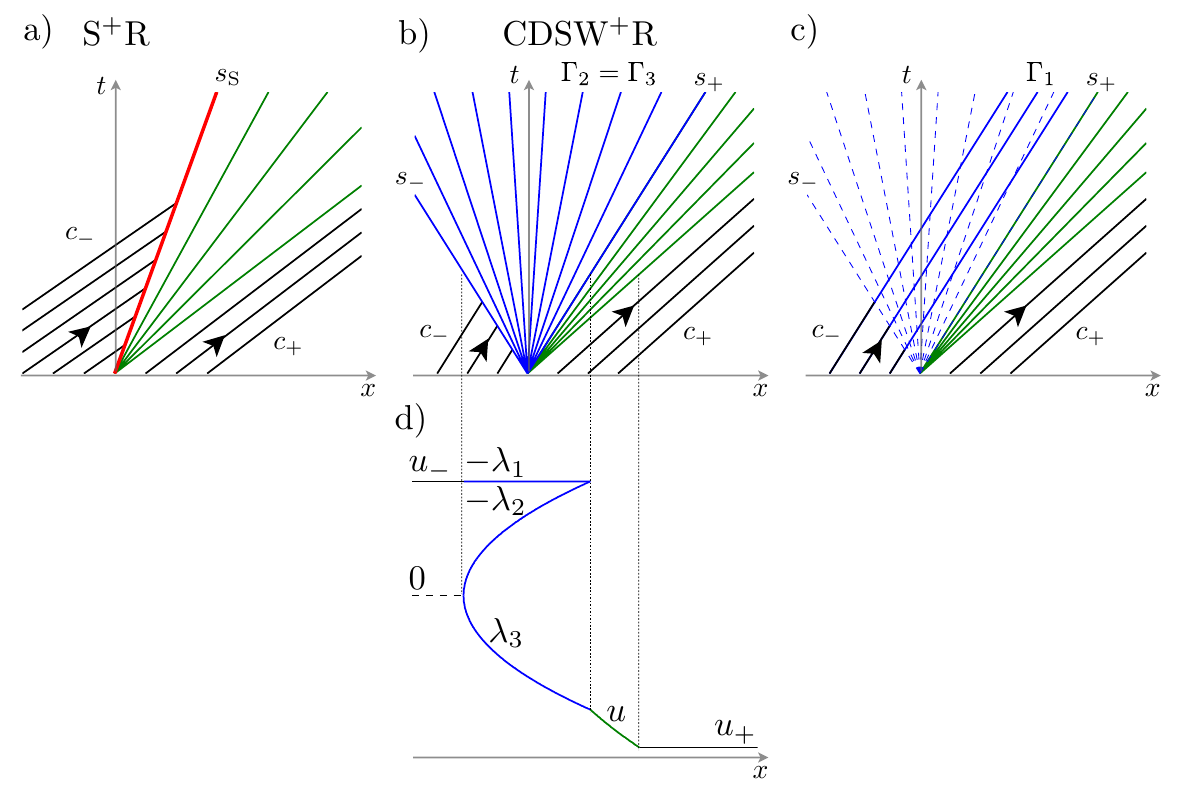}
  \caption{Characteristic diagrams for the composite shock-rarefaction
    (a) and contact DSW-rarefaction (b,c) solutions.  The modulation
    solution (d) and associated three characteristic families
    $\Gamma_1$, $\Gamma_2$, and $\Gamma_3$ (b,c) are shown.  The
    dashed curves correspond to $\Gamma_2 = \Gamma_3$.}
  \label{fig:charNegSR}
\end{figure}

\begin{figure}
  \centering
  \includegraphics{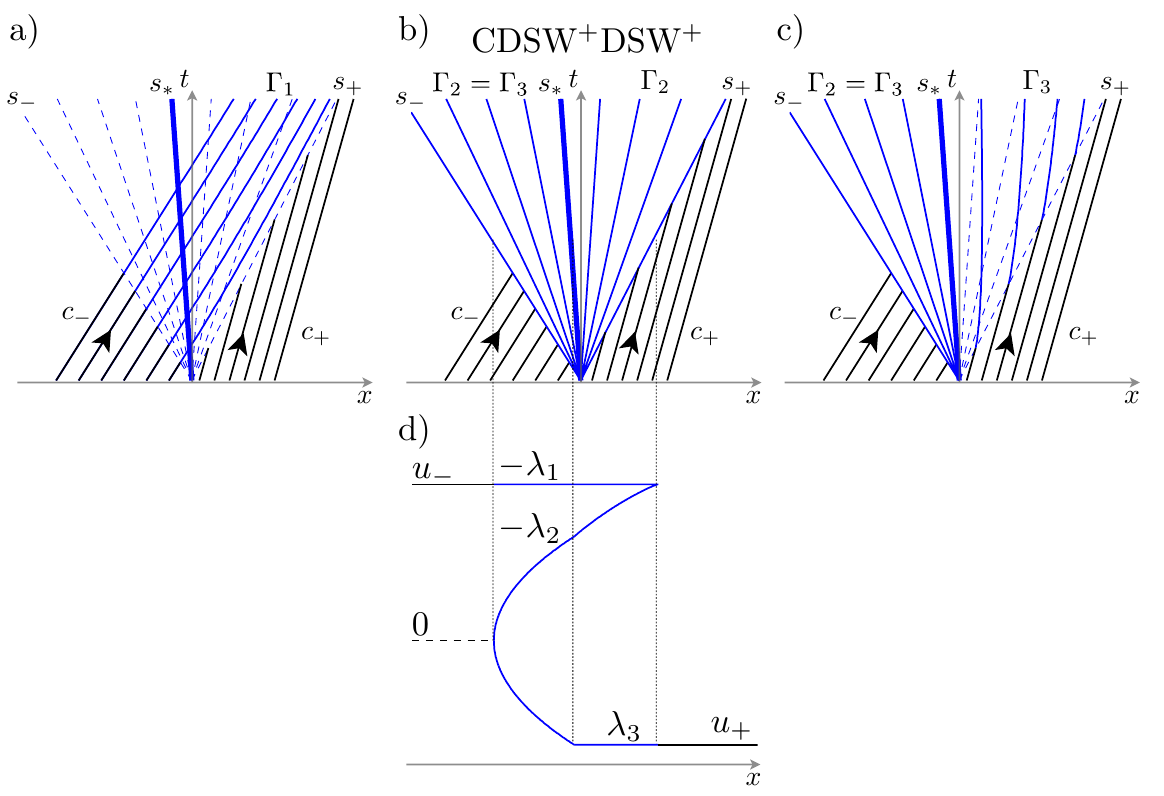}
  \caption{Characteristic diagrams for the three characteristic
    families $\Gamma_1$, $\Gamma_2$, and $\Gamma_3$ of the composite
    CDSW-DSW solution (a-c).  The modulation solution is in (d).  The
    dashed curves correspond to either $\Gamma_2 = \Gamma_3$ (between
    $s_-$ and $s_*$) or $\Gamma_2$ (between $s_*$ and $s_+$).}
  \label{fig:charNegCDSWDSW}
\end{figure}

\begin{figure}
  \centering
  \includegraphics{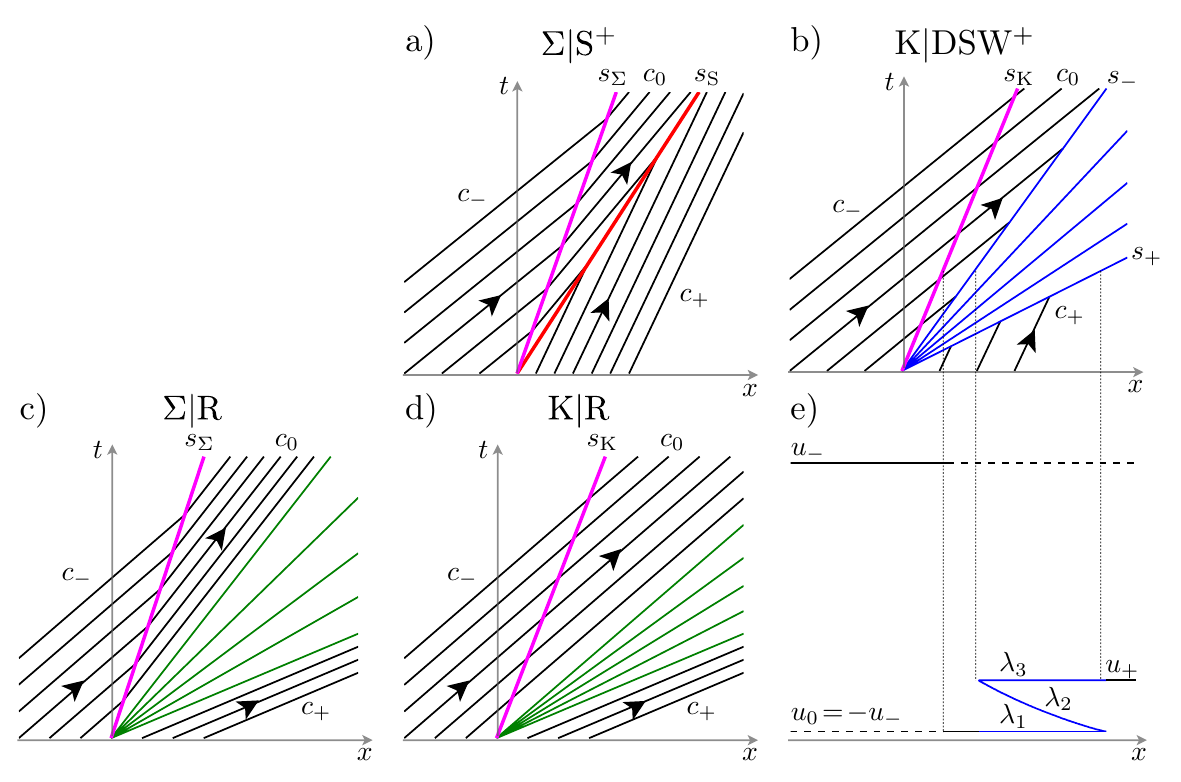}
  \caption{Characteristic diagrams for Riemann problem solutions of
    mKdVB (a,c) and mKdV (b,d) with $\mu > 0$.  The DSW modulation
    solution for Riemann invariants is shown in (e).  }
  \label{fig:charPos}
\end{figure}

Figures \ref{fig:charNegSR}, \ref{fig:charNegCDSWDSW}, and
\ref{fig:charPos} display characteristic diagrams and modulation
solutions for double wave structures in the mKdV(B) Riemann problem
classifications.  The mKdV-Whitham characteristic families are denoted
by $\Gamma_i$, $i = 1,2,3$ where $\Gamma_i$ is the \MH{set of
  characteristic curves $x = x(t)$ satisfying
  $\frac{\mathrm{d}x}{\mathrm{d}t} = W_i$, with $W_i$ the
  characteristic velocity} in (\ref{mkdvW}).  Although we display
generic diagrams, each characteristic is a legitimate one
computed for a specific Riemann problem, i.e., these are not sketches.
We now describe the implications of each of these diagrams.

Figure \ref{fig:charNegSR} compares the shock-rarefaction,
S$^+$\Mnote{R} in (a), and the contact DSW-rarefaction, denoted
CDSW$^+$R in (b-d), operable in the $\mu < 0$ regime. Both the shock
and the leading edge of the CDSW coincide with the trailing
characteristic in the rarefaction, with speed $s_+$.  Moreover, the
CDSW leading edge is a triple characteristic, i.e., $s_+=W_1=W_2=W_3$.
Here, all three modulation characteristics coincide, the reason for
the terminology {\em contact} DSW.  An additional feature of the CDSW
distinguishing it from the classical DSW (recall figure \ref{KdVxt})
is the double characteristic family $\Gamma_2 = \Gamma_3$, maintained
across the entire contact dispersive shock structure.  The remaining
characteristic family passes straight through the CDSW, unchanged from
the dispersionless characteristics with speed $c_-$, so that $c_-=s_+$
also.

Figure \ref{fig:charNegCDSWDSW} depicts a CDSW$^+$DSW$^+$ and allows
for a side-by-side comparison of the non-classical CDSW and classical
DSW characteristics.  In this double wave structure, the CDSW
modulation solution exhibits double characteristics \MH{$W_2 = W_3$ up
  to its interface with the DSW where $s_* = W_2 = W_3$}.  The
$\Gamma_1$ characteristic family carries information through the CDSW
with speed $c_-$ and into the DSW.  The CDSW modulation solution
(\ref{cdsw_mod}) is unfolded by considering the alternative Riemann
invariants $\lambda_3 \le -\lambda_2 \le - \lambda_1$ as in figures
\ref{fig:charNegSR}(d) and \ref{fig:charNegCDSWDSW}(d).  This choice
is natural for the CDSW$^+$ because its corresponding periodic TW
solution (\ref{eq13c}) satisfies $u_3 \le u \le u_4$ (recall figure
\ref{potential}).  Our original choice of Riemann invariants
(\ref{lu}) is more suited for TWs with $u_1 \le u \le u_2$.

When $\mu > 0,$ composite waves involve undercompressive shocks,
denoted $\Sigma$, for mKdVB, or kinks, denoted K, for mKdV.  Their
characteristics are shown in figure \ref{fig:charPos}.  In contrast to
the $\mu < 0$ double wave solutions, each wave in the composite
solution for $\mu > 0$ is separated by an intermediate constant state,
the separation indicated by a vertical line $|.$ The shock-rarefaction
solution S$^0$R for mKdVB, not displayed, is an exception in not
having an intermediate constant; the speed of S$^0$ coincides with the
characteristic speed $c_-$ and forms the trailing edge of the
rarefaction wave.  A defining property of undercompressive shocks is
the passage of a characteristic family through them.  Figures
\ref{fig:charPos}(a,c) show how characteristics enter the
undercompressive shock from the left and exit on the right with an
upward deflection.  In contrast, the corresponding kink solutions in
figures \ref{fig:charPos}(b,d) exhibit characteristics that pass
through with no deflection, just like one of the characteristic
families for the CDSW solutions in figures \ref{fig:charNegSR}(c) and
\ref{fig:charNegCDSWDSW}(a).  Through this similarity, the CDSW can
also be thought of as undercompressive, however without a diffusive
counterpart.  Unlike the DSW and CDSW, the kink is a genuine,
non-modulated TW solution of mKdV.  Its effect in the modulation
description of the K$|$DSW$^+$ of figure \ref{fig:charPos}(e) is a
discontinuous jump from $u_-$ to $-u_-$, the intermediate constant
connecting to the classical DSW.  The K$|$DSW$^\pm$ is the only mKdV
Riemann problem solution that incorporates both modulated (DSW) and
non-modulated (kink) TWs.

\subsection{Zero Diffusion Limit}
\label{sec:zero-diffusion-limit}

The limit $\nu \to 0^+$ for solutions of the mKdVB equation is
singular in that for every nonzero $\nu$, the long time shock behavior
of the Riemann problem is resolved into a heteroclinic orbit or TW
(\ref{ucshocksa}), (\ref{ucshocks2}).  Diffusion introduces a relaxation
mechanism to a steady configuration: TWs are attractors.  But when
$\nu = 0$, the only heteroclinic orbit available is the kink, valid
only for very specific left and right states when $\mu > 0$.  So, the
purely dispersive regularization \MH{for generic left and right
  states} results in dynamic, oscillatory coherent structures.  DSWs
are unsteady, multiscale attractors that combine a periodic TW with an
additional layer of self-similarity, the modulation solution.
Nevertheless, we can see from the numerical results in figures
\ref{fig:numericsPos} and \ref{fig:numericsNeg} a correspondence
between $\nu > 0$ and $\nu = 0$ in most cases.  It is therefore
natural to introduce a mapping of the $\nu > 0$ mKdVB solutions to
$\nu = 0$ mKdV solutions.  As we will see, this mapping is multivalued
and not one-to-one.

\begin{figure}
  \centering
  \includegraphics{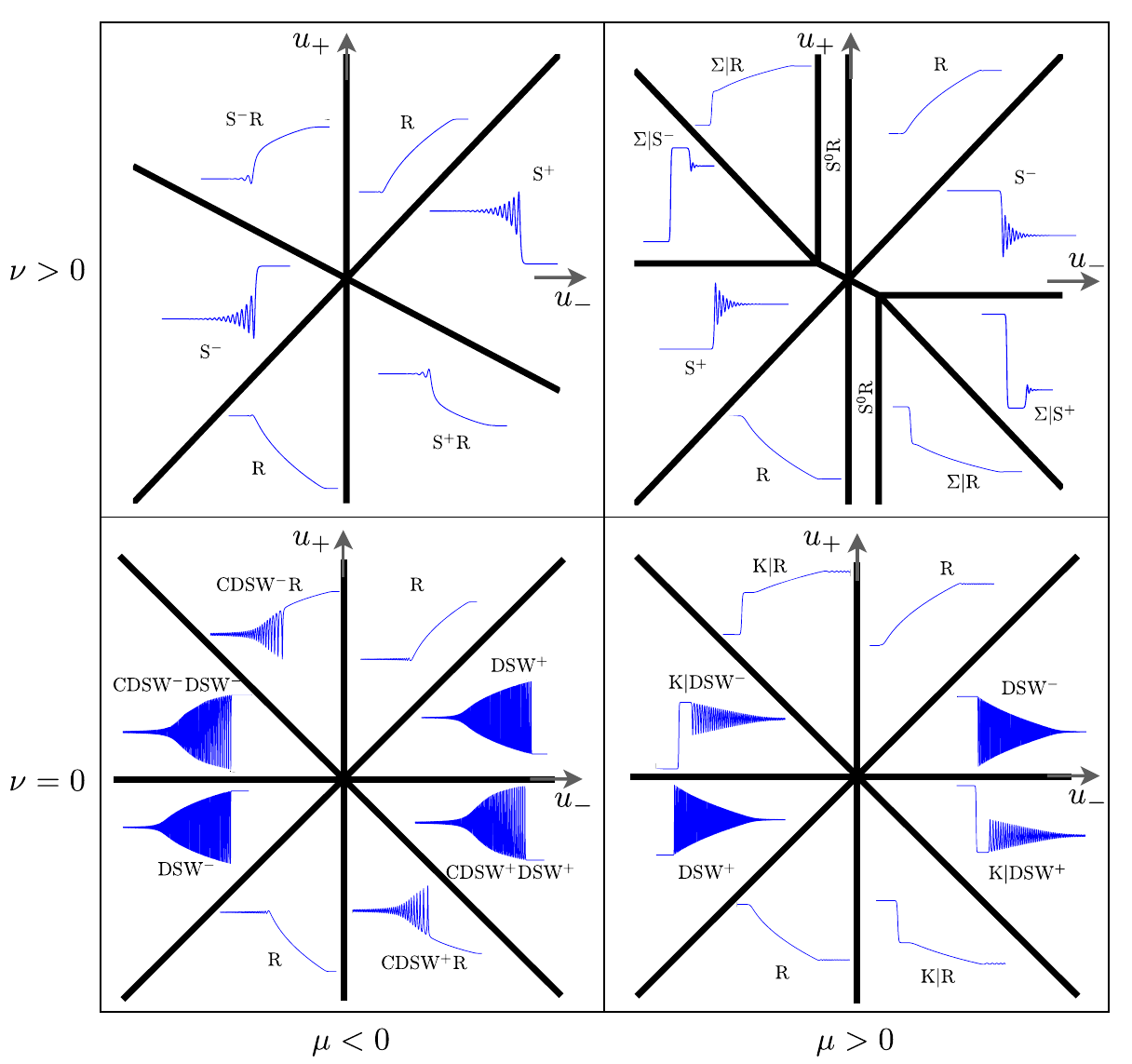}
  \caption{Replicas of figures \ref{riemann_fig1} and
    \ref{riemann_mkdv} with numerical solutions for interpretation of
    the zero diffusion mapping.}
  \label{fig:mapping}
\end{figure}
We define the zero diffusion mapping for the Riemann problem by the
classification figures \ref{riemann_fig1} ($\nu > 0$) and
\ref{riemann_mkdv} ($\nu = 0$) reproduced in figure \ref{fig:mapping}.
Common domains in the $u_-$-$u_+$ plane demonstrate how
diffusive-dispersive wave solutions map to the purely dispersive case.
The mapping is summarized in table \ref{tab:zero_diffusion}.
\begin{table}[ht]
  \centering
  \begin{tabular}{|c|c|}
    \hline
    negative dispersion $\mu < 0$ & positive dispersion $\mu > 0$ \\
    \hline
    R $\to$ R & R $\to$ R \\
    \hline
    $\textrm{S}^\pm \to \left \{ \begin{array}{c} \textrm{DSW}^\pm \\
        \textrm{or} \\
        \textrm{CDSW}^\pm\textrm{DSW}^\pm \end{array} \right .$ &
    $\textrm{S}^\pm \to  
    \left \{ \begin{array}{c} \textrm{DSW}^\pm \\ \textrm{or} \\
    \textrm{K}|\textrm{DSW}^\pm \end{array} \right .$
    \\
    \hline
    --- & $\Sigma|$S$^\pm \to $ K$|$DSW$^\pm$ \\
    \hline
    $\textrm{S}^\pm\textrm{R} \to \left \{ \begin{array}{c}
        \textrm{CDSW}^\pm\textrm{R} \\ \textrm{or} \\
        \textrm{CDSW}^\pm\textrm{DSW}^\pm\end{array} \right .$ &
    $\left . \begin{array}{c} \Sigma|\textrm{R} \\ \textrm{or} \\
        \textrm{S}^0\textrm{R} \end{array} \right \} 
    \to \left \{ \begin{array}{c} \textrm{K}|\textrm{R} \\
        \textrm{or} \\
        \textrm{K}|\textrm{DSW}^\pm \end{array} \right .$\\
    \hline
  \end{tabular}
  \caption{Zero diffusion limit mapping Riemann problem solutions of 
    mKdVB to those of mKdV.} 
  \label{tab:zero_diffusion}
\end{table}
The first and third quadrants of the $u_-$-$u_+$ plane admit a
single-valued, one-to-one mapping, comparable to the Riemann problem
classifications for the KdV and KdVB equations with convex flux
(recall figure \ref{KdVclass1}).  It is the remaining quadrants where
non-convexity introduces novel features.  The origin of the non
coinciding boundaries for the Riemann problem classifications of mKdVB
and mKdV can be traced back to the asymmetry introduced by the
Rankine-Hugoniot and Lax entropy conditions \eqref{laxm1} for mKdVB
and the symmetric causality and convexity conditions \eqref{adm-} for
mKdV.

A direct mapping is possible when $0 < u_- < u_+$ or $u_+ < u_- < 0$
for any fixed $\mu \ne 0$.  This implies that, as $\nu \to 0^+$,
S$^\pm \to $ DSW$^\pm$.  Diffusive-dispersive Lax shocks map to DSWs
in this regime.  This provides an explanation for the similarity
(within the regime considered) between the Lax entropy condition
(\ref{laxm}) and the DSW admissibility conditions (\ref{adm-}), the
new feature being the existence of two DSW speeds.  The rarefactions
are continuous and exhibit only weak discontinuities (i.e., in higher
derivatives), hence their regularizations are the same to leading
order.

In the case of positive dispersion $\mu > 0$, the diffusive-dispersive
classification approaches the same octant structure as the dispersive
classification in the zero diffusion limit because $\delta
=\nu/\sqrt{\mu} \to 0^+$ as $\nu \to 0^+$.  This leads to the mappings
$\Sigma|$S$^\pm \to \,$K$|$DSW$^\pm$ and $\Sigma|$R $\to \,$K$|$R.
Kinks are the purely dispersive analogue of diffusive-dispersive
undercompressive shocks.  This correspondence was similarly recognized
in the context of two-layer fluids \cite{klu2007}.  Numerical results
in figure \ref{fig:numericsPos} show that the intermediate constant
states associated with these double-wave structures attain different
values according as $\nu > 0$ or $\nu = 0$.  This is because a
dispersive kink's profile always admits the symmetry $x \to -x$, $u
\to -u$.  An undercompressive shock is not symmetric and the relation
between the two constant states that it connects changes with the
parameter $\delta = \nu/\sqrt{\mu}$, as described in \S
\ref{sec:mkdv_trav}.  Nevertheless, the qualitative features between the
diffusive-dispersive and dispersive solutions admit a natural
correspondence.  This correspondence with nonzero
diffusion % loses its
\MH{is neither} one-to-one \MH{nor} % and
single-valued.  There is the addition of the double-wave S$^0$R, which
does not exhibit an intermediate constant state nor does the shock
exhibit any oscillation.  Complicating the mapping further is the fact
that the boundaries between the different wave structures when $\nu >
0$ and $\nu = 0$ do not line up.  Then the mapping allows for
$\Sigma|$R \textit{or} S$^0$R $\to$ K$|$R \textit{or} K$|$DSW$^\pm$.

The negative dispersion $\mu < 0$ case is complicated by the fact
that, for any nonzero diffusion, the Riemann problem classification
divides the $u_-$-$u_+$ plane into only six regions whereas the zero
diffusion case involves eight.  The new feature associated with zero
diffusion is the existence of contact DSWs.  As we have shown in
previous sections, CDSWs have no ``natural'' diffusive-dispersive
correlate.  Nevertheless, the overlapping domains contained in the two
octants $0 < u_- < -u_+$, $-u_+ < u_- < 0$ of figure \ref{fig:mapping}
imply that S$^\pm$R $\to$ CDSW$^\pm$R as $\nu \to 0^+$ in this region.
Shock-rarefactions become CDSW-rarefactions in the zero diffusion
limit.  A careful examination of the numerical results shows that the
constant state connecting these double wave structures differ for the
diffusive and non-diffusive cases.  This is a manifestation of the
difference between the Rankine-Hugoniot relation and the CDSW
conditions.

The zero diffusion mapping S$^\pm \to \, $CDSW$^\pm$DSW$^\pm$ is
implied by the regions $0 < u_+ < -u_-/2$ and $-u_-/2 < u_+ < 0$ in
figure \ref{fig:mapping}.  Lax shocks map either to contact DSWs when
attached to a rarefaction or to a hybrid CDSW-DSW or to a pure DSW,
depending on the relation between $u_-$ and $u_+$.  This implies that
the zero diffusion mappings of S$^\pm$ and S$^\pm$R are multivalued.
The top two panels in figure \ref{fig:numericsPos}(a) show that, while
there may be some subtle features introduced by CDSWs (recall figure
\ref{fig:bordeaux_martini}), generally it is difficult to distinguish
them from pure DSWs.  Recall that CDSWs arise due to nonstrict
hyperbolicity at the level of the Whitham modulation equations, a
feature of the purely dispersive case.  Therefore it is perhaps not
surprising that diffusive-dispersive shocks (TWs) do not discern such
features.  The introduction of a non-convex flux implies that Lax
shocks map to either CDSWs or DSWs but their differences may be
difficult to identify in practice.

Aside from the ambiguity in the zero diffusion limit mapping of Lax
shocks introduced by CDSWs when $\mu < 0$, the other limits described
so far are natural.  However, the remaining, unexplored regions of the
$u_-$-$u_+$ plane with regards to the zero diffusion mapping are
$u_-/2 < -u_+ < u_-$ and $u_- < -u_+ < u_-/2$ when $\mu < 0$.  Figure
\ref{fig:mapping} shows S$^\pm$R $\to$ CDSW$^\pm$DSW$^\pm$ in this
region.  But this mapping is peculiar because there is no rarefaction
in the purely dispersive case.  The resolution of this apparent
contradiction can be understood by appealing to the small diffusion
regime and relevant asymptotic time scales, which we undertake in the
next subsection.

\subsection{Time scales and critical scalings}
\label{sec:time-scales}

\begin{figure}
  \centering
  \includegraphics{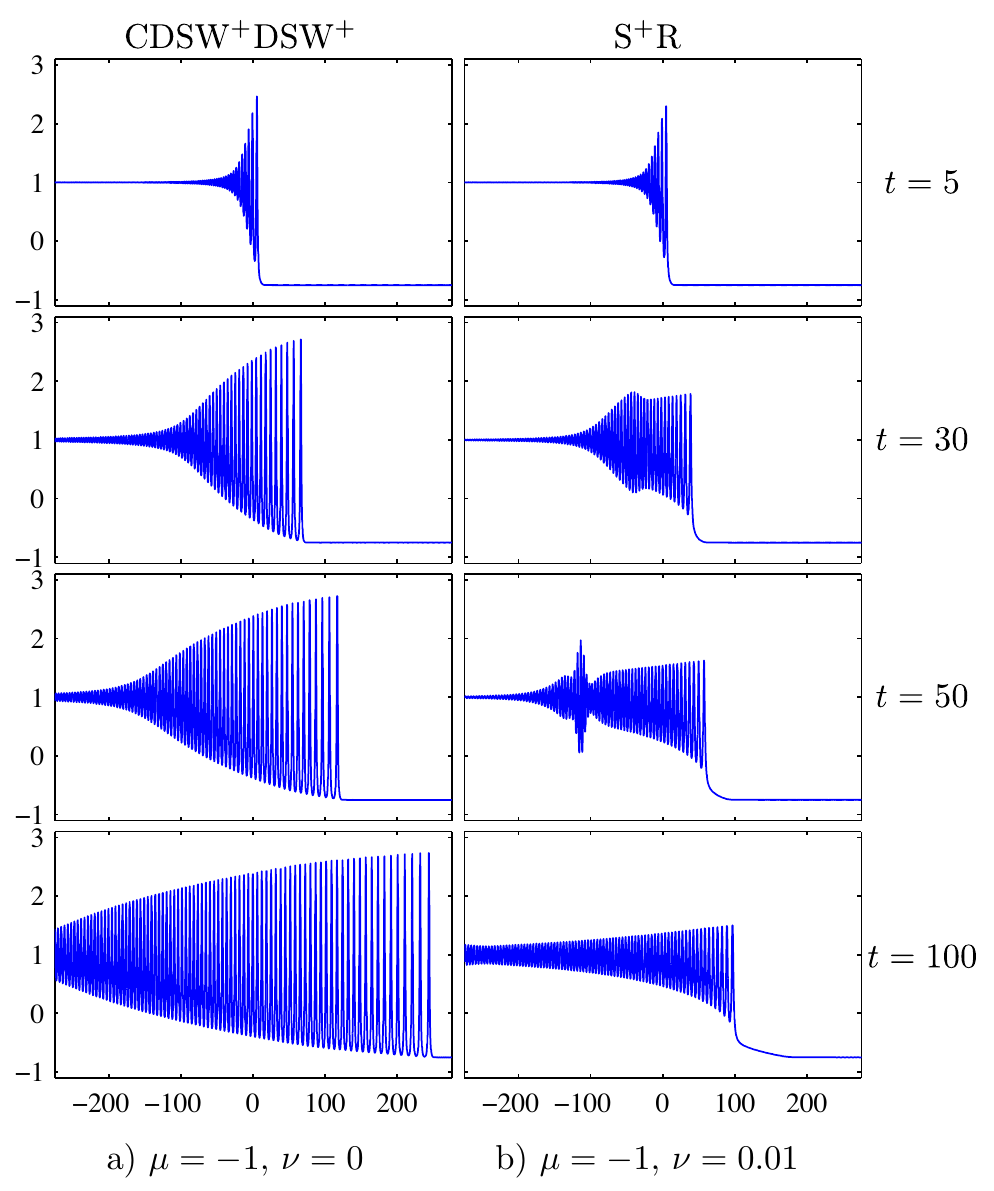}
  \caption{Numerical simulation of the Riemann problem for $u_- = 1$,
    $u_+ = -0.75$ comparing purely dispersive (left) and
    diffusive-dispersive (right) transient dynamics.}
  \label{fig:transient}
\end{figure}

Motivated by the singular transition S$^\pm$R $\to$
CDSW$^\pm$DSW$^\pm$ just described, we now investigate the dynamics
when the diffusion is small but nonzero.  The existence of the
parameter $\delta = \nu/\sqrt{|\mu|}$ in the TW solutions of \S
\ref{sec:mkdvb-non-classical} provides a clue as to the critical
scaling relationship between diffusion and dispersion.  By ``small
diffusion'', we really mean $0 < \nu \ll \sqrt{|\mu|}$.  We \MH{now}
rescale space and time in (\ref{mkdvb1}) to demonstrate the time
scales in which dispersion dominates the transient, intermediate
asymptotic behavior, and the time scale on which the long-time
behavior emerges, when diffusion and dispersion are in balance.  In
order to preserve the balance for transport on the left side of
(\ref{mkdvb1}), we scale space and time by the same
constant. Accordingly, let $t=a\tau, x= ay.$ Then equation
(\ref{mkdvb1}) transforms to 
\begin{equation}
  \label{scaled}
  u_\tau+(u^3)_y=\frac{\nu}{a} u_{yy}+\frac{\mu}{a^2} u_{yyy}.
\end{equation}
In order for the dispersive term to balance the transport term, we set
$a=\sqrt{|\mu|}.$ Then the diffusion coefficient of the first term
\MH{on the right hand side} becomes $\nu/\sqrt{|\mu|}$. If we take
this ratio to be small, then the diffusive term is negligible compared
to dispersion over the time interval \MH{where} % up to
$t \sim \sqrt{|\mu|}$.

Over long times, as the diffusive terms become significant, we expect
solutions to converge to a combination of traveling waves and
rarefactions, reflecting the combined effects of diffusion and
dispersion. The time scale for such behavior emerges when the
diffusive and dispersive terms are in balance.  From (\ref{scaled}) we
see that the coefficients on the right hand side are equal when
$a=|\mu|/\nu$.  The corresponding time scale is $t\sim |\mu|/\nu.$
Since $|\mu|/\nu= \sqrt{|\mu|} (\sqrt{|\mu|}/\nu) \gg \sqrt{|\mu|},$
we see that there is a transient regime $\sqrt{|\mu|} \ll t \ll
|\mu|/\nu$ over which the solution evolves from dispersion-dominated
to the diffusive-dispersive behavior analyzed here.  \MH{This
  transient evolution}
% , which
resembles the diffusion-dominated classical theory except in cases
where undercompressive waves appear.  

As $\nu$ approaches zero with fixed $\mu$, the diffusive effects take
longer to establish themselves, and in the limit $\nu\to 0^+,$ the
evolution is purely dispersive. The elongation of the transient
dynamics time interval is suggestive of the transition between
diffusive-dispersive solutions of the Riemann problem and purely
dispersive solutions. Figure \ref{fig:transient} showcases this
behavior for the numerical solution of a Riemann problem in the regime
$u_-/2 < -u_+ < u_-$ and $\mu < 0$ where apparently S$^+$R $\to$
CDSW$^+$DSW$^+$ as $\nu \to 0^+$.  We fixed $\mu = -1$ and evolved the
same initial data, a smoothed transition from $u_- = 1$ to $u_+ =
-0.75$ for $\nu = 0$ (figure \ref{fig:transient}(a)) and $\nu = 0.01$
(figure \ref{fig:transient}(b)).  As predicted by the scaling analysis,
the initial dynamics, e.g., $t = 5 = \mathcal{O}(\sqrt{|\mu|})$, for
both the dispersive and diffusive-dispersive models are essentially
the same.  The models initiate the formation of DSWs.  But in the
transient regime $1 = \sqrt{|\mu|} \ll t \ll |\mu|/\nu = 100$, we
observe the dynamics corresponding to a transition from a DSW to a
shock-rarefaction.  A large wave packet propagates backward, through
the initiated, approximate DSW, leaving a steady diffusive-dispersive
shock configuration adjacent to a rarefaction in its wake.  Also
predicted by the scaling analysis, at $t = 100 = |\mu|/\nu$, the
diffusive-dispersive behavior has set in for the solution profile.
\MH{This example indicates} 
%We see
that the resolution of the singular limit $\nu \to 0^+$ involves
nontrivial, transitory dynamics.

The transition S$^\pm$R $\to$ CDSW$^\pm$DSW$^\pm$ is a clear example
of the singular behavior of the zero diffusion limit, demonstrating
how to properly interpret the zero diffusion mapping in table
\ref{tab:zero_diffusion} as a dynamic process.  First, nonlinearity
alone approximately describes the dynamics up to \MH{wavebreaking}.
Then, dispersion leads to the generation of oscillations and finally
diffusion and dispersion together resolve the solution into a TW.  As
noted in \S \ref{sec:riemann-problem}, the resulting TW can be
interpreted as an underdamped oscillator.  The damping of this
oscillator goes to zero with $\delta$, hence approaches a continually
expanding dispersive shock wave.

%\section{Extension to bi-directional systems}
% \begin{table}[h!]
% \begin{center}
% \begin{tabular}{|l|l|l|l|l|l|}
%   \hline
%   % after \\: \hline or \cline{col1-col2} \cline{col3-col4} ...
%         & S &  R & DSW & $\Sigma$ & Kink \\
%    \hline
%    R-H &  Yes & No & No& Yes & Yes \\
%    \hline
%    RI &  No & Yes & Yes & No & Yes \\
%    \hline
%    E & Yes &  N/A & Yes & No & No \\
%   \hline
%   \end{tabular}
% \end{center}
% \caption{Conditions across the single-wave structure in a Riemann problem for non-convex dispersive Eulerian hydrodynamics: R-H --- Rankine-Hugoniot relations; RI --- Riemann invariant condition; E -- entropy conditions.}
% \end{table}

\subsection{Small-diffusion regime and modulation equations} 

As we have seen, the analytical approaches used in the descriptions of
diffusive-dispersive shocks and DSWs are quite different.  The former
is based on the analysis of an ODE for TW solutions of the mKdVB
equation, while the latter employs nonlinear modulation theory,
resulting in a system of homogeneous first order quasilinear
PDEs. These two approaches can be reconciled in the case of
small-diffusion $0 < \nu \ll \sqrt{|\mu|}$. The reconciliation can be
understood in the framework of an appropriate modification of Whitham
modulation theory, in which the diffusive term in the mKdVB equation
(\ref{mkdvb1}) is treated as a small perturbation of the mKdV
equation. In this section, we briefly describe this approach for the
well-studied case of the KdVB equation (\ref{kdvb1}) and then outline
its implications for the mKdVB case leaving a more careful analysis to
future publications.

In `perturbed' modulation theory, the balance laws of the perturbed
equation (e.g., the KdVB equation) are averaged over a family of
periodic solutions for the {\it unperturbed} equation (i.e., the KdV
traveling wave solutions (\ref{cnoidal}) \cite{forest_mclaughlin}; see
also \cite{dn89}). The result of this procedure is a non-homogeneous
modulation system
\begin{equation}
  \label{perturbed-whitham}
  \frac{\partial r_j}{\partial t} + V_j({\bf r})\frac{\partial
    r_j}{\partial x} = \delta R_j({\bf r}), \quad  j=1,2,3,  \quad
  \delta = \frac{\nu}{\sqrt{|\mu|}}\ll 1, 
\end{equation}
where $V_j({\bf r})$ are the KdV-Whitham characteristic speeds
(\ref{vi}) and the right-hand sides $R_j({\bf r})$ arise from
averaging the diffusive terms in the perturbed KdV balance laws.  See
\cite{AKN87}, \cite{GP87} for the explicit expressions in terms of
complete elliptic integrals.  The modulation system
(\ref{perturbed-whitham}) for the KdVB equation with small diffusion
can also be obtained using a direct multiple scales procedure,
equivalent to Whitham averaging (see \cite{MG}).  An effective general
method for the derivation of modulation systems for perturbed
integrable equations was developed in \cite{kamch2004}.

Unlike the unperturbed modulation system (\ref{kdvW}), the perturbed
system (\ref{perturbed-whitham}) possesses TW solutions
$r_j(x-st)$. It was shown in \cite{AKN87}, \cite{GP87} that the TW
solution of the KdVB-Whitham system connects two constant, disparate
states: $u \to u_\pm$ as $x \to \pm \infty$, $u_->u_+$, and satisfies
the Rankine-Hugoniot condition $s=(u_- + u_+)/2$.  Thus, this solution
describes slow modulations of the classical diffusive-dispersive shock
wave in the small diffusion regime.  As a matter of fact, this
outlined construction of the TW modulation solution is equivalent to
a direct multiple scales perturbation analysis of the ODE
(\ref{tw2}) with $\delta \ll 1$ for the TW solutions of the KdVB
equation itself (see \cite{johnson70}).  However, modulation theory
provides a much broader platform, enabling one to describe different
stages of diffusive-dispersive shock development (see \cite{AKN87,dn89}).

The general method \cite{kamch2004}  for the derivation of perturbed
modulation equations  via spectral finite-gap theory, and its
applications to the mKdV and Gardner equations in \cite{ksk-04},
\cite{kamch2012}, suggest that the perturbed modulation systems for
mKdVB and  KdVB are related by the same surjective mapping
(\ref{rla-}), (\ref{V-}) for $\mu<0$, and (\ref{rla+}), (\ref{V+}) for
$\mu>0$, as  their unperturbed, purely dispersive, counterparts. This
suggests that the modulation description of classical oscillating
mKdVB shocks can be derived from known solutions of the KdVB
modulation system.  However, the description of oscillatory contact
shocks in shock-rarefaction complexes for $\mu<0$ (see figure
\ref{fig:transient}(b), $t=100$) cannot be obtained from the KdVB
modulation solutions and requires further analysis. 

\section{Conclusions} \ \ The analysis of Riemann problems for the
KdV(B) and mKdV(B) models reviewed here provides an explicit link
between the theories of hyperbolic conservation laws and nonlinear
dispersive wavetrains.  These are universal, asymptotic models of
diffusive-dispersive Eulerian hydrodynamics.  When the Eulerian
pressure law is non-convex, the modified KdV-Burgers equation emerges.
The prototypical phenomena arising from the competition between
diffusion, dispersion, and non-convex flux include non-classical wave
features such as undercompressive shock waves, and double wave
complexes such as shock-rarefactions.  The purpose of this work has
been to compare the singular, zero diffusion limits of these waves
with their purely dispersive counterparts, identifying notable
similarities and differences.

An essential difference between diffusive-dispersive Lax shocks and
classical dispersive shock waves is that the former have fixed,
traveling wave profiles whereas the latter waves have expanding,
oscillating profiles.  Consonant to this difference are the dissimilar
shock jump conditions and their derivation.  Nevertheless, comparing
KdVB to KdV, there is a relatively simple, bijective zero diffusion
mapping from Lax shocks and rarefactions with diffusion to DSWs and
rarefactions without diffusion. The introduction of a non-convex flux
in the mKdVB and mKdV equations leads to a more complex, multivalued
mapping.  Now, the sign of the dispersion term plays an essential
role.  For negative dispersion and nonzero diffusion, the classical
theory of conservation laws prevails.  The new feature resulting from
a non-convex flux is the shock-rarefaction, not essentially reliant
upon the dispersive term in the equation.  In this regard, the small
diffusion regime provides limited clues to the purely dispersive
dynamics, where a new and unique feature arises, the contact DSW.
These solutions occur due to the nonstrict hyperbolicity of the
Whitham modulation equations.  A CDSW is non-classical in the sense
that either one characteristic family passes through it, as in a
CDSW-DSW, or one characteristic family becomes tangential to the
characteristics of an adjacent wave, as in a CDSW-rarefaction.  The
former behavior is analogous to an undercompressive shock, which also
has characteristics passing through, but occurs only in the positive
dispersion regime.  This behavior of characteristics in CDSWs
contrasts sharply with the fact that all diffusive-dispersive shocks
in the negative dispersion regime satisfy the Lax entropy conditions.
As a result, there is some ambiguity in the zero diffusion limit of
Lax shocks.  We explain and resolve the ambiguity for small diffusion
by identifying relevant time scales that include a transient window
during which the dispersive-dominated dynamics transition to a
diffusive-dispersive balance.

For positive dispersion, the diffusive-dispersive dynamics are
fundamentally different from the negative dispersion regime.  Here, the role of
dispersion is essential for the nonzero diffusion theory.  This is due
to the existence of non-classical TW solutions, namely undercompressive
shocks, that are not available in the dispersionless regime.  The zero
diffusion mapping identifies an undercompressive shock with a kink or
undercompressive DSW.  Kinks are monotone traveling waves satisfying
the Rankine-Hugoniot shock conditions, and consequently do not exhibit the
expansionary and oscillatory structure associated with classical DSWs.
The Lax entropy conditions are not fulfilled for kinks because all
characteristics pass directly through, without any deflection.

 This work  emphasizes   the universal scalar dynamics occurring in diffusive - dispersive modifications (\ref{eulerd1}) of non-convex  Eulerian systems.  Physically interesting dispersive and dispersive-diffusive systems exhibiting complex structure of solutions due to non-convex fluxes 
%include a broad range of systems arising 
arise in fluid dynamics, plasma physics, magnetohydrodynamics, nonlinear optics, Bose-Einstein condensates and  magnetization dynamics. As examples of  important  bi-directional non-convex dispersive systems we mention  Miyata-Choi-Camassa system for fully nonlinear internal waves \cite{choi_camassa1999}, \cite{esler2011} and the cubic-quintic nonlinear Schr\"odinger equation  arising in nonlinear optics and Bose-Einstein condensate dynamics \cite{trillo}.   

The universality of the mathematical description developed in this paper is  due to the generality of the %mathematical model,
 mKdVB equation (\ref{mkdvb1}), which is obtained via the well-established procedure of multiple scale expansions, and captures all essential properties of uni-directional, weakly nonlinear, long-wave diffusive-dispersive dynamics occurring in the neighborhood of the linear degeneracy point (\ref{critical}) for the background flow. While the mKdVB description of the diffusive-dispersive dynamics of classical and non-classical shocks is based on the explicit calculations of TWs \cite{jacobs-mckinney-shearer95}, the  construction of the zero-diffusion DSW theory employs subtle integrability properties of the mKdV equation which are manifested, in particular, in the availability of Riemann invariants for the associated system of Whitham modulation equations. The extension of the present work to the fully nonlinear, vector (bi-directional propagation)  case described by full non-convex Eulerian systems  (\ref{eulerd1}) or systems of similar structure will be a major development of the existing theory. 

We note that system (\ref{eulerd1})  can be re-written using Lagrangian co-ordinates as a diffusive-dispersive modification of a general $p$-system (see, e.g., \cite{dafermos}). The theory of undercompressive shocks for non-convex $p$-systems was constructed in \cite{LeFloch,schulze-shearer99}.   The counterpart zero-diffusion theory is not available at present and its development  will require an extension of the {\em non-integrable} methods of \cite{el05}, \cite{hoefer14} to the non-convex flux case. The inclusion of the possibility of non-convexity in the linear dispersion relation will introduce additional complexity to the Riemann problem classifications and their comparisons for  diffusive-dispersive and pure dispersive cases.

% This comparison demonstrates the link between the well-established
% field of hyperbolic conservation laws and the growing field of
% dispersive hydrodynamics.
It is hoped that this work will inspire further examination and
  exploitation of the bridge between the well-established field of
  hyperbolic conservation laws and the growing field of dispersive
  hydrodynamics.

% \ack

% This work was supported by a Royal Society International Exchange
% Scheme (all authors) and NSF CAREER award DMS-1255422 (Hoefer). The authors thank G. Esler, E. Ferapontov and A. Kamchatnov for valuable discussions.

\appendix 

\setcounter{section}{0}

\section{Traveling wave solutions of the mKdV equation}

We present the classification of the  traveling wave solutions of the mKdV equation which are used in the construction of various DSWs  in Section \ref{sec:dsw_theory}. 
These solutions are obtained by integrating the ODE (\ref{eq3})  for two cases $\mu>0$ and $\mu<0$. In describing traveling wave  solutions we refer to two basic configurations of the potential curve $Q(u)$ shown in figure \ref{potential}.

If $\mu>0$ (positive dispersion), then the oscillations occur in the interval $u_2 \le u \le u_3$, where $Q(u) \ge 0$ (see figure~\ref{potential}(a)), and the solution of (\ref{eq3}) is expressed in terms of Jacobi elliptic functions as
\begin{equation}\label{el1}
    u=u_2+\frac{(u_3-u_2)\mathrm{cn}^2(\theta,m_1)}{1-
    \frac{u_3-u_2}{u_4-u_2}\mathrm{sn}^2(\theta,m_1)} \, ,
\end{equation}
where
\begin{equation}\label{el1a}
    \theta=\sqrt{(u_3-u_1)(u_4-u_2)} \ \eta/2 \, ,  \quad  \eta=\frac{x-Ut}{\sqrt{2|\mu|}} 
    \end{equation}
and
\begin{equation}\label{el2}
    m_1=\frac{(u_3-u_2)(u_4-u_1)}{(u_4-u_2)(u_3-u_1)},
\end{equation}
is the modulus, $0 \le m_1 \le 1$.
%The wavelength is given by the formula
%\begin{equation}\label{el3a}
%    L=\frac{4\sqrt{2 \mu}\K(m_1)}{\sqrt{(u_3-u_1)(u_4-u_2)}},
%\end{equation}
%where $\K(m_1)$ is the complete elliptic integral of the first kind.
The soliton limit $m_1 \to 1$ can be achieved in one of two ways: when $u_2 \to u_1$ or when $u_3 \to u_4$.

When $u_2 \to u_1$, we obtain the ``bright'' soliton of elevation with amplitude $a=u_3-u_1$ propagating against a constant background $u=u_1$,
\begin{equation}\label{eq70}
    u=u_1+\frac{u_3-u_1}{\cosh^2\theta-
    \frac{u_3-u_1}{u_4-u_1}\sinh^2\theta} \, .
\end{equation}
The soliton speed $U$ is found from (\ref{el3}), (\ref{constr})
\begin{equation}\label{speedbright+}
U= \frac12(3u_1^2 +3u_1u_4+u_4^2) \, .
\end{equation}

Analogously, for $u_3 \to u_4$ one obtains a ``dark'' soliton of depression
\begin{equation}\label{eq7a}
    u=u_4-\frac{u_4-u_2}{\cosh^2\theta-
    \frac{u_4-u_2}{u_4-u_1}\sinh^2\theta},
\end{equation}
having the amplitude $a=u_4-u_2$ and propagating against the constant background $u=u_4$ with the speed
\begin{equation}\label{speeddark+}
U= \frac12(3u_4^2 +3u_1u_4+u_1^2)\, .
\end{equation}
Note that, due to invariance of the mKdV equation with respect to the reflection transformation $ u \mapsto -u$ formulas (\ref{eq7a}), (\ref{speeddark+}) can be obtained from (\ref{eq70}), (\ref{speedbright+}) by replacing
\begin{equation}\label{invariance}
u \to -u,  \quad u_i \to - u_{5-i}\, , \quad i=1,2,3,4.
\end{equation}
Thus the mKdV equation with fixed $\mu>0$ admits the existence of solitons of both polarities. This is sharply distinct from the KdV dynamics where polarity of admissible solitary waves is uniquely related to the dispersion sign (see \S \ref{sec:kdv-equation-dsws}). This property of the mKdV equation is due to the quartic nature of the potential curve $Q(u)$ and is ultimately related to non-convexity of the hydrodynamic flux  of the dispersionless limit (\ref{inviscid_pde}).

If both $u_2 \to u_1$ and $u_3 \to u_4$ then the polynomial $Q(u)$ in the right-hand side of (\ref{eq3})
has two double roots, which implies that the solution assumes the form of a {\it kink/antikink}, which can be obtained directly by integration of (\ref{eq3}) with $u_1=u_2$ and
$u_3=u_4$. Noting that due to (\ref{constr})  the double roots  satisfy $u_1+u_4=0$ and choosing the  constant of integration  so that $u=0$ at $\eta=0$ we obtain:
\begin{equation}\label{kink}
u=   \pm \frac{1}{2}  (u_4-u_1) \tanh[ \frac12 (u_4-u_1) \eta] \, .
\end{equation}
The lower sign corresponds to the kink with $u \to u_4$ as $\eta \to-\infty$ and $u \to u_1$ at $\eta \to\infty$;
the upper sign yields the ``anti-kink'' with $u \to u_1$ as $\eta \to-\infty$ and $u \to u_4$ at $\eta \to\infty$.
The speed of kink  propagation in both cases is found from (\ref{el3}) to be $U= u_1^2$, which agrees with
the classical  shock speed (\ref{rh2}),  where $u_+=u_1$, $u_-=u_4=-u_1$.

We also present the small-amplitude asymptotics of the solution (\ref{el1}) when $m_1 \to 0$ ($u_2 \to u_3$). In this limit,
 the cnoidal wave (\ref{el1}) asymptotically transforms into a linear harmonic wave
\begin{equation}\label{eq7b}
  \begin{split}
    u&\cong u_2+\frac12(u_3-u_2)\cos(k\eta), \quad  (u_3-u_2) \ll u_2 ,\\
    k&=\sqrt{(u_2-u_1)(u_4-u_2)},  \quad  U=\frac{1}{2}(u_1^2 + u_4^2). \quad \\
  \end{split}
\end{equation}

\medskip
Now we turn to the case of negative dispersion $\mu < 0$. The potential curve configuration is shown in figure~\ref{potential}(b), and one can see that finite, real-valued motion can occur either in the interval
$u_1\le u \le u_2$ or  in $u_3 \le u \le u_4$. We only consider here the second case; the results for the first one can be obtained by applying the transformation (\ref{invariance}).

\medskip
Let $u_3\le u \le u_4$. Integration of (\ref{eq3}) yields
\begin{equation}\label{eq11}
    u=u_3+\frac{(u_4-u_3)\mathrm{cn}^2(\theta,m_2)}{1+
    \frac{u_4-u_3}{u_3-u_1}\mathrm{sn}^2(\theta,m_2)},
\end{equation}
where $\theta$ is defined by (\ref{el1a}) 
%\begin{equation}\label{el1b}
%    \theta=\sqrt{(u_3-u_1)(u_4-u_2)}  \ \eta/2,
%\end{equation}
and the modulus (cf. (\ref{el2}))
\begin{equation}\label{eq12}
    m_2=\frac{(u_4-u_3)(u_2-u_1)}{(u_4-u_2)(u_3-u_1)}.
\end{equation}
%Now the wavelength is given by
%\begin{equation}\label{eq3a}
%    L=\frac{4\K(m_2)}{\sqrt{|\al|(u_3-u_1)(u_4-u_2)}}.
%\end{equation}

In the soliton limit $u_3\to u_2$ ($m_2 \to 1$), we get
\begin{equation*} %\label{eq13}
u = u_2+\frac{u_4-u_2}{\cosh^2\theta+
    \frac{u_4-u_2}{u_2-u_1}\sinh^2\theta} . \\
\end{equation*}
This is a ``bright'', elevation soliton with amplitude $a=u_4-u_2$. Its propagation speed is found from (\ref{el3}), (\ref{constr}) to be
\begin{equation*} %\label{vdep2}
U=\frac12(u_4^2 + 2u_2u_4 + 3u_2^2).
\end{equation*}

\medskip
\noindent The limit $m_2\to 0$ can be reached in two ways.

{\it (1)} If $u_3\to u_4$, we get asymptotically
\begin{equation*}%\label{eq13b}
  \begin{split}
    u&\cong u_3+\frac{1}{2}(u_4-u_3)\cos(k\eta), \quad u_4-u_3 \ll u_3, \\
    k&=\sqrt{(u_3-u_1)(u_3-u_2)}, \quad
    U= \frac{1}{2}(3u_4^2 +2u_2u_4+u_2^2).
  \end{split}
\end{equation*}
This is a small-amplitude harmonic limit analogous to (\ref{eq7b}) obtained for $\mu>0$.

{\it (2)} If $u_2=u_1$, but $u_3 \ne u_4$  then we arrive at the nonlinear trigonometric solution
\begin{equation}\label{eq13c}
%    u=u_3 + \frac{(u_4-u_3)\cos^2\theta}{1+
%    \frac{u_4-u_3}{u_3-u_1}\sin^2\theta}
u=u_3+\frac{u_4-u_3}{1+\frac{u_4-u_1}{u_3-u_1}\tan^2 \theta},
\end{equation}
where
\begin{equation*}%\label{eq13d}
  \begin{split}
    \theta=\sqrt{(u_3-u_1)(u_4-u_1)} \ \eta/2, \quad  U=
    \frac18(3u_3^2 +3u_4^2+2u_3 u _4)\, ,
  \end{split}
\end{equation*}
and the amplitude $a=u_4-u_3$ is generally order unity. The nonlinear trigonometric solution (\ref{eq13c}) has no analogs in the KdV theory. Indeed, the very existence of the nonlinear trigonometric solution is possible due to the quartic nature of the potential function $Q(u)$, i.e. due to the non-convex hyperbolic flux of the mKdV equation.

If $u_3\to u_2=u_1$, then $\theta \to 0$ and solution
 (\ref{eq13c})  transforms into an {\em algebraic soliton of elevation,}
\begin{equation}\label{eq13e}
    u=u_1+\frac{u_4-u_1}{1+(u_4-u_1)^2 \eta^2/4}, \quad U=3u_1^2.
 \end{equation}
Note that the algebraic soliton propagates with the characteristic speed of the hyperbolic limit (\ref{inviscid_pde}) evaluated at the background $u=u_1$. The counterpart ``dark'' solutions corresponding to the interval $u_1 \le u \le u_2$ are obtained from (\ref{eq11}) -- (\ref{eq13e}) by applying the reflection transformation (\ref{invariance}).

Finally, we mention a family of periodic solutions corresponding to
the potential curve configuration where the roots $u_2$ and $u_3$ are
complex conjugates. Such solutions are modulationally unstable
\cite{dron76} but they do not arise in the Riemann problem so we do
not describe these solutions here.  See \cite{ercolani} for the proof
of hyperbolicity of the zero dispersion limit of the mKdV equation
with $\mu<0$.

%\appendix
\section{Numerical method}
\label{sec:numerical-method}
\setcounter{section}{2}

Here we describe the numerical method of solution for the KdV(B) and
mKdV(B) equations.  The stiffness of the third order derivative term,
the long time integration required, and the large domain needed to
properly resolve the wave solutions make this a challenging
computational problem.  A high order, integrating factor (IF),
pseudospectral Fourier method that incorporates differing, constant
boundary values is developed, motivated by Trefethen's approach to
solving KdV \cite{trefethen_spectral_2000}.  The IF-RK4 method, utilizing the fourth-order Runge-Kutta algorithm (RK4), has
been shown to perform only slightly slower than other state-of-the-art
methods for the KdV and Burgers equations
\cite{kassam_fourth-order_2005}.  Its ease of formulation and
implementation make IF-RK4 attractive.

Interest here is in the long time behavior of the Riemann Problem
\begin{align}
  \label{eq:2}
  &u_t + c_0 u^p u_x = \nu u_{xx} + \mu u_{xxx}, \quad x \in \mathbb{R}, \quad
  t > 0, \\
  \label{eq:11}
  &u(x,0) =
  \begin{cases} u_- & x < 0,\\
    u_+ & x > 0 .\\
  \end{cases}
\end{align}
The parameters include $c_0 \in \{1,3\}$, $p \in \{1,2\}$, $\nu \ge
0$, $\mu \in \mathbb{R}$.

We consider a truncated domain $x \in [-L,L]$ and assume that, to an
appropriate level of approximation, $u(\pm L,t) = u_\pm$ and
$\partial_x u(x,t)$ is smooth and compactly supported on $[-L,L]$ for
each $t \in [0,T]$, some $T > 0$.  Then, the function $v(x,t)$ defined
by
\begin{equation}
  \label{eq:3}
  v(x,t) = u_x(x,t), \quad u(x,t) = \int_{-L}^x v(y,t) \ud y + u_-,
\end{equation}
is smooth and compactly supported in this domain.  Differentiating
(\ref{eq:2}) gives a nonlocal equation for $v$
\begin{equation}
  \label{eq:4}
  v_t + c_0(u^p v)_x = \nu v_{xx} + \mu v_{xxx} .
\end{equation}
Since $v(x,t)$ is localized within $[-L,L]$, we can treat its periodic
extension with the Fourier series expansion
\begin{equation*}
%  \fl
  v(x,t) = \sum_n \widehat{v}_n(t) e^{i k_n x}, \quad
  \widehat{v}_n(t) = \frac{1}{2L} \int_{-L}^L v(x,t) e^{-ik_n x},
  \quad k_n = n \pi/L .
\end{equation*}
The Fourier coefficients $\widehat{v}_n$ satisfy the differential
equations
\begin{equation*}
  \frac{\ud \widehat{v}_n}{\ud t} + i \mu k_n^3 \widehat{v}_n = -i c_0
  k_n\,\widehat{(u^p v)}_n - \nu k_n^2 \widehat{v}_n.
\end{equation*}
Letting
\begin{equation*}
  \widehat{V}_n(t) = e^{i \mu k_n^3t} \widehat{v}_n(t),
\end{equation*}
then
\begin{equation}
  \label{eq:7}
  \frac{\ud \widehat{V}_n}{\ud t} = -i c_0 k_n e^{i \mu k_n^3t}
  \widehat{(u^p v)}_n - \nu k_n^2 \widehat{V}_n. 
\end{equation}
A suitable truncation of the Fourier series is chosen, $|n| < N$, and
the corresponding eqs.~(\ref{eq:7}) are then stepped forward in time
by the standard RK4 method.  The stiffest, third derivative term from
(\ref{eq:4}) is absorbed into the transformation to
$\widehat{V}_n$, thus allowing for stable timestepping with larger
$\Delta t$.  We do not incorporate the second derivative term into the
phase transformation because it leads to exponentially large/small
terms during the calculation and causes numerical instability.

A calculation shows that the accumulation function in (\ref{eq:3})
satisfies
\begin{eqnarray}
  \label{eq:10}
 % \fl
  u(x,t) &= \sum_{n\ne 0} \frac{\hat{v}_n(t)}{i k_n} e^{i k_n x} -
  \frac{1}{2L} \int_{-L}^L x v(x,t) \ud x + (x+L)\frac{u_+ - u_-}{2
    L} + u_- .
\end{eqnarray}
Equation (\ref{eq:10}) is suitable for numerical computation and is
used for the reconstruction of $u(x,t)$ where all Fourier series
coefficients are computed using the FFT.  The integral in
(\ref{eq:10}) is computed with the trapezoidal rule.  So long as the
solution $v(x,t)$ remains localized within $[-L,L]$, this method is
spectrally accurate.  

The initial condition (\ref{eq:11}) is smoothed by a hyperbolic
tangent profile with width one.

For the computations presented in this work, the numerical method was
thoroughly tested and validated using known, exact traveling wave
solutions.  Generally, the purely dispersive computations required
more care due to the radiation of small amplitude dispersive waves to
the boundary. We monitored the conserved quantity $\int_{-L}^L v(x,t)
\ud x$ to be within $10^{-13}$ its nominal value $u_+ - u_-$ across
all simulations.  The deviation of $u$ at the boundaries from $u_\pm$
was maintained below $10^{-4}$ (usually well below that).  The
numerical parameters were informed by the validation studies and
chosen based upon the particular Riemann problem considered.
Typically, we used $\Delta x \in \{0.05,0.1,0.2\}$, $\Delta t \in
\{10^{-3},10^{-4},5\cdot 10^{-5}\}$, and $L \in \{400,800,1600\}$.


\begin{thebibliography}{99}

\bibitem{abeyaratne} R. Abeyaratne and J.K. Knowles, Kinetic relations and the propagation of phase boundaries in solids. Arch. Rational Mech. Anal.,  {\bf 114} (1991), 119--154.

\bibitem{ablowitz} M.J. Ablowitz, {\it Nonlinear Dispersive Waves. Asymptotic Analysis and Solitons} (Cambridge University Press, 2011).

\bibitem{ablowitz2013} M. J. Ablowitz and D. E. Baldwin, Dispersive
  shock wave interactions and asymptotics, Phys. Rev. E, {\bf 87},
  022906 (2013). 

\bibitem{AKN87} V.V. Avilov, I.M.  Krichever and  S.P. Novikov, 
Evolution of Whitham zone in the theory of Korteweg-de Vries, 
Sov. Phys. Dokl. {\bf 32} 564--566 (1987).


\bibitem{benjamin_lighthill}  T.B. Benjamin and M.J. Lighthill,  On cnoidal waves and
bores {\it Proc. Roy. Soc., A}  {\bf 224} (1954) 448--460.

\bibitem{benzoni2013} S. Benzoni-Gavage, Planar traveling waves in capillary 
fluids, Differential and integral equations, Khayyam Publishing {\bf 26} (2013) 433-478.

\bibitem{benzoni2014} S. Benzoni-Gavage, P. Noble  and L. M. Rodrigues, Slow modulations of periodic waves in Hamiltonian
PDEs, with application to capillary fluids, Journ. Nonlin. Sci. {\bf 24} (2014) 711- 768.

\bibitem{bertozzi_thinfilms} A.L. Bertozzi, A. Munch and
M. Shearer, Undercompressive shocks in thin
film flows. Physica D, {\bf 134} (1999),
431--464.

\bibitem{bona} J.L. Bona and M. E. Schonbeck, Travelling wave solutions to the Korteweg - de Vries-Burgers equation, Proc. Roy. Soc. Edinburgh, {\bf 101 A} (1985)  207 - 226.

   \bibitem{bressan}
A.~Bressan.
\newblock {\em Hyperbolic systems of conservation laws: the one-dimensional
  Cauchy problem},  {\em Mathematics and its Applications}  vol.~20.
\newblock Oxford Univ. Press, 2000. 

\bibitem{chanteur1987} G. Chanteur and M. Raadu, Formation of shocklike modified Korteweg-de Vries solitons:  Application to double layers, Phys. Fluids {\bf 30} 2708 (1987).

\bibitem{choi_camassa1999} W. Choi and R. Camassa, Fully nonlinear internal waves in a
two-fluid system.,  J. \ Fluid \ Mech.\  {\bf 396} (1999) 1-36.
\bibitem{christie89} D.R. Christie, Long nonlinear waves in the lower atmosphere, J. Atmos. Sci., 46,  1462 -1491 (1989).

\bibitem{trillo} M. Crosta, S. Trillo, and A. Fratalocchi, Crossover dynamics of dispersive shocks in
Bose-Einstein condensates characterized by two- and three-body interactions, Phys. Rev. A {\bf 85} (2012) 043607.

\bibitem{dafermos} C.M. Dafermos, {\it Hyperbolic Conservation Laws in Continuum Physics}, Springer (2009) New York.

\bibitem{deift1997} P. Deift, S. Venakides, and X. Zhou, New results
  in small dispersion KdV by an extension of the steepest descent
  method for Riemann-Hilbert problems, Int. Math. Res. Notices, {\bf
    6}, 285--299 (1997).

 \bibitem{diperna} R.J. DiPerna,   Decay and asymptotic behavior of solutions to nonlinear hyperbolic systems of conservation laws.
Indiana Univ. Math. J. 24 (1974/75), no. 11, 1047--1071. 

\bibitem{dron76} C.F. Driscoll and T. M. O'Neil,
Modulational instability of cnoidal wave solutions of the modified Kortweg-de Fries equation, J. Math. Phys. {\bf 17,} 1196 (1975).

\bibitem{dubrovin2006} B. A. Dubrovin, On Hamiltonian perturbations of
  hyperbolic systems of conservation laws, II: Universality of
  critical behavior, Comm. Math. Phys., {\bf 267}, 117--139 (2006).

\bibitem{dn89}
B.A. Dubrovin and S.P. Novikov,   Hydrodynamics of weakly
deformed soliton lattices. Differential geometry and Hamiltonian
theory, {Russian  Math. Surveys}, {\bf 44}  35--124 (1989).

\bibitem{teschl_2013}  I. Egorova, Z. Gladka, V. Kotlyarov, G. Teschl,  Long-time asymptotics for the Korteweg--de Vries equation with step-like initial data, Nonlinearity, {\bf 26} 1839 --1864 (2013).

\bibitem{el05} G.A. El, Resolution of a shock in hyperbolic systems modified by weak dispersion, Chaos, {\bf 15} (2005) 037103 .

\bibitem{el2001} G. A. El, R. H. J. Grimshaw, and M. V. Pavlov,
  Integrable shallow-water equations and undular bores,
  Stud. Appl. Math., {\bf 106}, 157--186 (2001).
  
\bibitem{el_hoefer_2016}  G.A. El and M.A. Hoefer, Dispersive shock waves and modulation theory,  submitted to Physica D.

\bibitem{ercolani} N. M. Ercolani, S. Jin, C. D. Levermore, and W. D. MacEvoy,
The zero-dispersion limit for the odd flows in the focusing Zakharov-Shabat hierarchy
{ Int. Math. Res. Notices,} {\bf 47,}  2529 (2003).


\bibitem{esler2011} J.G. Esler and J. D. Pierce, Dispersive dam-break and lock-exchange flows in a two-layer fluid,
J. Fluid Mech., {\bf 667,} 555 (2011)

\bibitem{evans}
L.C. Evans.
\newblock {\em Partial Differential Equations}, vol.~19.
\newblock Graduate Series in Mathematics, Amer. Math. Soc., 2010.

\bibitem{flaschka1980} H. Flaschka, M. G. Forest, and
  D. W. McLaughlin, Multiphase averaging and the inverse spectral
  solution of the Korteweg-de Vries equation, Comm. Pure
  Appl. Math. {\bf 33}, 739--784 (1980).
  
  \bibitem{forest1986} M.G. Forest and J. Lee, Geometry and modulation
  theory for the periodic nonlinear Schr\"{o}dinger equation, in {\it
    Oscillation Theory, Computation, and Methods of Compensated
    Compactness}, Ed. by C. Dafermos, J. L. Ericksen, D. Kinerlehrer,
  and M. Slemrod, {\it IMA Volumes on Mathematics and Its
    Applications} {\bf 2}, 35--69, Springer, New York (1986).

\bibitem{forest_mclaughlin} M.G. Forest and D.W. McLaughlin, Modulations of perturbed KdV wave trains, SIAM J. Appl. Math. {\bf 44}, 278--300 (1984).

\bibitem{glimm} J. Glimm,   Solutions in the large for nonlinear hyperbolic systems of equations. Comm. Pure Appl. Math., {\bf 18} (1965), 697--715.

 \bibitem{glimm2} J. Glimm and P.D. Lax,  {\em Decay of Solutions of Systems of Nonlinear Hyperbolic Conservation Laws.}    Memoirs Amer. Math. Soc.
    vol. 101,    1970.


\bibitem{grad_hu} H. Grad and P.N. Hu,  Unified shock profile in a plasma, Phys. Fluids {\bf 10} 2596 (1967).

\bibitem{grava_klein2007} T. Grava and C. Klein, Numerical solution of the small dispersion limit of Korteweg--de Vries and Whitham equations, Comm. Pure Appl. Math. {\bf 60}, 1623 --1664 (2007).

\bibitem{grava_klein2008}  T. Grava and C. Klein,  Numerical study of a multiscale expansion of the Korteweg--de Vries equation and Painlev\'e-II equation, Proc. Roy. Soc. A {\bf 464} 733-757 (2008).


\bibitem{grava_pierce_tian_CH} T. Grava, V.U. Pierce and F.-R. Tian, Initial value problem of the Whitham equations for the Camassa-Holm equation, Physica D {\bf 238},  55-66 (2009).

\bibitem{zumbrun} O. Gu\`{e}s, G. M\'{e}tivier, M.  Williams, and K. Zumbrun,  Nonclassical multidimensional viscous and inviscid shocks,     Duke Math. J.
    {\bf 142} (2008), 1-110.
    
\bibitem{gurevich90} A.V. Gurevich, A.L. Krylov, and G.A. El, Nonlinear modulated waves in dispersive hydrodynamics, Sov. Phys. JETP, {\bf 71}, 899--910 (1990).

\bibitem{gurevich91} A.V. Gurevich, A.L. Krylov, and G.A. El, Breaking
  problem in dispersive hydrodynamics, JETP Lett., {\bf 54}, 102--107 (1991).

\bibitem{gp74} A.V. Gurevich and L.P. Pitaevskii,  Nonstationary structure of a collisionless shock wave
{Sov. Phys. JETP} {\bf 38} 291 (1974).

\bibitem{GP87} A.V. Gurevich,   and L.P. Pitaevskii,  Averaged description
of waves in the Korteweg-de Vries-Burgers equation, Sov.
Phys. JETP, {\bf 66} 490 (1987).

\bibitem{hayes-leFloch97} B.T. Hayes and P.G. LeFloch, Nonclassical shocks and kinetic relations: Scalar conservation laws, Arch. Rational Mech. Anal.,  {\bf 139}, 1--56 (1997).

\bibitem{hayes-shearer99} B. Hayes and M. Shearer, Undercompressive shocks and Riemann problems for scalar conservation laws with non-convex fluxes, Proc. Roy. Soc. Edinburgh {\bf A129}, 733 -- 754 (1999).

%\bibitem{hm-2006} K.R. Helfrich and W.K. Melville,
%Long nonlinear internal waves, Annu. Rev. Fluid Mech. {\bf 38,} 395 (2006).


\bibitem{hoefer14} M.A. Hoefer, Shock waves in dispersive Eulerian fluids, Journ. Nonlin.Sci. {\bf 24} (2014) 525--577.

\bibitem{scholarpedia} M. Hoefer and M. Ablowitz,  Dispersive shock waves. Scholarpedia, 4(11) 5562 (2009).


\bibitem{ha2006}  M.A. Hoefer, M.J. Ablowitz, I. Coddington, E.A. Cornell,
P. Engels, and V. Schweikhard, Dispersive and classical shock waves in Bose-Einstein condensates and gas dynamics, Phys. Rev.  A {\bf 74,} 023623 (2006).


\bibitem{jacobs-mckinney-shearer95} D. Jacobs, B. McKinney, and M. Shearer, Traveling wave solutions of the modified Korteweg -- de Vries-Burgers equation, Journ. Diff. Equations {\bf 116,}  448 -- 467 (1995).

\bibitem{jin1999} S. Jin, C. D. Levermore, and D. W. McLaughlin, The
  semiclassical limit of the defocusing NLS hierarchy, Comm. Pure
  Appl. Math., {\bf 52}, 613--654 (1999).

\bibitem{johnson70} R.S. Johnson, A non-linear equation incorporating damping and dispersion, J. Fluid Mech., {\bf 42}, 49--60 (1970).

\bibitem{johnson_book} R.S. Johnson, {\it A Modern Introduction to the Mathematical Theory of Water Waves}  (Cambridge University Press, 1997). 

\bibitem{kamch2000}  A.M. Kamchatnov, {\it Nonlinear Periodic Waves and
Their Modulations---An Introductory Course}, World Scientific (2000).
Singapore.

\bibitem{kamch2004} A.M. Kamchatnov, On Whitham theory for perturbed integrable
equations, {\it Physica} D {\bf 188,} 247--261 (2004).

\bibitem{kamch2014}  A. M. Kamchatnov, Y. V. Kartashov,1 P.-\'E. Larr\'e, and N. Pavloff,  Nonlinear polarization waves in a two-component Bose-Einstein condensate, Phys. Rev. A {\bf 89}, 033618 (2014)

\bibitem{kamch2012} A.M. Kamchatnov, Y.-H. Kuo, T.-C. Lin, T.-L. Horng, S.-C. Gou,
  R. Clift, R., G.A. El \& R.H.J. Grimshaw, Undular bore theory for the
 Gardner equation, Phys. Rev. E {\bf 86} (2012) 036605.

\bibitem{kamch2013} A.~M. Kamchatnov, Y.-H. Kuo, T.-C. Lin, T.-L. Horng, S.-C. Gou,
  R. Clift, R., G.~A. El \& R. H.~J. Grimshaw, Transcritical flow of a stratified fluid over topography: analysis of the forced Gardner
equation, Journ. Fluid Mech. {\bf 736}, 495 - 531 (2013).

\bibitem{ksk-04} A.M. Kamchatnov, A. Spire, V.V. Konotop,
On dissipationless shock waves in a discrete nonlinear Schr\"odinger equation,
J. Phys. A: Math. Gen. {\bf 37,} 5547 (2004).


\bibitem{kassam_fourth-order_2005} A.-K. Kassam and L.N. Trefethen, Fourth-Order Time-Stepping for Stiff PDEs, {SIAM J. Sci. Comput.} {\bf 26,} 1214--1233 (2005).


\bibitem{krichever1988} I.M. Krichever, Method of averaging for two-dimensional "integrable" equations, Func. Anal. Appl. {\bf 22,} 200 -- 2013 (1988).

\bibitem{klu2007} A. Kluwick, S. Scheichl, and E.A. Cox, Near-critical hydraulic flows in two-layer fluids, J. Fluid
  Mech. {\bf 575} 187 (2007).

\bibitem{kruzkov}
S.N. Kruzhkov.
\newblock First order quasilinear equations in several independent variables.
\newblock {\em Math. USSR Sb.}, 10(2):217--243, 1970.

\bibitem{kod2008} Y. Kodama, V.U. Pierce and F.-R. Tian, On the Whitham equations for the defocusing complex modified KdV equation, SIAM J. Math. Anal. {\bf 40} 1750 (2008).

\bibitem{kudashev1991} V.R. Kudashev, ``Wave-number conservation'' and succession of symmetries during a Whitham averaging, JETP Lett. {\bf 54} (1991) 175 - 178

\bibitem{lax1}
P.~D. Lax.
\newblock Hyperbolic systems of conservation laws II.
\newblock {\em Comm. Pure   Appl. Math.}, {\bf 10} (1957), 537--566.

\bibitem{lax2} P.D. Lax,   \textit{Hyperbolic Systems of Conservation Laws
and the Mathematical Theory of Shock Waves.} Conference Board of
the Mathematical Sciences, Regional Conference Series in Applied
Mathematics (SIAM, Philadelphia, 1973).


\bibitem{lax_levermore}  P.D. Lax \& C.D. Levermore,  The small dispersion limit of
the Korteweg -- de Vries equation I, II, III. Comm. Pure
Appl. Math.  {\bf 36} (1983) 253--290, 571--593, 809--829.

\bibitem{leach2012} J.A. Leach, An initial-value problem for the defocusing modified Korteweg -- de Vries equation, Journ. Diff. Eq. {\bf 252} 1032 (2012).

\bibitem{leach-needham2008} J.A. Leach and D.J. Needham, The large-time development of the solution to an initial-value problem for the Korteweg de Vries equation: I. Initial data has a discontinuous expansive step, Nonlinearity, {\bf 21}, 2391--2408 (2008).

\bibitem{LeFloch} P.G. LeFloch, {\it Hyperbolic systems of conservation laws} (Birkhauser, 2002)


\bibitem{shearer_lefloch}  P.G. LeFloch and M. Shearer, Nonclassical Riemann solvers with nucleation,   Proc. Roy Soc Edinburgh., {\bf 134A}, 961--984 (2004). 

\bibitem{lefloch_mishra} P.G. LeFloch and S. Mishra, Numerical methods
  with controlled dissipation for small-scale dependent shocks, Acta
  Numerica, {\bf 23}, 743--816 (2014).

\bibitem{levermore88} C.D. Levermore,  The hyperbolic nature of the zero dispersion
KdV limit. \textit{Comm. Partial Differential Equations } {\bf 13}, 495--514 (1988). 

\bibitem{conduit} N.K. Lowman and M.A. Hoefer, Dispersive shock waves in viscously deformable media, Journ. Fluid Mech. {\bf 718}, 524--557 (2013).

 \bibitem{majda1} A.J. Majda {\em The Existence of Multi-dimensional Shock Fronts.}  Memoirs Amer. Math. Soc.
    vol. 281,    1983.
 
 \bibitem{majda2} A.J. Majda {\em The Stability of Multi-dimensional Shock Fronts.}  Memoirs Amer. Math. Soc.
    vol. 275,    1982.

\bibitem{march2008} T.R. Marchant,
Undular bores and the initial-boundary value problem for the modified Korteweg-de Vries equation,
Wave Motion, {\bf 45,} 540 (2008).

\bibitem{miura} R.M. Miura, Korteweg -- de Vries equation and generalizations. I. A remarkable explicit non-linear transformation, J. Math. Phys. {\bf 9} 1202 - 1204 (1968).

\bibitem{MG} S. Myint and R.H.J. Grimshaw, The modulation of nonlinear
periodic wavetrains by dissipative terms in the Korteweg -- de Vries
equation,  Wave Motion, {\bf 22,} 215--238 (1995).

\bibitem{ts84} 
S.P. Novikov, S.V. Manakov, L.P. Pitaevskii, and V.E. Zakharov, 
 {\it The Theory of Solitons: The Inverse Scattering Method}.
Consultants (1984)  New York.

\bibitem{Oleinik1} O. Oleinik,  On the uniqueness of the generalized solution of the Cauchy
problem for a nonlinear system of equations occurring in mechanics, (in Rus-
sian), Usp. Mat. Nauk (N.S.) {\bf 12} (1957)169Ð176.

\bibitem{Oleinik2} O. Oleinik,  Discontinuous solutions of nonlinear differential equations,
Amer. Math. Soc. Transl. Ser. {\bf 26} (1963) 95Ð172.

\bibitem{pavlov1987} M. V. Pavlov, Nonlinear Schr\"{o}dinger equation
  and the Bogolyubov-Whitham method of averaging, Teor. Mat. Fiz.,
  {\bf 71}, 351--356 (1987).

\bibitem{pav95} M.V. Pavlov, Double Lagrangian representation of KdV and the general solution of the Whitham equations,
Russian Acad. Sci. Dokl. Math. {\bf 50,} 400 (1995).

\bibitem{pierce_tian2006} V.U. Pierce and F.-R. Tian, Self-similar solutions of the non-strictly hyperbolic Whitham equations, Comm. Math. Sci. {\bf 4} 799 (2006)

\bibitem{rothenberg1989}  J.E. Rothenberg and D. Grischkowsky, Observation of the formation of an optical intensity shock and wave breaking in the nonlinear propagation of pulses in optical fibers, Phys. Rev. Lett., {\bf 62}, 531--534 (1989).

\bibitem{ruderman2008} M. Ruderman, T. Talipova, E. Pelinovskii, Dynamics of modulationally unstable
ion-acoustic wavepackets in plasmas with negative ions, J. Plasma Physics, {\bf 74} (2008) 639--656

\bibitem{sagdeev62} R.Z. Sagdeev,  The fine structure of a shock-wave front propagated across a magnetic field in a
rarefied plasma, Sov. Phys. Tech. Phys. {\bf 6}  867 (1962).

% \bibitem{moiseev_sagdeev63} S.S. Moiseev and R.Z. Sagdeev, Collisionless shock waves in a plasma, (Journ. Nuclear Energy Part C) {\bf 5} 43 (1963)

\bibitem{sagdeev66} R.Z. Sagdeev, Cooperative phenomena and shock waves in collisionless plasmas, in {\it Reviews of Plasma Physics,   ed. M.A. Leontovich} Consultants Bureau, New York {\bf 4} 23 (1966) 

\bibitem{schaeffer1} D.G. Schaeffer and M. Shearer, Riemann problems for nonstrictly hyperbolic $ 2\times 2$ systems of conservation laws. Trans. Amer. Math. Soc. 304 (1987), 267--306.

\bibitem{schulze-shearer99} M.R. Schulze and M. Shearer, Undercompressive shocks for a system of hyperbolic conservation laws with cubic nonlinearity, Journ. Math. Anal. Appl. {\bf 229}, 344 -- 362 (1999).

\bibitem{smyth_holloway} N.F. Smyth and P. E. Holloway, Hydraulic jump and undular bore formation on a shelf break, J. Phys. Oceanog., {\bf18}, 947--962 (1988).

 \bibitem{spayd}   K. Spayd and M. Shearer, The Buckley-Leverett equation with dynamic capillary pressure.
 SIAM J. Appl. Math.,  {\bf 71}  1088--1108 (2011).
%\bibitem{sagdeev79} The 1976 Oppenheimer lectures: Critical problems in plasma astrophysics. II. Singular layers and reconnection, Rev. Mod. Phys. {\bf 51} 11 (1979)


\bibitem{trefethen_spectral_2000} L.N. Trefethen, {\it Spectral Methods in Matlab}, SIAM,  Philadelphia (2000).

\bibitem{tian93} F.-R. Tian, Oscillations of the zero dispersion limit of the Korteweg-de Vries equation, Comm. Pure Appl. Math., {\bf 46},  1093-1129 (1993).


\bibitem{venakides1985} S. Venakides, The zero-dispersion limit of the
  Korteweg-de Vries equation with non-trivial reflection coefficient,
  Comm. Pure Appl. Math., {\bf 38}, 125--155 (1985).

\bibitem{wan_etal_2007} W. Wan, S. Jia, and J. W. Fleischer, Dispersive superfluid-like shock waves in nonlinear optics, Nat. Phys., {\bf 3},  46--51 (2007).

\bibitem{watanabe84} S. Watanabe, Ion acoustic soliton in plasma with negative ion. J. Phys. Soc. Japan {\bf 53},  950 - 956 (1984).

\bibitem{wh65}
G.B. Whitham,  Non-linear dispersive waves,  Proc. Roy. Soc., {\bf A283},   238--291 (1965).

\bibitem{wh74} G.B. Whitham, {\it Linear and Nonlinear Waves}, Wiley-Interscience, New York (1974).



%\bibitem{dm-2011} E. Demler, A. Maltsev,
%Semiclassical solitons in strongly correlated systems of ultracold bosonic atoms in optical lattices,
%Ann. Phys. {\bf 326,} 1775 (2011).


\bibitem{wu} C.C. Wu,  
 Magnetohydrodynamic Riemann problem and the structure of the magnetic reconnection layer,
 J. of Geophys. Res.: Space Physics, {\bf 100},
 5579--5598 (1995).

 


\end{thebibliography}
\end{document}